\documentclass[aps,superscriptaddress,amsmath,amssymb,eqsecnum,nofootinbib,11pt]{revtex4}

\usepackage{amsmath,amssymb}
\usepackage{url,amsfonts}
\usepackage{mathrsfs}
\usepackage{graphicx,bm}
\usepackage{epstopdf}
\usepackage{ulem}
\usepackage[usenames]{color}
\usepackage{float}
\usepackage{multirow}

\allowdisplaybreaks

\newcommand{\sbar}[1]{\ooalign{\hfil/\hfil\crcr$#1$}}

\def\Tr{\rm Tr}

\def\L{{\cal L}}
\def\la{\langle}\def\ra{\rangle}
\def\be{\begin{eqnarray}}\def\ee{\end{eqnarray}}
\def\lsim{\mathrel{\rlap{\lower3pt\hbox{\hskip1pt$\sim$}}
     \raise1pt\hbox{$<$}}} 
\def\gsim{\mathrel{\rlap{\lower3pt\hbox{\hskip1pt$\sim$}}
     \raise1pt\hbox{$>$}}} 
\def\le{ \begin{array}{ll}}\def\re{\end{array}}

\def\lear{ \left( \begin{array}{cc}}\def\rear{\end{array} \right)}

\def\le{ \left( \begin{array}{cc}}\def\re{\end{array} \right)}

\def\bi{\bibitem}

\def\del{\partial}

\def\Z{{\cal Z}}
\def\bchi{\bar{\chi}}

\begin{document}

\title{Towards the Hadron--Quark Continuity Via a Topology Change
\\
in Compact Stars}
\author{Yong-Liang Ma}
\email{yongliangma@jlu.edu.cn}
\affiliation{Center for Theoretical Physics and College of Physics, Jilin University, Changchun, 130012, China}

\author{Mannque Rho}
\email{mannque.rho@cea.fr}
\affiliation{Universit\'e Paris-Saclay, CNRS, CEA, Institut de Physique Th\'eorique, 91191, Gif-sur-Yvette, France }

\date{\today}



\begin{abstract}
We construct a generalized effective field theory approach to dense compact-star matter that exploits the Cheshire Cat Principle for hadron-quark continuity at high density, adhering only to hadronic degrees of freedom, hidden topology and  hidden symmetries of QCD. No Landau--Ginzburg--Wilsonian-type phase transition is involved in the range of densities involved. The microscopic degrees of freedom of QCD, i.e., quarks and gluons, possibly intervening at high baryonic density are traded in for fractionalized topological objects.  Essential in the description are symmetries invisible in QCD in the matter-free vacuum: Scale symmetry, flavor local symmetry and parity-doubling. The partial emergence of scale symmetry is signaled by a dilatonic scalar in a ``pseudo-conformal" structure. Flavor gauge symmetry manifests with the $\rho$ meson mass going toward a Wilsonian RG fixed point identified with  the ``vector manifestation fixed point (VMFP)" at which the flavor gauge boson mass goes to zero. Parity doubling is to take place as the quasi-nucleon mass converges to the chiral invariant $m_0$. The theory with a few controllable  parameters accounts satisfactorily  for {\it all} known properties of normal nuclear matter and makes certain predictions that are drastically different from what is available in the literature. In particular, it provides a  topological mechanism, argued to be robust, for the cross-over from soft-to-hard equation of state that predicts the star properties in overall agreement with the presently available data, including the maximum star mass $M_{max}\sim 2.3 M_\odot$ and the recent LIGO/Virgo gravity-wave data. What is most glaringly different from {\it all other} approaches known, however, is the prediction for the {\it rapid convergence} to a sound velocity of star  $v_s^2\approx 1/3$ {(in unit $c=1$)} at a density $n\gsim 3 n_0$, far from the asymptotic density $\gsim 50n_0$ expected in perturbative QCD. We interpret this to signal the onset of {\it albeit} approximate conformal symmetry in dense compact-star matter. We argue  that models that properly implement quark degrees of freedom at high densities in the sense of the hadron-quark continuity  should, with the parameters fine-tuned, arrive at a  qualitatively similar  pseudo-conformal structure. The model developed in this paper, if validated by observations, could bring out a new paradigm in nuclear/hadron physics, exploiting ideas ubiquitous in various areas of physics,  condensed matter physics, nuclear and particle physics and astrophysics.

\end{abstract}

\maketitle

\allowdisplaybreaks
\newpage
\tableofcontents



\section{Introduction}

\subsection{Objective}
There have been some remarkable developments in nuclear astrophysics with accurate measurements of maximum mass of compact stars~\cite{Demorest:2010bx,Antoniadis:2013pzd,Cromartie:2019kug}  and the observation of gravity waves from coalescing neutron stars~\cite{TheLIGOScientific:2017qsa,Abbott:2018exr}. This is of course a great event for astrophysical science, but perhaps more significantly it has strong impacts on the most fundamental issue of nuclear physics that has defied theorists since decades in their efforts to understand the state of matter under extreme conditions, namely at high density. It also links ``matter in heaven to matter on earth"~\cite{heaven-earth}.

There are currently two main thrusts in research on this matter from the point of view of nuclear astrophysicists. One is to build the models,  phenomenological or effective field theoretic, that purport to explain the data provided by  astrophysical observations complemented by terrestrial experiments~\cite{Lattimer-Prakash,Holt-Rho-Weise,Drews-Weise}. The other is to decipher, in terms of a framework anchored on a precisely defined theory, new physics from  on-going observations. The former exploits, to construct models, the constraints already available at low density, typically at normal nuclear matter density $n\sim n_0\approx 0.16$ fm$^{-3}$, from theory and experiments, such as the well-established nuclear structure information, i.e., normal nuclear matter properties, and whatever experimental information available at higher densities. The models so constructed inevitably contain a number of parameters to be adjusted so as to accommodate on-coming more precise data. The latter, much less successful in confronting Nature, but aiming at uncovering hitherto unexplored aspects of the strongly-interacting state of matter, is to exploit ubiquitous techniques developed in all areas of physics, not limited to strong interactions and particle physics, but also encompassing condensed matter and chemical physics.

 In this review, we focus on the second line of development resorting to a framework which is as unified as feasible that espouses effective field theories of strong interactions. The spirit here is quite different from the main stream of activities in this field as our formalism resorts to a variety of ideas coming from different fields of physics with overlapping concepts, i.e., condensed matter, particle and nuclear, up to date largely unexplored in the nuclear and astro-nuclear physics community. The objective of this review is to bring them all together in a unified framework and confront Nature with it and gauge how it fares. What we have obtained at high baryonic densities constitutes the predictions of the theory, eschewing mere fitting or limiting to post-dictions. The advantage of the approach is that it involves relatively  small number of arbitrary parameters, and they are more or less constrained by the premise of the theory.  One cannot obviously expect such an approach to enjoy as good an agreement with available experiment data as do phenomenological approaches equipped with a number of adjustable parameters.

The extreme condition we are zero-ing in here is high density well beyond the normal nuclear matter density $n_0$ but far below what we shall identify as the asymptotic density, $\gsim 50 n_0$. The relevant density regime in massive compact stars that we shall focus on is ${\sim} (2-7) n_0$. Up to the normal nuclear matter density, the strong interactions governing nuclear interactions are fairly well studied both phenomenologically and in effective field theory based on chiral dynamics, aided by experiments up to the  lab. energy $\sim 350$ MeV. But beyond that density, up to date, there is no model-independent theoretical tool known that can be trusted.  Lattice QCD, the only nonperturbative tool known for QCD, is currently moot at high density because of the sign problem and the perturbative QCD approach is inapplicable at the relevant density.

In the absence of controllable/reliable theoretical tools that can access the whole range of densities involved, the strategy currently employed by nuclear theorists is what is generically called  ``energy density functional approach" (EDF approach for short)\footnote{A most up-to-date conceptual discussion on EDF is found in \cite{EDF-furnstahl}.}. In  this class belong, among others,  ``relativistic mean field (RMF) theory," ``Skyrme potential approach,"  ``chiral effective field theory ($\chi$EFT)" and others --- including our approach we will describe below. As it stands, EDF is considered to be generally successful for treating nuclear dynamics up to the density of normal nuclear matter $n_0$ or in some cases, slightly above. There is an enormous number of literature on this, with some schemes more successful than others depending on the process and the number of adjustable parameters available {(see, e.g., Refs.~\cite{Drews-Weise,EDF-furnstahl,weise} and references therein)}. { Instead of going into details of or comparison with all the approaches available in the literature,  we pick in what follows  the $\chi$EFT approach to bring out our main points.} It will also be appropriate in elucidating what distinguishes our approach from other EDF approaches.

What will be referred to as ``standard $\chi$EFT" (S$\chi$EFT for short)~\cite{weise} -- to be distinguished from our approach to be developed here -- takes only the nucleon and the pion as the relevant degrees of freedom in nuclear dynamics and organizes them in setting up the chiral (power) counting series. Being an EFT with pions only (apart from the nucleons), the theory is defined with the cutoff set typically at $\Lambda\sim (400-500)$ MeV. The rational for this scale  is that the experimental data are available to $\sim 350$ MeV and hence resonances above that energy scale, e.g., the vector-meson  channels $\rho$ and $\omega$ and the possible scalar\footnote{This will be identified later as a Nambu-Goldstone mode, ``dilaton," of  scale symmetry.}  $\sigma$ (or alternatively $\chi$ which will be used mostly in this review), are to be integrated out from the EFT Lagrangian. The power counting is currently made up to next--next--next-to-the-leading order (``N$^3$LO") or in some cases, albeit partially, up to N$^4$LO. In this S$\chi$EFT approach,  three-nucleon potentials figuring at N$^2$LO and N$^3$LO play a crucially important role not only for nuclear matter stability but also for finite nuclear structure properties. In the approach we will develop, some of the higher-order terms can be absorbed into lower-order terms in modified counting schemes.  A very  illuminating case is the description of the highly suppressed Gamow--Teller transition in the C-14 dating~\cite{C14}. Here what corresponds to the short-range three-body potential, an  N$^2$LO effect, can be mostly, if not entirely, incorporated into the coefficient of an NLO two-body potential. A similar situation occurs in the scalar channel where higher-chiral order terms can be captured in lower-order terms involving a dilaton scalar in a scale-chiral symmetric scheme.

It is safe to say that generally nuclear structure is ``well" reproduced in the N$^3$LO treatments in S$\chi$EFT up to $\sim n_0$.

There are, however,  several reasons to believe that this expansion must break down as density increases beyond $n_0$.

One reason is that nuclear matter at saturation density can be identified as the Landau Fermi liquid at its fixed point~\cite{shankar} with $1/\bar{N} \ll 1$ --- where $\bar{N}=k_F/(\tilde{\Lambda}-k_F)$ with $\tilde{\Lambda}$ the cutoff on top of the Fermi sea. This means that the  $k_F$-power expansion made in S$\chi$EFT, possibly valid near $n_0$, must effectively go over to an $1/k_F$ expansion at some density above $n_0$.

Another reason, which is more crucial for compact-star physics, is that at some higher density, say,  $\gsim 2n_0$, quarks,  triggered by increasing {density}, could start percolating between overlapping nucleons, thereby changing the state of matter. This means that new degrees of freedom, not present in S$\chi$EFT, must enter in the game.

In this review, we implement both features mentioned above by introducing at densities exceeding $n_0$: (I) Two symmetries not visible in QCD in the vacuum, referred herewith to as ``hidden symmetries," and (II) a topology change signaling the emergence of new degrees of freedom. As for (I), the hidden gauge (or local) symmetry  associated with the vector mesons $\rho$ and $\omega$ -- and possibly the infinite towers as in holographic QCD -- as well as the hidden scale symmetry associated with the dilaton $\sigma$ ($\chi$). As for (II), there takes place a topology change in the baryonic matter that encodes the putative hadron-quark continuity considered to be present in QCD at densities $\gsim 2n_0$. The topology involved is not ``visible" in QCD proper, hence the topology change is a hidden process in dense medium. In the absence of direct nonperturbative access to QCD, the only tool available -- and justifiable for low-energy processes -- is anchored on the notion of effective field theory for the strong interactions, which is best expressed in general -- but more appropriately for nuclear physics -- by Weinberg in his Folk Theorem~\cite{weinberg}. Our strategy is to follow the line of the Folk Theorem applied to nuclear physics implementing the elements (I) and (II) in addition to what is the basis for S$\chi$EFT. We may refer to this approach as a generalized nuclear EFT (``$Gn$EFT" for short). For the reason explained in detail below, our formalism for  $Gn$EFT will be based on a Lagrangian called $bs$HLS where $b$ stands for baryon, $s$ for scalar (dilaton) and HLS for hidden local symmetry.

It should be stressed that the topology change we are proposing as a trade-in for the hadron-quark continuity does not involve phase transitions per se in the sense of the Landau--Ginzburg--Wilsonian paradigm but involves a drastic modification in the equation of state (EoS) of the baryonic matter at high density.  Therefore we eschew the common practice of adhering to low-density properties of the EoS for, say, $n\sim n_0$, as constraints for high density EoS for $n\sim (5-7)n_0$. In our approach, such constraints have no meaning.

\subsection{Compact star properties}\label{compact-star-bounds}
In confronting the predictions made by our approach with Nature, we will avoid dwelling too much on comparing in detail with available data, both from terrestrial and space laboratories. There are a number of works dealing with such statistical approach as Bayesian and machine-language etc. establishing correlations between various observables. Our focus will be mainly on what could be considered as {\it robust} features of the measurements and to assess how our predictions fare with the available bounds and constraints. It suffices for our purpose to  make in this section a brief summary of the possible constraints imposed by nuclear and astronomical measurements presently available with which our predictions should be compatible.

Among various constraints, we consider those constraints arrived at by Bayesian approaches using nuclear data, inferred masses, various properties of neutron stars along the line of Ref.~\cite{Miller-Chirenti-Lamb}. In particular, the relevant quantities are  nuclear symmetry energy, maximum mass, tidal deformability, radius measurements etc.

We must say there are some serious differences among the workers in the field. For example, one of the most important quantity, generally agreed in the field, is the symmetry energy. However there are some arguments that it could constrain high density properties~\cite{BAL} while some argue that it constrains low-density properties but not high density~\cite{Miller-Chirenti-Lamb}.  We will find indeed that the symmetry energy is one of the key issues in this matter, manifesting basically differently in high density from in low density.

The strategy we will adopt in this review is, while being consistent with what is available in observations at low density, both terrestrial and astronomical, to uncover possible new physics buried in dense hadronic matter that cannot be accessed by QCD proper. The presently available observational constraints at $n \gsim 2n_0$ are as follows.
\begin{itemize}
\item While the lower mass stars are more accurately given, we will simply adopt as an indication the maximum mass~\cite{Miller-Chirenti-Lamb}
\begin{subequations}
\be
M &=& 1.908\pm 0.016 M_\odot\ {\rm for}\ {\rm PSR}\ J1614-2230\mbox{~\cite{Demorest:2010bx}}
,\label{M3}\\
&=& 2.01\pm 0.04 M_\odot\ {\rm for}\ {\rm PSR}\ J0348+0432\mbox{~\cite{Antoniadis:2013pzd}},\label{M2}\\
&=& 2.17^{+0.11}_{-0.10} M_\odot\ \ {\rm for}\ {\rm PSR}\ J0740+6620\mbox{~\cite{Cromartie:2019kug}}.\label{M1}
\ee
\end{subequations}

\item From GW170817, we have the upper bound for the {\it dimensionless} tidal deformability  $\Lambda_{1.4} < 800$ for a $1.4 M_\odot$ neutron star ~\cite{Tsang}. And, with 90\% confidence, the radius $R_{1.4}$ of a neutron star of mass $1.4 M_\odot$, it is argued~\cite{R-Oezel-P},  cannot exceed $\sim 13.6$ km.  Multiwavelength analyses of the EM counterpart of GW170817 (``AT2017gfo") indicates a constraint on the mass weighted tidal deformability $\tilde{\Lambda}$ as $\tilde{\Lambda} = 300^{+500}_{-190}$ for low spin binary stars~\cite{radicetal}. Combining the two we will consider the presently available bounds as
\be
400 < \tilde{\Lambda} < 800.\label{gwbound}
\ee
There are discussions in the literature that the upper bound could be tightened to a lower value. We will consider this possibility in comparing our result with the bound (\ref{gwbound}).

\item Apart from the heavy ion data for lower densities,  the GW170817 gives pressure bounds at $2n_0$ and $6n_0$
\be
P(2n_0)&=&3.5^{+2.7}_{-1.7}\times 10^{34}{\rm dyn/cm}^2,\label{P1}\\
P(6n_0) &=& 9.0^{+7.9}_{-2.6}\times 10^{34}{\rm dyn/cm}^2.\label{P2}
\ee
The pressure bound at $2n_0$ is of little use. Practically every reasonable EoS consistent with nature at $n_0$ would satisfy it. That at $6n_0$ will be found to be of relevance.
\end{itemize}
\subsection{A brief summary of principal results}\label{summary}
Here we give an overview of  the main results obtained in the work reviewed here. They are accompanied by ``Propositions" because the chain of arguments developed and the consequences therefrom require further considerations to be confirmed.

The principal actor  in our work is, conceptually,  the Cheshire Cat principle. It is applied, using a chain of reasoning analogous to what one does in condensed matter, to highly dense matter relevant to the interior of compact stars, which is inaccessible at present by non-perturbative QCD. How the Cheshire Cat enters in the problem is via a skyrmion crystal simulation exploiting hidden symmetries of QCD, namely, hidden gauge symmetry and hidden scale symmetry.

It is found to bare a cusp singularity in the equation of state of dense matter due to an interplay of the dilaton of scale symmetry and the vector mesons of hidden gauge symmetry, and the hitherto undiscovered role of nuclear tensor force, leading to the emergence of parity doubling and quasiparticle degrees of freedom at high density in the form of fractionalized skyrmions.

What transpires is that topology in hadronic variables neatly captures the physics of hadron-quark continuity in QCD variables, what one might call ``duality" in nuclear physics, with {{\it no}} phase transitions involved.\footnote{Similar in spirit to our approach aiming at the hadron-quark continuity is one anchored on the gauge-gravity dual (holographic) Sakai--Sugimoto model which exploits instanton interactions~\cite{SS-model} with different results, including phase transitions. There are quite a few approaches in the literature that hybridize hadronic models at low density and quark models at high density, with inevitable phase transitions and consequently different predictions~\cite{alford}. It is difficult to make comparisons with them, so we do not discuss them in this review.} The range of densities involved for the former match that for the latter in the range where strongly-coupled quarks figure, say, $\sim (2-7)$ times normal nuclear matter density.

One of the novel predictions of the theory that begs to be confirmed or refuted is the precocious emergence of partial conformal symmetry in compact stars at a density $\gsim 3 n_0$ with the sound velocity of the star converging to the conformal value $v_s^2=1/3$, a feature not shared by any other theories or models in the field.

An interesting spin-off of the work is a possible link between the cusp in the nuclear symmetry energy  (at high density) and the possible renormalization-group invariance of the nuclear tensor force in the monopole matrix element in the structure of exotic nuclei (at low density). This could be checked in  RIB (rare-ion-beam) experiments\footnote{Just to give a few examples where such experiments could be done, HIAF (High Intensity Heavy-ion Accelerator Facility) in Huizhou, China~\cite{HIAF} and RAON in Daejeon, Korea~\cite{RAON}}.

\section{Hidden Symmetries of QCD}\label{hidden-symmetries}
To construct a  $Gn$EFT, the first ingredient is the cut-off scale involved. Once the cutoff scale is defined, one can then specify the relevant degrees of freedom.

What governs nuclear dynamics at low energies is the chiral symmetry reflecting the small masses for the up and down (and strange if needed)  quarks. The $Gn$EFT relies on the chiral (power) expansion. Both scale symmetry and chiral symmetries are to enter together in the power expansion. In this section, we limit to the chiral expansion, with the scale expansion being brought in later.   The coefficients of the expansion, referred to as low-energy constants, are to be fixed for a {\it given scale} mainly by experiments. What sets the scale of the cutoff is then the energy/momentum to which {\it precise} experimental data are available. In nuclear processes, the scale is set by the lab momentum $p\sim 350$ MeV. Therefore the typical value for the cutoff is
\be
\Lambda_{{\rm S}\chi{\rm EFT}}\sim (400-500)\ {\rm MeV}.
\ee
This requires that excitations above the cutoff scale be integrated out of the $Gn$EFT Lagrangian. This means that the vector mesons $\rho$ and $\omega$ as well as the scalar $\chi$ need not figure {\it explicitly} in the effective Lagrangian. This leads to the S$\chi$EFT Lagrangian used predominantly in nuclear physics community. Here, apart from the nucleon necessary for nuclear dynamics, the only relevant degree of freedom is the pion\footnote{Skyrmions generated as solitons in mesonic Lagrangian can also -- and will later -- figure as a relevant degree of freedom. For the moment we put the nucleon as an explicit degree of freedom in the EFT.}. The nucleon mass $\sim 1$ GeV is much greater than the cutoff, but what is involved in low-energy nuclear processes is ``soft" and hence chiral perturbation expansion could make sense. Indeed it does in certain nuclear process and in some cases very accurately if soft-pion effects dominate~\cite{CNDIII}. Nuclear dynamics, both in infinite matter and in finite nuclei, in those channels that are integrated out, i.e., vector and scalar, can and do appear at higher orders in chiral expansion via loop corrections and higher derivative terms. At low density, up to, say, $n_0$,  the expansion involving the Fermi momentum $k_F$ is considered to be fairly successful, a beautiful support for the Folk Theorem. However at higher densities, as stated above, there are reasons to believe that the straightforward extrapolation in the chiral series, presently feasible in practice up to N$^4$LO, is questionable. We explain why this is so and suggest how to go about resolving the problem.

First  we introduce the notion of hidden symmetries of QCD, which is well-known to the particle physics community but may be foreign to the nuclear community.
\subsection{Local flavor symmetry}
As will be seen,  the $\rho$ meson  plays an extremely important role in our formalism in compact-star structure. In fact  how it figures at high density is one of the key points in this review.  The point is that QCD has no local flavor symmetry and hence if it were to appear as a local gauge field, it could appear {\it only} as an {\it emergent} degree of freedom.

Consider the two-flavor chiral symmetry $SU(2)_L\times SU(2)_R$. We consider two flavors for the time being. Later in consideration of scale symmetry  we need to extend it to three flavors including the strangeness. Now the chiral field $U=e^{2i\pi/f_\pi}$ transforming as $g_LU(x)g_R^\dagger$ under chiral symmetry can be written as a product of L and R fields
\begin{eqnarray}
U(x) & = & \xi_L^\dagger(x)  \xi_R(x).
\end{eqnarray}
This has a redundancy, obvious when sandwiched with $h(x)h^\dagger=1$. This redundancy can be elevated to a gauge symmetry by introducing a local gauge field $V_\mu=(\rho_\mu,\omega_\mu)$ through the covariant derivative $D_\mu \xi_{L,R} = (\partial_\mu - i V_\mu)\xi_{L,R}$. As long as the field $V_\mu$ does not propagate, there is no new physics in this ``gauging." It is just a redundancy. However it can happen to become dynamical due to some strong correlations. In fact such gauge fields are generated in condensed matter physics and play a crucial role in such phenomena as deconfined quantum critical phenomena, fractional quantized Hall effects and many other phenomena. Those are ``emergent fields" generated by strong correlations in electrons. That this can happen also in strong interactions can be shown using a Grassmannian action~\cite{yamawaki-grassman}.

The resulting chiral Lagrangian, i.e., non-linear sigma model, suitably gauged with the kinetic energy term, is~\cite{HY-VM,HY:PR}
\begin{eqnarray}
{\cal L}_{\rm HLS} & = & f_\pi^2 {\rm Tr}\left[\hat{\alpha}_{\perp \mu}\hat{\alpha}_{\perp}^\mu\right] + a f_\pi^2 {\rm Tr}\left[\hat{\alpha}_{\parallel \mu}\hat{\alpha}_{\parallel}^\mu\right] - \frac{1}{2g_V^2}{\rm Tr}\left[V_{\mu\nu}V^{\mu\nu}\right] +\cdots\label{HLS}
\end{eqnarray}
written in terms of the Maurer-Cartan 1-forms which transform covariantly,
\begin{eqnarray}
\hat{\alpha}_{\perp \mu} & = & \frac{1}{2i}\left(D_\mu \xi_R\cdot \xi_R^\dagger - D_\mu \xi_L\cdot \xi_L^\dagger\right) , \nonumber\\
\hat{\alpha}_{\parallel \mu} & = & \frac{1}{2i}\left(D_\mu \xi_R\cdot \xi_R^\dagger + D_\mu \xi_L\cdot \xi_L^\dagger\right) . \label{MC1form}
\end{eqnarray}
The Lagrangian is given to the leading ($O(p^2)$) order with the ellipsis standing for higher order terms, which are easy to write down in terms of the covariant 1-forms. In reality there are also chiral-symmetry-breaking terms that we will not write down explicitly. Baryons can also be suitably coupled in, which we will do later. For the moment we continue with the leading hidden local symmetry Lagrangian (\ref{HLS}).

The hidden gauge coupling $g_V$ stands for $V=(\rho, \omega)\in U(2)$  if not stated otherwise. Later on we will distinguish them as $g_\rho$ and $g_{\omega}$ because at high density the $U(2)$ symmetry breaks down to $SU(2)\times U(1)$.

There are two remarkable aspects of the Lagrangian (\ref{HLS}) that figure crucially in what follows.

One is that limited to the leading order in the power counting, it captures extremely well certain strong interaction dynamics even at tree order. For instance it encodes vector dominance and the vector mass formula $m_V^2 = a g_V^2 f_\pi^2$  that capture Nature very closely~\cite{komargodskiHLS}. In fact the KSRF mass formula holds to all orders of loop corrections~\cite{HKY-KSRF}. Also surprisingly chiral perturbation series with the vector mass taken on the same footing as the pion mass, works even at the next-to-leading order power counting~\cite{HY:PR}.

The other is that treated at one-loop order in Wilsonian renormalization group, one unearths what is called ``vector manifestation (VM) fixed point" where the hidden gauge coupling  goes to zero.   Though not verified explicitly, it is considered  valid to higher orders. It has been shown that this fixed point is arrived at when the quark condensate $\Sigma\equiv\la\bar{q}q\ra$ is driven to zero~\cite{HY:PR}. We assume it is the density that does the driving, and as we will see the critical density, known neither theoretically nor experimentally, comes out in our formalism to be at $n_{\rm vm}\gsim 25n_0$. What this means is that via the KSRF relation $m_V^2\sim f_\pi^2 g_V^2$ which holds to all loop orders with (\ref{HLS}), the vector mass will drop to zero as $\Sigma\to 0$. Since higher-power corrections to (\ref{HLS}) go as $O(m_V^2/\Lambda_\chi)$ where $\Lambda_\chi$ is the chiral symmetry scale $\sim 1$ GeV and the vector mass drops rapidly, the argument holds better as the VM fixed point is approached. Note that the vector meson mass can go to zero independently of what $f_\pi$ is. Hence at high density, the in-medium mass does not scale with the in-medium pion decay constant. {\it It is  not the pion decay constant, as often assumed, but it is the gauge coupling that drives the vector mass drop to zero at high density.} This prediction, highly robust,   will be found {\it crucial} for the role of hidden gauge fields in compact-star matter.

\subsubsection{Inevitability of composite gauge field and the vector manifestation fixed point}
The presence of the VM fixed point  has a highly important implication on S$\chi$EFT that contains pion fields only.  This aspect is not widely recognized by the aficionados of S$\chi$EFT. According to the Suzuki theorem~\cite{suzuki}  which states ``when gauge-invariant local field theory is written in terms of matter fields alone, a composite gauge boson or bosons must be formed dynamically."   {\it What this means in S$\chi$EFT is that  gauge boson or bosons should  \underline{inevitably} be formed dynamically from pion fields if and only if there were vector mesons whose mass is driven to zero -- in the chiral limit -- by density or temperature.} The VM scenario is the specific way the zero mass boson could be bared exposing the hidden gauge symmetry. Whether this excitation reflecting the composite local gauge symmetry exists or not in nature is not yet settled by experiments\footnote{There seems to be consensus among heavy-ion physicists based on theoretical considerations in phenomenological models or along the line of S$\chi$EFT that the dilepton data in the NA60 experiment rules out the $\rho$ mass going to zero at the chiral restoration temperature. This conclusion is premature as explained in Ref.~\cite{dilepton-fiasco}. That applies also to high density.}. What we will do in what follows is to show that  this feature encoded in the VM is crucial for certain novel observations (such as ``pseudo-conformal" sound speed to be described below)  in compact stars. The important implication is that should there be a signal toward the vector manifestation fixed point, regardless of where the fixed point lies, then thanks to the Suzuki theorem, there must exist a local gauge field that emerges from strong correlations among the composites. In the case of the $\rho$ meson, it would be interactions among pions that would generate the gauge particle. This could be one of the most important ``new physics" inputs brought by nuclear physics.

$\bullet$ {\bf Proposition I: \it Hidden local symmetry can emerge in nuclear dynamics with the vector meson mass driven to zero at the vector manifestation fixed point  by high density.}
\subsubsection{Baryonic HLS}
As mentioned,  baryons can be brought into the EFTs in two ways: they can be generated as solitons in mesonic theories or put in by hand as matter fields. The former will be seen to play an important role in bringing in topological structure to hadronic interactions.  For addressing many-nucleon processes, it is more convenient to have explicit baryonic fields.

Given baryon fields, it is straightforward to couple them hidden-local invariantly to the meson fields. Take
 the baryon doublet $\psi(x)$ in the iso-space
\begin{eqnarray}
\psi(x) & = & \left(
                \begin{array}{c}
                  p \\
                  n \\
                \end{array}
              \right)
\end{eqnarray}
transforming under hidden flavor symmetry as
\begin{eqnarray}
\psi(x) & \to & h(x)\psi(x).
\end{eqnarray}
Then, using  the Maurer-Cartan 1-forms \eqref{MC1form},  we have
\begin{eqnarray}
{\cal L}_{\rm bHLS} & = & \bar{\psi}\left( iD \hspace{-0.25cm}\slash ~ - m_N + \frac{g_A}{2} \gamma^\mu \gamma_5 \hat{\alpha}_{\perp \mu} + \frac{g_V}{2}\gamma^\mu\hat{\alpha}_{\parallel \mu}  \right)\psi ,
\label{bHLS}
\end{eqnarray}
with  the covariant derivative $D_\mu \psi = (\partial_\mu - i V_\mu)\psi$.  We have written only the leading-order terms. It is straightforward to write higher-order terms.
\subsection{Scale symmetry}
In the particle data booklet, a low-lying scalar is listed as $f_0(500)$. In nuclear physics, a scalar of mass around $\sim 600$ MeV is introduced both for deriving nuclear force and for doing relativistic mean-field calculations. Although a scalar of that mass is observed with a huge width comparable to the mass, it has been invoked as a local field and has been found to work  successfully in both finite nuclei as well  as infinite matter up to the normal nuclear matter density $n_0$. It has been widely incorporated in what is referred to as Relativistic Mean Field (RMF) Theory  and applied with some success to compact-star matter. The rationale behind for the latter is that the RMF formulation could be made equivalent to Landau Fermi liquid theory as first pointed out by Matsui~\cite{matsui}. The equivalence must, however, cease as density increases beyond $n_0$, but that a scalar must figure in nuclear dynamics is without doubt.

But what is this scalar in EFT?

In S$\chi$EFT, scalar excitations in the corresponding channel could be generated in higher loop effects in $\pi$-$\pi$ interactions, accounting for the necessary attraction for nuclear binding. But at high density, this procedure is problematic because such a scalar excitation can become infrared sensitive and turn unstable. Perturbation approach cannot access such excitations.

In our approach, we introduce a dilaton field $\sigma$ (a.k.a. $\chi$) as a Nambu-Goldstone boson arising from spontaneous breaking of scale symmetry. For this we follow Crewther and Tunstall {(CT for short)}~\cite{CT,CCT}. There  is a long-standing controversy as to whether this scheme which assumes the existence of an IR fixed point for two or three flavors we are concerned with makes sense in nuclear dynamics as discussed in Ref.~\cite{scalar-conundrum}.  It is far from settled. There seems to exist a general consensus among particle theorists that an IR fixed point in QCD (for $N_f\lsim 3$) is not tenable. There are several arguments for this conclusion, among which the most prevailing argument is that there is no evidence for NG boson of low mass of scalar comparable to that of the pion.\footnote{ This is in some sense moot since $f_0 (500)$ may be located at as low as $\sim 440$ MeV, even lower than the kaon mass $\sim 500$ MeV which is in the $SU(3)$ chiral symmetry framework. Furthermore the $\rho$ meson is treated in hidden local symmetry on the same footing as the pion which is gauge-equivalent to nonlinear sigma model of chiral symmetry.}

There are several reasons to believe, however,  that our approach does indeed make a good sense in going to dense matter and  in fact has more predictive power over the S$\chi$EFT approach. First is that we are dealing with scale symmetry arising as emergent from nuclear dynamics, and there are  indications that there are ``soft" modes in the scalar channel at high density --- that we will associate with what is referred to later as approaching  the ``dilaton limit fixed point (DLFP)." This means that at increasing density, the effective scalar mass in medium can fall with density and hence becomes an explicit degree of freedom in the space defined by the cutoff.  For instance, at the normal nuclear matter density $n_0$, the  scalar of free-space mass of $\sim 600$ MeV can fall below the cutoff $\Lambda\approx 500$ MeV, hence  should not be integrated out. Secondly it could be treated on the same footing as the pion into a scheme of scale-chiral symmetry with the scalar exchange entering at the leading tree order instead of at loop-correction orders. This allows a systematic counting rule that can be set up for systematic high-order calculations~\cite{CNDIII}.\footnote{It should be pointed out that there is an extremely subtle issue in formulating scale-chiral expansion when baryons with their mass scale higher than, say, the scalar are present~\cite{CT,CCT}.  In fact it is a lot more serious than in baryon chiral perturbation theory, which was resolved in a variety of ways. In this aspect, the presently available formulation~\cite{LMR} is incomplete and remains to be improved on. This caveat is avoided in the LOSS approximation applied in what follows. }    Furthermore there are some indications that an infrared fixed point might exist for QCD with 2 or 3 flavors~\cite{IRFP}\footnote{In this article, numerical stochastic perturbation theory is applied to the calculation of the $\beta$ function including fermionic contributions up to four loops and in the Pad\'e approximation.  The IR fixed point for $N_f=2$ is found at 4 loops and in the Pad\'e approximation with a frozen $\alpha_s$. To the best of our knowledge this calculation has not been given a subsequent support, not to mention improvement, so remains unconfirmed. There are no known {\it nonperturbative} calculations in the literature.}.
%
%
A perhaps relevant information on the possible existence of an IR fixed point in QCD comes from a recent thermal lattice calculation~\cite{alexandru} where a possible new phase of thermal QCD is observed.  Analysis of the lattice results indicates the onset of changes toward IR scale invariance in conjunction with chiral symmetry restoration. Roughly three temperatures are involved, $T_A < T_{IR} < T_{UV}$.  $T_A$ is found to be approximately  150 MeV $\lsim T_c$ where $T_c$ is the chiral crossover temperature that marks the ``onset" of IR scale invariance. In this region chiral condensate plays an important role. The IR phase proper arises at $T_{IR}\gsim 200$ MeV.  At $T_{UV} \sim 1$ GeV,  the asymptotic scale symmetry sets in.  In between $T_{IR}$ and $T_{UV}$ the IR and UV scale symmetries coexist. What may be significant is the possible zero-mass glueball excitation which may or may not be a dilaton. It is however unclear whether this observation can be given an interpretation in terms of the {CT} theory~\cite{CT}\footnote{We would like to thank Rod Crewther for a comment on this matter.}.
 \subsubsection{Scale symmetry as a hidden symmetry}
 With the above remarks taken into account, we take the point of view that scale symmetry may emerge and manifest in dense nuclear systems from strong nuclear correlations and can be associated with an IR fixed point that can be reached at high density. Whether or not such scale symmetry is in QCD in the vacuum is not crucial to the issue concerned. In fact a way to look at it is that scale symmetry is hidden in QCD like the hidden local symmetry discussed above and can be made to emerge in nuclear dynamics.

\subsubsection{Going from nonlinear sigma model to scale-symmetric model}
The aspect most relevant to our approach in nuclear processes is that scale symmetry is actually present, invisible or hidden, even in nonlinear sigma model but can be ``exposed" by dialing a parameter of the model. One can see this starting with a linear sigma model that captures standard Higgs model~\cite{hiddenscalar}.

Let us consider two extreme limits in the Lagrangian~\eqref{LagM}: strong coupling limit and weak coupling limit.
 \begin{enumerate}
\item    In the strong coupling limit, $\lambda\rightarrow \infty$, $\la\sigma\ra\rightarrow f=f_\pi$, so one simply gets the familiar non-linear sigma model
 \begin{equation}
 {\cal L}_{L\sigma M}  \mathop{\longrightarrow}^{\lambda \rightarrow \infty}  {\cal L}_{NL\sigma}  =
   \frac{f_\pi^2}{4}\cdot {\rm Tr} \left(\partial_\mu U \partial^\mu U^\dagger\right)    \,.
   \label{NL}
    \end{equation}
Note that the  breaking of the scale invariance gets shoved into the kinetic term for the pion,
which being of scale dimension 2 is no longer scale invariant. The constant $f$ is identified with the pion decay constant. The Lagrangian (\ref{NL}) is the leading term in the chiral expansion for effective field theory for nuclear physics, say, S$\chi$EFT or $bs$HLS defined precisely later, applicable in the vicinity of nuclear matter density $n_0$.

\item Now we turn to the weak coupling limit $\lambda\rightarrow 0$.
Define the scale-dimension-1 and mass-dimension-1 field $\chi$,  what is often referred in the particle physics community working on dilatonic Higgs to as ``the conformal compensator field"\footnote{In what follows we will work with  the dilaton field represented by  $\chi$, not by $\sigma$. In the CT theory which we believe is relevant to our problem, the IR fixed point is in the Nambu-Goldstone (NG) mode and not in the Wigner-Weyle (WW) mode as is the case with the situation with large $N_f$ near a conformal window. This matter is discussed in detail in Ref.~\cite{CCT}. In the LOSS approximation that we are employing, the subtle issue involved does not seem to arise but we follow the notations of Ref.~\cite{CT}.}
\be
\chi =f_\chi e^{\sigma/f_\chi}\,.
\ee
Under  scale transformation, $\chi$ transforms linearly while $\sigma$ transforms nonlinearly
 \begin{equation}
\delta \chi=(1+x^\mu \partial_\mu) \chi\,, \qquad
\delta \sigma= f_\chi+x^\mu \partial_\mu\sigma\,.
 \end{equation}
Here $f_\chi$ is the decay constant for the scalar $\sigma$. Expressed in terms of the field $\chi$, the Lagrangian (\ref{LagM}) can be written as
\be
  {\cal L}_{L\sigma M} &=& {\cal L}_{\rm sinv} - V(\chi)\,\label{L-dilaton}
\ee
with
\be
{\cal L}_{\rm sinv} &=& \frac{1}{2} \left(\partial_\mu \chi \right)^2+ \frac{f_\pi^2}{4}\left(\frac{{\chi}}{f_\chi}\right)^2\cdot {\rm Tr} \left(\partial_\mu U \partial^\mu U^\dagger\right)\,, \label{s-inv}\\
 V(\chi) &=& \frac{\lambda}{4} f_\chi^4 \left[\left(\left(\frac{\chi}{f_\chi}\right)^2 -1\right)^2-1\right]  \,,\label{potv}
 \ee
with $\frac{\del}{\del\chi} V(\chi)|_{\la\chi\ra=f_\chi}=0$.
The first term of (\ref{L-dilaton}) is scale-invariant with scale breaking lodged entirely in the potential (\ref{potv}). It is important to note that scale invariance is obtained in the limit $\lambda\rightarrow 0$ from a linear sigma model. 

Under the scale transformation, the potential transforms
\begin{equation}
\delta V(\chi) ={} +\lambda f_\chi^4 \left(\frac{\chi}{f_\chi}\right)^2
+ \textrm{total derivative}
\,,
\label{Vtrans}
\end{equation}
which yields
\begin{equation}
\partial^\mu {\bf{D}}_\mu=\theta^\mu_\mu={} - \delta V(\chi)={} - \lambda f_\chi^4 \left(\frac{\chi}{f_\chi}\right)^2\,,
\label{trace}
\end{equation}
where ${\bf D}_\mu$ is the dilatation current and $\theta^\mu_\mu$ is the trace of the energy-momentum tensor. Then this leads to the partially conserved dilatation current (PCDC) analogous to PCAC
\begin{eqnarray}
m_\chi^2 f_\chi^2 & = &{} - \langle 0| f_\chi \partial^\mu {\bf D}_\mu|\phi\rangle ={} - d_\theta \langle \theta^\mu_\mu\rangle = 2\lambda f_\chi^4 \left\langle \left(\frac{\chi}{f_\chi}\right)^2\right\rangle= 2 \lambda f_\chi^4\,,
\label{PCDC}
\end{eqnarray}
where $\theta^\mu_\mu$ has a scale dimension $d_{\theta} =2$ as one can see from Eq.~(\ref{trace}).
The potential (\ref{potv}) is the first term in a more general potential that encodes the trace anomaly,
\begin{equation}
V(\phi)\Bigg|_{\rm anomaly}= \frac{m_\chi^2 f_\chi^2}{4} \left(\frac{\chi}{f_\chi}\right)^4 \left(\ln \frac{\chi}{f_\chi} -\frac{1}{4}\right)
\label{anomaly}
\end{equation}
which yields $\langle \delta V\rangle = {} -\langle \theta^\mu_\mu\rangle= m_\chi^2 f_\chi^2\langle \chi^4\rangle/(4f_\chi^4)=m_\chi^2 f_\chi^2/4$ and  has a minimum at $\langle \chi\rangle=f_\chi$.

\end{enumerate}

$\bullet$ { \bf Proposition II: \it Baryonic matter can be driven by increasing density from Nambu-Goldstone mode in scale-chiral symmetry to the dilaton-limit fixed point  in pseudo-conformal mode.}

\subsubsection{Scale-chiral expansion}

A key point we will develop is that baryonic density does the dialing $\lambda=\infty$ to $\lambda=0$ in going from low density -- in the vicinity of finite nuclei and normal nuclear matter at density $\sim n_0$  -- to high density relevant to the interior of compact stars, approaching what is called ``dilaton-limit fixed point (DLFP)." This limit is either in the vicinity of, or coincident with, the vector manifestation fixed point.

\begin{figure}[h]
\includegraphics[scale=0.4]{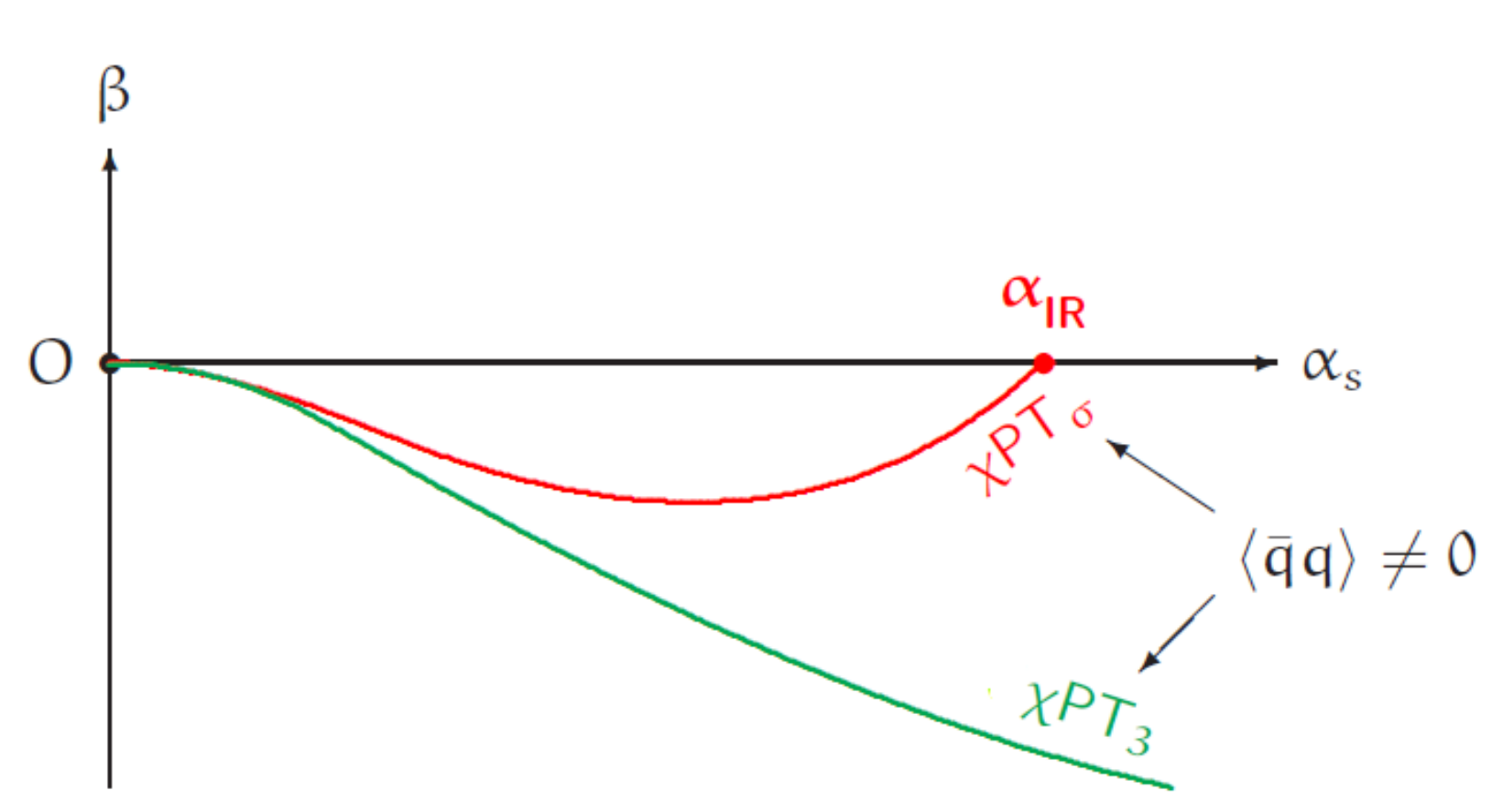}
\caption{The $\beta$ function for the IR structure in three-flavor QCD \`a la Crewther and Tunstall~\cite{CT}. }
\label{CT}
\end{figure}

The scheme we are proposing is the notion of an IR fixed point in QCD proposed by Crewther and Tunstall (CT for short)~\cite{CT,CCT}.  It is represented in Fig.~\ref{CT} by the IR structure of  the $\beta$ function as a function of the QCD gauge coupling (to be denoted $g_s$ to be distinguished from hidden gauge coupling) separate from the known ultraviolet (asymptotic freedom) structure.
There are other approaches in the literature ~\cite{GS} which differ at NLO in scale-chiral counting, explained briefly below,  because their $N_c$ and $N_f$ countings differ. The difference reflects different physics involved, specifically whether one is dealing with QCD with $N_f\sim 2, 3$ or for large $N_f$ for the physics of Beyond the Standard Model (BSM). However to the LO, which will be denoted hereon as ``LOSS" (leading order scale symmetry), there is no difference. Furthermore the same low-energy theorems that we consider to be the crucial element are shared by them at the LO~\cite{LMR,CNDIII}.

Consider setting up the counting rule for scale symmetry in conjunction with chiral symmetry, that is, scale-chiral symmetry. The key element is the trace of  the energy--momentum tensor (TEMT) of QCD
\be
 \theta_\mu^\mu = \frac{\beta(\alpha_s)}{4\alpha_s}G_{\mu\nu}^a G^{a\mu\nu} + \sum_{q}\left(1 + \gamma_m^q(\alpha_s)\right)m_q \bar{q}q,
\label{Trace-anomaly}
\ee
where $\beta$ is the QCD beta function given in terms of the fine-structure constant $\alpha_s=g_s^2/4\pi$ and  $\gamma_m^q = \mu \partial \ln m_q/\partial\mu$ is the anomalous dimension of the quark mass operator. Equation (\ref{Trace-anomaly}) says that there can be an exact scale invariance if, in the chiral limit $m_q\rightarrow 0$, there is an IR fixed point $\beta(\alpha_{IR})=0$. We adopt the suggestion by CT that such an IR fixed point with a {\it non-vanishing chiral condensate}, nonperturbative in character, is highly plausible and that far below from the chiral scale $4\pi f_\pi\sim 1$ GeV,  the $\beta$ function can flow along the trajectory leading to the IR fixed point. The scale symmetry associated with the vanishing of $\theta_\mu^\mu$ is then assumed to be spontaneously broken, giving rise to a NG boson, namely the dilaton. Note in this scheme that the chiral symmetry is spontaneously broken at the IR fixed point as long as the quark condensate $\la\bar{q}q\ra$ is non-vanishing.

That the two spontaneous broken symmetries are intimately locked to each other, we should stress, is the key point of our development.

The scalar $f_0(500)$ is identified with that scalar NG boson {associated with the spontaneous scale symmetry breaking} with the mass generated by {\it ``explicit" symmetry breaking encoded in both the departure of $\alpha_s$ from $\alpha_{IR}$ (with a non-zero gluon condensate) { and} the current quark mass}. Thus the dilaton $\sigma$ joins the pseudo-scalars,  pions and kaons, to form the pseudo-NG multiplet. What ensues is then a potentially powerful EFT that combines both chiral symmetry and scale symmetry with the possibility of doing systematic expansions both in the chiral counting and in the scale counting, i.e., ``scale-chiral" counting.

There are advantages in this approach in particle physics, such as the  simple explanation of the $\Delta I=1/2$ rule for kaon decays that is accomplished by elevating next-to-leading order (loop) terms in three-flavor chiral perturbation theory {$\chi$PT$_3$} into the leading tree order in terms of the $\sigma$ field in $\chi$PT$_\sigma$. What we are particularly interested in is what this scheme with a NG scalar put together with the NG pseudo-scalars does in nuclear phenomena, particularly at high density.

We recall that for applying to nuclear matter, the key degrees of freedom are the nucleons and the pions, the degrees of freedom figuring in the usual 2-flavor chiral perturbation theory $\chi$PT$_2$. Implementing the scalar $\sigma$ in the CT scheme, however on the other hand, requires three flavors including the strangeness. This is because the mass of the scalar is comparable rather to the kaon mass $\sim 500$ MeV than to the pion mass in the matter-free space.  In what follows, however, we will be focusing on non-strange phenomena that take place in nuclear systems, so we will be projecting out the two-flavor sector from $SU(3)$ for most of the consideration, apart from the structure of the $\sigma$. This becomes more justified as the effective mass of the scalar drops as density increases whereas the kaon mass gets affected less due to the strange quark mass. Of course to apply to strange hadrons, hyperons and kaons need to be, and can be straightforwardly, incorporated.

To access the baryonic matter lying near but -- not on -- the IR fixed point  $\beta (\alpha_{IR})=0$, one expands the $\beta$ function to the linear order in $\Delta\alpha_s=\alpha_{IR}-\alpha_s$,
 \be
 \beta (\alpha_s)=\beta^\prime \Delta \alpha_s+\cdots
 \ee
 where
 \be
 \beta^\prime= \frac{\del}{\del \alpha_s}\beta|_{\alpha_{IR}}
 \ee
 is the anomalous dimension of the gluonic tensor operator $G_{\mu\nu}^a G^{\mu\nu a}$. This $\Delta\alpha_s$  and the dilaton mass add to the usual chiral counting as
 \be
 m_\chi^2\sim \Delta\alpha_s\sim O(p^2).
 \ee
The combined scale and chiral counting provides a systematic scale-chiral expansion.  We call the perturbative approach based on the resulting scale-chiral Lagrangian $\chi$PT$_\sigma$, to be distinguished from the chiral perturbation theory $\chi$PT.

We will not need the full expression of the scale-chiral Lagrangian for the calculation that will be given below. But we write down the detailed expression to the leading order (LO) in the scale-chiral expansion because some of the intricate observations that will be made in this review in studying dense matter will depend on the specific structure. To the leading order, i.e., $O(p^2)$, it has the form Ref.~\cite{CT}
\begin{eqnarray}
\cal L_{\chi {\rm PT}_\sigma}^{\rm LO} & = & {\cal L}_{{\rm inv}}^{d=4} + {\cal L}_{{\rm anom}}^{d > 4} + {\cal L}_{{\rm mass}}^{d < 4},\label{eq:CTL}
\end{eqnarray}
with
\begin{subequations}
\begin{eqnarray}
{\cal L}_{\rm inv}^{d=4} & = & c_1 \frac{f_\pi^2}{4} \left( \frac{\chi}{f_\chi}\right)^2 {\rm Tr}\left( \partial_\mu U \partial^\mu U^\dagger \right) + \frac{1}{2} c_2 \partial_\mu \chi \partial^\mu \chi + c_3 \left( \frac{\chi}{f_\chi}\right)^4, \label{eq:CTL40}\\
{\cal L}_{\rm anom}^{d > 4} & = & (1 - c_1)\frac{f_\pi^2}{4} \left( \frac{\chi}{f_\chi}\right)^{2+\beta^\prime} {\rm Tr}\left( \partial_\mu U \partial^\mu U^\dagger \right)\nonumber\\
& &{} + \frac{1}{2}(1 - c_2) \left( \frac{\chi}{f_\chi}\right)^{\beta^\prime} \partial_\mu \chi \partial^\mu \chi + c_4 \left( \frac{\chi}{f_\chi}\right)^{4+\beta^\prime},\label{eq:CTLg40}\\
{\cal L}_{\rm mass}^{d < 4} & = &{} \frac{f_\pi^2}{4} \left( \frac{\chi}{f_\chi}\right)^{3-\gamma_m} {\rm Tr}\left( \mathcal{M}^\dagger U + U^\dagger \mathcal{M} \right),\label{eq:CTLm40}
\end{eqnarray}
\end{subequations}
where the superscript $d$ stands for scale dimension, $\mathcal{M}$ stands for  the current quark matrix with $\mathcal{M} = {\rm diag}(m_\pi^2,m_\pi^2, 2m_K^2 - m_\pi^2)$,  $\gamma_m$ is the anomalous dimension of the quark mass operator $\bar{q}q$, $c_i$'s are unknown constants of scale-chiral order  $O(p^0)$ for $i=1,2$ and $O(p^2)$ for $i=3,4$ but mass {dimension-zero}   for $i=1,2$ and mass dimension-four for $i=3,4$. It is important to note that the coefficient $c_3$ figuring in the leading order scale-invariant term is implicitly $O(p^2)$ in scale-chiral counting. This is because it plays a role for the dilaton mass similarly to $\mathcal{M}$ for the pseudo-scalar NG mesons. However differently from $\mathcal{M}$, setting to zero of which corresponds to turning off explicit chiral symmetry breaking, i.e., going to the chiral limit,  $c_3$ does not turn off the explicit scale symmetry breaking. This difference should be kept in mind in keeping track of scale-chiral order and explicit symmetry breaking for the dilaton.
\subsubsection{Scale-chiral Lagrangian at LOSS}\label{loss-lagrangian}
One can set up a systematic higher-order expansion, and it has been explicitly worked out to next-to-leading order (NLO)~\cite{LMR}. At first sight, with so many unknown parameters, it looks daunting even at the leading scale-chiral order to make sense of the Lagrangian, not to mention going to higher orders.  It turns however that one can make a substantial progress and arrive at a manageable form. For this purpose let us look at the LO Lagrangian in the chiral limit. In the matter-free vacuum, the dilaton potential can be written as
\begin{eqnarray}
V(\chi) & = & {} - (4 + \beta^\prime)c\left( \frac{\chi}{f_\chi}\right)^4 + 4 c \left( \frac{\chi}{f_\chi}\right)^{4 + \beta^\prime},
\label{eq:potentialchictLO}
\end{eqnarray}
where
\be
c={} -\frac{1}{4}c_4 = \frac{1}{4+\beta^\prime}c_3.
\ee
We see that, with $\beta^\prime \neq 0$, the dilaton potential is in the Nambu-Goldstone (NG)  mode, i.e., the minima of the potential appears at $\langle \chi \rangle = f_\chi$. However, if $\beta^\prime = 0$, $V(\chi) = 0$ so the scale symmetry cannot be broken. This simple observation illustrates that the spontaneous breaking and explicit breaking of scale symmetry are correlated and the spontaneous breaking is triggered by explicit breaking, which agrees with that unlike chiral symmetry, spontaneous breaking of scale symmetry cannot take place in the absence of explicit symmetry breaking~\cite{Freund-Nambu}.  How this explicit symmetry breaking enters is an intricate issue that still is controversial. A nonperturbative mechanism for explicit symmetry breaking is discussed in Ref.~\cite{CCT}. This issue might be highly relevant to how to approach the dilaton-limit fixed point where the dilaton mass goes to zero in the chiral limit. In what follows, we will be far from this limit, so we do  not believe our calculations are seriously affected by this problem.

In what follows, we adopt what we call ``leading-order scale symmetry (LOSS)" that corresponds to
\be
c_1\approx c_2\approx 1
\ee
with
\be
\beta^\prime >0.
\ee
In this LOSS approximation,  scale symmetry breaking -- in the chiral limit -- is lodged entirely in the dilaton potential $V(\chi)$. What results is
\begin{eqnarray}
{\cal L}_{{\rm LO}}^{\chi{\rm limit}}  =  \frac{f_\pi^2}{4} \left( \frac{\chi}{f_\chi}\right)^2 {\rm Tr}\left( \partial_\mu U \partial^\mu U^\dagger \right) + \frac{1}{2} \partial_\mu \chi \partial^\mu \chi
+V(\chi).
\label{LOSS}
\end{eqnarray}
This Lagrangian is what results from the linear sigma model in the limit $\lambda\to 0$ in Yamawaki's argument for hidden scale symmetry~\cite{hiddenscalar}. It is also of the form obtained in the leading order in scale-chiral symmetry in Ref.~\cite{GS}.

Perhaps the best argument in favor of scale-chiral symmetry adopted here for dense matter comes from the low-energy theorems for the dilaton that parallel to the low-energy theorems for the pions~\cite{ellis}. The point is that in dense matter, the scalar becomes light and goes toward -- although not too close, as will be explained, to -- the dilaton-limit fixed point with the scalar mass dropping and hence joins the pions in the NG modes.  For instance, it is instructive to derive the analog for the dilaton to the PCAC for the pion. For this we can work in the scale-invariant limit for the dilaton as in the chiral limit for the pion. In chiral symmetry, a spurion field is introduced as ${\cal M}$ that transforms as the $U$ field and write down {\it formally} chiral invariant Lagrangian. At the end of the day, one sets ${\cal M}\sim m_q\sim O(p^2)$. This approach leads to the well-known low-energy theorems such as the Goldberger--Treiman relation, PCAC, Gell-Mann--Oakes--Renner relation etc. Now we can do the same treatment for the dilaton. The only scale-symmetry breaking term in Eq.~(\ref{LOSS}) is in the dilaton potential. Introduce the spurion field ${\cal S}$ that has the scale dimension $d_{\cal S}=1$ and mass dimension zero. Then a unique form having scale dimension four, hence formally scale-invariant, that correctly reproduces the gluonic scale anomaly can be written as Ref.~\cite{matsuzaki-yamawaki}
\begin{equation}
V(\chi) = \frac{m_\chi^2 f_\chi^2}{4} \left(\frac{\chi}{f_\chi}\right)^4 \left(\ln \frac{\chi/{\cal S}}{f_\chi} -\frac{1}{4}\right).
\label{anomaly1}
\end{equation}

This yields the scale Ward-Takahashi identity for  ${\cal S}=1$
\be
\la\theta_\mu^\mu\ra=\la\del_\mu D^\mu\ra ={} - \frac{m_\chi^2}{4f_\chi^2}\la\chi^4\ra
\ee
which is the partially conserved dilatonic current (PCDC) relation, the counterpart to the PCAC for the pion. Other low-energy theorems, some with baryons incorporated, e.g., the Goldberger--Treiman relation, can be derived.
\subsection{Dilaton-limit fixed point and parity-doubling symmetry}\label{dlfp}
For accessing dense matter, both scale symmetry and hidden local symmetry need to be implemented in baryonic chiral Lagrangians. Making scale-symmetric baryonic HLS Lagrangian -- that we shall refer to as $bs$HLS -- in the LOSS approximation is straightforward. Going beyond the LOSS is also feasible, but at present, it is practically impossible to do any realistic higher-order calculations in scale-chiral expansion~\cite{LMR}. It is however feasible to do a mean-field calculation which corresponds, with a suitable matching to QCD correlators,  to  Landau Fermi-liquid fixed point theory~\cite{CNDIII}. This will be done in the coming chapters.

In this subsection, we bare another symmetry which is not visible in QCD in the vacuum but emerges in dense matter, namely, the parity-doubling in the nucleon structure.

We first show the approach to the dilaton-limit fixed point~\cite{beane-vankolck} at which the dilaton mass goes to zero. This corresponds to taking the  weak-coupling limit in the hidden scale symmetry Lagrangian discussed above. We do this first with a chiral-invariant nucleon mass $m_0$ ``put by hand"  which will be exposed at the dilaton limit. We will also restrict ourselves to  the hidden local symmetry $h(x) = SU(2)_V \times U(1)$ instead of $U(2)$ to indicate that the global $U(2)$ symmetry for $V=(\rho, \omega)$ breaks down at high density.
The $bs$HLS Lagrangian to the leading order in scale-chiral counting can be written as~\cite{DLFP-PD}\footnote{Various coupling constants such as $a_{\rho,\omega}$ that figure in HLS Lagrangians, not essential for discussions, appear in this formula. They can be looked up in Ref.~\cite{DLFP-PD} if desired.}
\begin{eqnarray}
{\cal L} &=& {\cal L}_N + {\cal L}_M + {\cal L}_\chi\,,
\label{Dlfplag}
\end{eqnarray}
where
\begin{eqnarray}
\mathcal{L}_{N}
&=& \bar{Q}i\gamma^{\mu}D_{\mu}Q - g_{1}f_{\pi}\frac{\chi}{f_{\chi}}\bar{Q}Q
{}+ g_{2}f_{\pi}\frac{\chi}{f_{\chi}}\bar{Q}\rho_{3}Q - im_{0}\bar{Q}\rho_{2}\gamma_{5}Q
+ g_{v\rho} \bar{Q}\gamma^{\mu}\hat{\alpha}_{\parallel \mu}Q\nonumber\\
&&{}+ g_{v0} \bar{Q}\gamma^{\mu}\mbox{Tr}\left[\hat{\alpha}_{\parallel \mu} \right]Q + g_{A}\bar{Q} \rho_{3} \gamma^{\mu}\hat{\alpha}_{\perp \mu}\gamma_{5} Q\,,
\label{NchiLargrangian} \\
{\mathcal L}_M
& = & \frac{f_{\pi}^2}{f_{\chi}^2} \chi^2\mbox{Tr}\left[ \hat{\alpha}_{\perp\mu}
  \hat{\alpha}_{\perp}^{\mu} \right]
{}+ \frac{a_\rho f_{\pi}^2}{f_{\chi}^2} \chi^2\mbox{Tr}\left[ \hat{\alpha}_{\parallel\mu}
  \hat{\alpha}_{\parallel}^{\mu} \right] + \frac{(a_\omega-a_\rho) f_{\pi}^2}{2f_{\chi}^2} \chi^2\mbox{Tr}\left[ \hat{\alpha}_{\parallel\mu} \right]
  \mbox{Tr}\left[ \hat{\alpha}_{\parallel}^{\mu} \right] \nonumber\\
 &&{}- \frac{1}{2}\mbox{Tr}\left[ \rho_{\mu\nu}\rho^{\mu\nu} \right]
{}- \frac{1}{2}\mbox{Tr}\left[ \omega_{\mu\nu}\omega^{\mu\nu} \right]\,,
 \\
{\mathcal L}_\chi
&=& \frac{1}{2}\partial_\mu\chi\cdot\partial^\mu\chi {}-V(\chi).
\end{eqnarray}
Here  $Q$ is the nucleon doublet
\begin{eqnarray}
Q & = & \left(
          \begin{array}{c}
            Q_1 \\
            Q_2 \\
          \end{array}
        \right)
\end{eqnarray}
which transforms as $Q \to h(x) Q$, the covariant derivative $D_\mu = \partial_\mu - i V_\mu$, $\rho_i$ are the Pauli matrices acting on the parity-doublet, $g_{v0}=\frac 12 (g_{v\omega}-g_{v\rho})$, $a_\omega, a_\rho$,  $g_A$ and $g_{v\rho,v\omega}$
are all dimensionless parameters.

To move toward a chiral symmetric Gell-Mann--L\'evy (GML)-type linear sigma model, we do the field re-parameterizations $\Z=U\chi f_\pi/f_\chi = s+i\vec{\tau}\cdot \vec{\pi}$,  defining the scalar $s$ and write  (\ref{Dlfplag}) composed of  two parts, one that is regular, ${\cal L}_{\rm reg}$, and the other that is singular, ${\cal L}_{\rm sing}$, as {$\mbox{Tr}(\Z\Z^\dagger)\equiv\kappa^2 = 2\left( s^2 + \pi^{a\,2}\right) \rightarrow 0$}.\footnote{Note that this limiting process is equivalent to dialing $\lambda$ to 0 to go from nonlinear sigma model to scale-symmetric theory via linear sigma model as was done with (\ref{potv}) discussed above. { In later sections where we discuss the half-skyrmion phase in skyrmion crystal simulation of dense matter, we invoke the space averaged quark condensate $\overline{\Sigma}$ which goes to zero in the half-skyrmion phase. There the pion decay constant remains non-zero, so the vanishing $\overline{\Sigma}$ does not imply chiral symmetry restoration. Here in analogy, setting  {$\mbox{Tr}(\Z\Z^\dagger)\to 0$}, going to the DLFP,  does not mean that $f_\pi\to f_\chi \sim \la\chi\ra$ goes to zero. This point is important for understanding the emerging pseudo-conformal structure  we predict in compact stars.}} The singular part that arises solely from the scale invariant part of the original Lagrangian (\ref{Dlfplag}) takes the form
{
\begin{eqnarray}
\mathcal{L}_{\rm sing} & = &
\left( g_{v\rho} -g_A \right) {\cal A} \left( 1/\Tr \left[ \Z \Z^{\dagger} \right]\right) +  \left( \alpha -1\right) {\cal B} \left( 1/\Tr \left[ \Z \Z^{\dagger} \right]\right)\,,
\label{sing}
\end{eqnarray}
}
where $\alpha \equiv f_\pi^2/f_\chi^2$ and
\be
{\mathcal A}
&=&{} -
\frac{i }{4} \Tr \left( \Z \Z^{\dagger} \right)^{-2} \bar{\psi} \Big[  \Tr\left( \sbar{\partial} \left(\Z \Z^{\dagger} \right)  \right)\left\{ \Z, \Z^{\dagger} \right\} \nonumber\\
&&{} - 2 \Tr\left( \Z \Z^{\dagger} \right) \left( \Z \sbar{\partial} \Z^{\dagger} + \Z^{\dagger} \sbar{\partial} \Z \right) \Big] \psi \nonumber \\
&&{} - \frac{i }{2} \Tr\left( \Z \Z^{\dagger} \right)^{-1} \bar{ \psi } \rho_{3} \gamma_{5} \left( \Z \sbar{\partial} \Z^{\dagger} - \Z^{\dagger} \sbar{\partial} \Z \right) \psi \, , \\
{\mathcal B} &=&{} - \frac{1}{16 \alpha } \Tr \left( \Z \Z^\dagger \right)^{-1} \Tr \left[ \partial_\mu \left( \Z \Z^\dagger \right)\right] \Tr \left[ \partial^\mu \left( \Z \Z^\dagger \right)\right] \,,
\ee
where
\begin{eqnarray}
\psi & = &  \frac{1}{2}\left[ \left( \xi_R^{\dagger}+ \xi_L^{\dagger} \right)
+ \rho_{3} \gamma_5 \left( \xi_R^{\dagger} - \xi_L^{\dagger}  \right) \right] Q.
\end{eqnarray}
That ${\mathcal L}_{\rm sing}$ be absent leads to the conditions that
\be
g_{v\rho}-g_A\rightarrow 0\,,
\quad
\alpha -1 \to 0\,.
\ee
The second condition is precisely the locking of $f_\pi $ and $f_\chi$ mentioned above\footnote{It is perhaps worth pointing out here that  in addressing the $N_f=8$ dilatonic EFT framework for dilatonic Higgs problem, very similar low-energy theorems are discussed. There the ratio $f_\pi/f_\chi$ comes out to be $\sim 0.1$~\cite{appelquist}  whereas in our case where the scale symmetry is emergent, the two are locked to each other, essentially due to how the scale-chiral symmetry \`a la CT manifests in nuclear medium.}.
Using large $N_c$ sum-rule and renormalization-group arguments~\cite{beane-vankolck}, we infer
\be
g_A-1\rightarrow 0\,.
\ee
In the density regime where GML-type linear sigma model is valid, the nucleon mass can be given as
\begin{equation}
m_{N_\pm} = {} \mp g_2 \langle s \rangle + \sqrt{\left( g_1 \langle s \rangle \right)^2 +  m_0^2}\,,\label{nmass}
\end{equation}
where $\langle s \rangle$ is the vacuum expectation value of $s$.
As the chiral symmetry restoration point is approached, $\langle s \rangle\rightarrow 0$, so in the limit {$\mbox{Tr}(\Z \Z^\dagger) \rightarrow 0$}, we expect
\begin{equation}
m_{N_\pm} \rightarrow m_0\,.\label{m0}
\end{equation}
These are the constraints that lead to the dilaton limit as  announced above. It follows then that
\be
g_{\rho NN}=g_\rho(g_{v\rho}-1)\rightarrow 0.\label{vmhere}
\ee
We thus find that in the dilaton limit, the $\rho$ meson decouples from the nucleon\footnote{This $\rho$ decoupling from the nucleon that takes place in conjunction with the dilaton limit may happen most likely before the vector manifestation fixed point at which the $\rho$ mass goes to zero with $g_\rho\to 0$. Here we are dealing with high density. We cannot say whether or not something similar takes place in temperature.}. In contrast, the limiting $\mbox{Tr}(\Z \Z^\dagger)\rightarrow 0$ {\it does not} give any constraint on $(g_{v\omega}-1)$. The $\omega$-nucleon coupling remains non-vanishing in the Lagrangian which in  unitary gauge  and in terms of fluctuations $\tilde{s}$ and $\tilde{\pi}$
around their expectation values, takes the form
\begin{eqnarray}
{\mathcal L}_N
& = &
\bar{N}i\sbar{\partial}{ N} - \bar{ N}\hat{M}{ N}
{}- g_1\bar{ N}\left(
\hat{G}\tilde{s} + \rho_3\gamma_5 i\vec{\tau}\cdot\vec{\tilde{\pi}}
\right)  N  + g_2\bar{ N}\left(
\rho_3 \tilde{s} + \hat{G}\gamma_5 i\vec{\tau}\cdot\vec{\tilde{\pi}}
\right) {N} \nonumber\\
& &{}
+ \left(1-g_{v\omega} \right) g_\omega{ N}  \frac{\sbar{\omega}}{2}  {N}\,,
\end{eqnarray}
where $N$ is in parity eigenstate. This Lagrangian is the same as the one given in Ref.~\cite{detar-kunihiro} except for the $\omega$-nucleon interaction. This is just the nucleon part of the linear sigma model in which the $\omega$ is minimally coupled to the nucleon, applicable infinitesimally below the (putative) chiral restoration critical density $n_c$ with the effective nucleon mass replacing $m_0$.

$\bullet$ {\bf Proposition III: \it Moving toward to the dilaton-limit fixed point, the fundamental constants in scale-chiral symmetry get transformed as $f_\pi\to f_\chi$, $g_A\to g_{v\rho}\to 1$ and the $\rho$ meson decouples while the $\omega$ remains coupled, breaking the flavor $U(2)$ symmetry for $(\rho,\omega)$.}

\subsection{Emergent parity doubling}\label{paritydoubling}

Here we show that the parity doubling arises strictly by nuclear correlations from $bs$HLS that contains no chiral-invariant $m_0$ term~\cite{interplay}. We can do this in the mean-field approximation using the simplified $bs$HLS Lagrangian which is obtained from Eq.~(\ref{NchiLargrangian}) with $m_0$ set to zero and put in parity eigenstates,
\begin{eqnarray}
\mathcal{L}
&=& \bar{N}i\gamma^{\mu}D_{\mu}N - hf_{\pi}\frac{\chi}{f_{\chi}}\bar{N}N + g_{v\rho} \bar{N}\gamma^{\mu}\hat{\alpha}_{\parallel \mu}N\nonumber\\
&&
{}+ g_{v0} \bar{N}\gamma^{\mu}{\Tr \left[\hat{\alpha}_{\parallel \mu} \right]}N
{}+ g_{A}\bar{N}  \gamma^{\mu}\hat{\alpha}_{\perp \mu}
\gamma_{5} N\ + V(\chi),
\label{NchiLargrangian1}
\end{eqnarray}
where $V(\chi)$ is the dilaton potential that we take in the form of~\eqref{potv} {and $N = (p,n)^T$ being the baryon iso-doublet}.

We consider the $Gn$EFT Lagrangian effective in a vacuum modified by density and construct the thermodynamic potential.  To do the mean-field approximation for the
thermodynamic potential it is important to properly treat the density dependence of the bare parameters. Otherwise one loses the rearrangement terms and hence fails to conserve  the energy-momentum tensor~\cite{Song:1997kn}.

\begin{figure}[h]
\begin{center}
\includegraphics[width=8cm]{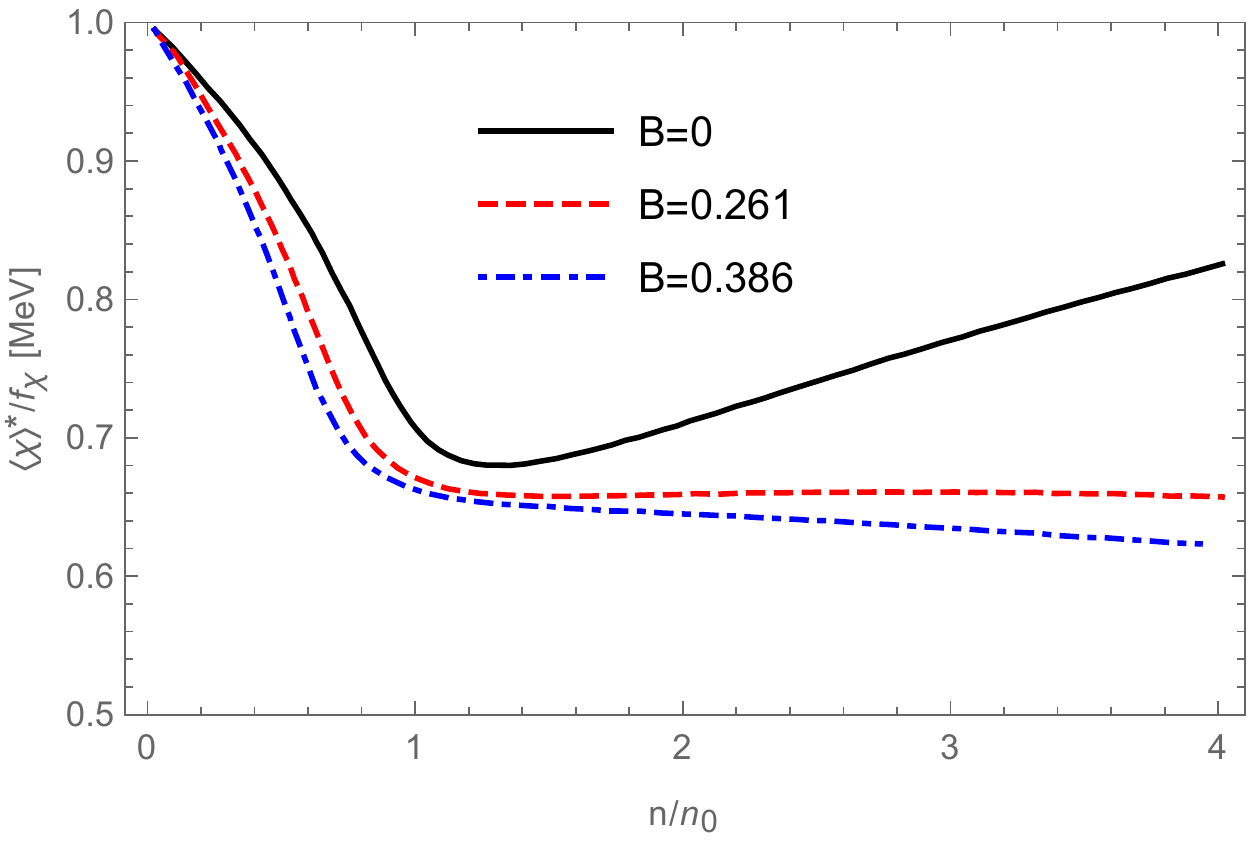}
\caption{The ratio $m_N^*/m_N\approx \la\chi\ra^*/\la\chi\ra_0$  as a function of
density for varying density dependence of $g_{v\omega}^\ast$.
Note that for a given $\omega$-nuclear coupling, the nucleon mass stops dropping at a density $n_A$
above nuclear matter density $n_0$ and stays  constant above that
density.
}
\label{mass_coupling}
\end{center}
\end{figure}

With the density dependence of the bare parameters of the Lagrangian indicated by the asterisk, the thermodynamic potential in the mean-field approximation takes the form
\begin{eqnarray}
\Omega(\chi,\, n)&=&
\frac{1}{4\pi^{2}} \left[ 2 E_{F}^{3} p_{F} - m_{N}^{\ast 2} E_{F} p_{F}
{}- m_{N}^{\ast 4} \ln \left( \frac{E_{F} + p_{F} }
{m_{N}^{\ast}} \right) \right]\nonumber\\
&&{} + \frac{\left(g_{v\omega}^\ast -1 \right)^2}
{2a_\omega f_{\pi}^2
{\chi}^2/f_{\chi}^2} n^2
{} -  V(\chi) -\mu(n) n\,,
\label{omega}
\end{eqnarray}
where $E_F = \sqrt{p_F^2 + m_N^{\ast\, 2}}$
and the chemical potential is given as a function of density $n$ by
\begin{eqnarray}
\mu(n) & = & E_F(n) + \frac{\left( g_{v\omega}^\ast -1\right)^2}
{a_\omega f_{\pi}^2{\chi}^2/ f_\chi^2} n
{}+ \frac{\left( g_{v\omega}^\ast -1\right)}
{a_\omega f_{\pi}^2{\chi}^2/ f_\chi^2}n^2
\frac{\partial \left( g_{v\omega}^\ast -1\right) }{\partial n}\,.
\end{eqnarray}
The nucleon mass is connected to the $\omega$-nucleon coupling by the
equation of the motion for $\chi$ and $\omega$, and the in-medium property
of the $\chi$ condensate -- equivalently the in-medium mass of the dilaton -- controls
the behavior of the nucleon mass at high density. The nucleon mass depends
on $\bchi=\la \chi\ra$ via
\be
m_N^\ast = h\bchi\,.
\ee
The gap equation for $\chi$ is
\be
&& \left[
\frac{ m_N^2}{\pi^2 f_\chi^2}\left(
p_F E_F - m_N^{\ast\, 2}\ln\left(\frac{p_F + E_F}{m_N^\ast}\right)
\right) \right. \nonumber\\
&&\left.\quad {}- \frac{\left(g_{v\omega}^\ast -1 \right)^2}
{a_\omega f_{\pi}^2 \chi^4/f_{\chi}^2} n^2
 {}+ \frac{ m_\chi^2}{2} \left( \frac{\chi^2}{f_\chi^2} \right)
\ln\left(\frac{\chi^2}{f_\chi^2}\right)\right] \chi = 0\,.
\label{gapchi}
\ee
In the mean field approach, the dilaton limit is reached as $\bchi \rightarrow 0$.  Suppose
the $\omega$-nucleon coupling drops slowly. This not only causes the nucleon
mass to drop slowly, but also delays the dilaton limit, $g_{A} = g_{v\rho}=1$, to higher density. This feature can be seen in Fig.~\ref{mass_coupling} given in  Ref.~\cite{interplay}.
Let us take the scaling of the $\omega$-nucleon coupling in the simple form
\begin{equation}
\frac{g_{v\omega}^\ast - 1}{g_{v\omega}-1}
= \frac{1}{1 + B n/n_0}\,. \label{omega_coupling}
\end{equation}
Here the scaling of the hidden gauge coupling $g_\omega$ is ignored, which is negligible. Thus only the scaling of the effective coupling $g_{v\omega}$ intervenes.

For a given constant $B$, the nucleon mass is calculated by fitting the
binding energy and the pressure of nuclear matter at $n_0$. The two density-dependent quantities involved are $m_\chi^*$ and
$g_{v\omega}^*$ that are determined by the binding energy and the pressure
at $n=n_0$ for given $B$.  The result is plotted in Fig.~\ref{mass_coupling}. Remarkably  the nucleon mass is found to drop almost linearly in density to about 70\% of the free-space mass at a density denoted $n_A$ above $ n_0$. Up to $\sim n_0$, the dilaton condensate, locked to the quark condensate,  is consistent with the empirical value of the quark condensate estimated from the in-medium pion decay constant measured in deeply bound pionic states. It then stabilizes to a constant for $n\gsim n_A$. This density will later be identified with the skyrmion-to-half-skyrmion topology change  density denoted $n_{1/2}$ we shall encounter below.

How this flattening at $n_A$ comes about is an intricate interplay between the nucleon mass and the $\omega$-NN coupling after $n\sim n_A$.
We should stress here that this mean-field calculation was made with  $m_0=0$. Nevertheless, we have found
$m_N^\ast \sim 0.7 m_N$ in high density, indicating that a non-vanishing $m_0$
emerges dynamically. Thus the appearance of $m_0$ signaling parity doubling is linked to (\ref{vmhere}), i.e.,  the decoupling of the $\rho$ meson from nucleons.  We will see later that the emergence of pseudo-conformal structure at high density is also closely tied to this property. The interplay between the nucleon mass and the $\omega$-nucleon coupling
as revealed in this way is  similar to what was found by the renormalization group equation analysis~\cite{interplay}  and consistent with what was phenomenologically
observed in nuclear EFT description  modified by the topology
change~\cite{PKLMR}. 

In brief, this analysis suggests that as density reaches $n_A\sim n_{1/2}$ the effective nucleon or quasi-nucleon mass goes via (\ref{vmhere}) as
\be
m_N^\ast\propto \la\chi\ra^\ast\propto m_0.
\ee
{\it Thus parity doubling emerges via an interplay between $\omega$-nuclear coupling -- with $U(2)$ symmetry strongly broken -- and the dilaton condensate.}

\subsection{Trace anomaly with parity doubling}\label{trace-anomaly}
Relying on {\bf Proposition III}, one can do  an extremely simple -- and what will turn out to be useful for compact stars -- calculation for the trace of the energy--momentum tensor (TEMT) in the mean-field approximation with  the Lagrangian Eq.~(\ref{NchiLargrangian1}) together with the dilaton potential \eqref{potv}.  The relevant quantities are the energy density $\epsilon$ and the pressure $P$ (at $T=0$)~\cite{PKLMR}\footnote{We ignore for simplicity affixing the asterisk on the medium-dependent  parameters other than $m_N^\ast$.}
\begin{eqnarray}
\epsilon
&=& \frac{1}{4\pi^2} \left[ 2E_F^3 k_F - m_N^{\ast\,2}E_F k_F - m_N^{\ast\,4} \ln\left( \frac{E_F + k_F}{m_N^\ast}\right) \right]  \nonumber\\
&&{} + g_\omega \left( g_{v\omega} -1 \right) \langle \omega_0 \rangle n -\frac{1}{2} a_\omega  f_{\pi}^2 g_\omega^2 \frac{\langle \chi \rangle^2}{f_\chi^2} \langle \omega_0 \rangle^2 +  V(\langle \chi \rangle)
\label{Eden}
\end{eqnarray}
and
\begin{eqnarray}
P &=& \frac{1}{4\pi^2} \left[ \frac{2}{3}E_F k_F^3 - m_N^{\ast\,2}E_F k_F + m_N^{\ast\,4} \ln\left( \frac{E_F + k_F}{m_N^\ast}\right) \right] \nonumber\\
&&{} + \frac{1}{2} a_\omega f_{\pi}^2 g_\omega^2 \frac{\langle \chi \rangle^2}{f_\chi^2} \langle \omega_0 \rangle^2 -  V(\langle \chi \rangle)\,.\label{Pre}
\end{eqnarray}
Using the solutions of the gap equations for $\chi$ and $\omega$ that follow from extremizing Eq.~(\ref{omega}), i.e.,
\begin{eqnarray}
&\frac{m_N^2\langle \chi \rangle}{\pi^2 f_\chi^2} \left[ k_F E_F - m_N^{\ast\,2} \ln \left( \frac{k_F + E_F}{m_N^\ast} \right) \right] -\frac{a_\omega f_{\pi}^2}{f_\chi^2} g_\omega^2 \langle \omega_0\rangle^2 \langle \chi \rangle + \left. \frac{\partial\, V(\chi)}{\partial \chi} \right|_{\chi = \langle \chi \rangle} =0\,,
 \label{gap1}\\
&g_\omega \left(g_{v\omega}-1 \right)n -a_\omega f_{\pi}^2 g_\omega^2 \frac{\langle \chi \rangle^2}{f_\chi^2} \langle \omega_0 \rangle = 0\,, \label{gap2}
\end{eqnarray}
it is straightforward to derive from (\ref{Eden}) and (\ref{Pre}) the VEV of the TEMT $\theta_\mu^\mu$ (we work in the chiral limit)
\begin{eqnarray}
\langle \theta^\mu_\mu \rangle
&=& \langle \theta^{00} \rangle - \sum_i \langle \theta^{ii}\ra = \epsilon - 3 P = 4V(\langle \chi \rangle) - \langle \chi \rangle \left. \frac{\partial V( \chi)}{\partial \chi} \right|_{\chi = \langle \chi \rangle}. \label{TEMT1}
\ee
What is significant of this result is that in the mean field of $bs$HLS, the TEMT is given solely by the dilaton condensate. This is in the chiral limit, but we expect this relation to more or less hold for small pion mass.

$\bullet$ {\bf Proposition IV: \it Going toward the dilaton-limit fixed point with the $\rho$ decoupling from the nucleons, the parity doubling emerges and   $m_N^\ast\to \la\chi\ra^\ast\to m_0$.  Consequently the  trace of  energy -- momentum tensor $\theta_\mu^\mu$ in medium in $V_{lowk}$ RG theory  is a function of only $m_0$,  $\langle \theta^\mu_\mu \rangle^\ast\approx F(m_0)$, which is independent of density. This leads to the ``pseudo-conformal" sound velocity $v_s^2\approx 1/3$ in compact stars (to be discussed below).}

As will be shown below, the mean-field treatment of $bs$HLS amounts to doing Landau Fermi-liquid fixed point approach ignoring corrections of $O(1/\bar{N})$ (where $\bar{N}=k_F/(\Lambda - k_F)$ with $\Lambda$ being the cutoff above the Fermi sea). In Ref.~\cite{PKLMR}, the corrections to the Fermi-liquid fixed-point approximation were included in the so-called ``$V_{lowk}$ RG" formalism which will be described later for compact-star physics.

\section{Topology in Nuclear Interactions}
As explained in Introduction, our approach is anchored on Weinberg's Folk Theorem and what is necessary is the identification of relevant degrees of freedom for given cutoff scale.  For low-density nuclear physics, the S$\chi$EFT, with the nucleons and pions as relevant degrees of freedom, expanded to N$^k$LO for $k\leq 4$, is having a certain success validating the notion of EFT embodied in the Folk Theorem.  For density relevant to the compact-star interior, {$n \gsim 2 n_0$}, as argued above, higher degrees of freedom integrated out in S$\chi$EFT become indispensable. Their role then will necessitate going beyond the limited premise of the underlying ``soft-pion" kinematics. For this we opt to resort to topology, believed to be encoded but not visible in the matter-free vacuum in QCD. This is because, as we are learning from condensed matter physics, topology can provide robust insight into strongly correlated phenomena presumably taking place in dense baryonic matter.
\subsection{Power of topology}\label{poweroftopology}
A recent development on topological structure of the nucleon renders the notion of skyrmions as baryons in nuclear matter stronger than before.

In the large number of color limit, QCD is dominated by planar
diagrams, with infinitely many weakly interacting mesons and glueballs~\cite{THOOFT}. In this limit, baryons are heavy solitons made out of the interacting mesons. The coupling of the mesons is weak and of order $1/N_c$, while the coupling of the baryons is strong and of order $N_c$~\cite{WITTEN}. Chiral solitons made solely of non-linearly interacting pions are prototype of these solitons, an idea put forth
decades ago by Skyrme~\cite{SKYRME} well before the advent of QCD. Chiral solitons are topologically protected in $3+1$
dimensions, and their quantum numbers emerge through semi-classical quantization.  There have been tremendous developments on the impact of skyrmions in practically in all areas of physics, particularly in condensed matter and also in string theory~\cite{multifacet}.

There was however a serious stumbling block to the skyrmion theory being  a macroscopic description of QCD. The problem was that there was no skyrmion for one flavor system, i.e., $N_f=1$. This is because the homotopy $\pi_3 (U(1))=0$ whereas for $N_f> 1$, say, {$N_f=2$, $\pi_3(SU(2))=Z$.  This is at odds with QCD which supports baryons for any value of $N_f$.

This problem has been recently resolved by Komargodski's discovery~\cite{ZOHAR} that due to the $U_A(1)$ axial anomaly for the $\eta^\prime$ meson, the baryon for $N_f = 1$ can be regarded as a massless edge excitation of $1 + 2$ dimensional charged sheets carrying topology current $J_{\alpha\beta\gamma} = \epsilon_{\alpha\beta\gamma\lambda}\partial^\lambda \eta^\prime/2\pi$ and analogous to a fractional quantum Hall droplet described by (2+1) dimensional Chern-Simons topological field theory, a ``pancake" different from the spherical (3+1) dimensional skyrmion.
The reasoning leading to this description involves a highly intricate topology that is outside of the structure of this review. What is, however, eminently relevant is that baryonic interactions in {\it all} regimes of density  can be given a robust description in terms of topology that can access the highly nonperturbative dense baryonic matter relevant to the core of compact stars, which cannot be accessed directly by QCD. Furthermore there exists a direct map between the microscopic QCD degrees of freedom, i.e., quarks and gluons, and the macroscopic degrees of freedom, i.e., hadrons, given in terms of topological object. The mapping which provides the link between QCD and topology is made via what is known as ``the Cheshire Cat Principle"~\cite{CHESHIRE} that exploits the ``chiral bag"~\cite{chiral-bag}.
\subsection{The Cheshire Cat}
\subsubsection{From the chiral bag to the  Cheshire Cat Principle}
The chiral bag was constructed in such a way that quarks are confined inside a bag (e.g., the MIT bag) of volume $V$ and the bag is clouded by meson fields outside of the bag in  volume $\bar{V}$. As will become obvious, we need to consider the three-flavor case, with up, down and strange. For simplicity we will ignore the quark masses in writing down the action. The action
for $SU(N_f)_L\times SU(N_f)_R$ (for $N_f=3$) (in Minkowski space) is
 \be
S&=& S_V+S_{\overline{V}}+S_{\delta V},\label{cheshire}
\ee
where
\be
S_V&=& \int_V d^4x \left(\bar{\psi}i\not\!\!{D}\psi -\frac{1}{2}
{\rm tr}\ G_{\mu\nu}G^{\mu\nu}\right)+\cdots \, ,\nonumber\\
S_{\overline{V}}&=&\frac{f_\pi^2}{4}\int_{\overline{V}} d^4x \Big({\rm Tr}\
\del_\mu U^\dagger \del^\mu U +\frac{1}{4N_f} m^2_{\eta^\prime}({\rm Tr}{\rm
ln}U-{\rm Tr}{\rm ln}U^\dagger)^2
\Big) +S_{WZ}+\cdots \, ,\nonumber\\
S_{\delta V}&=& \frac{1}{2}\int_{\delta V} d\Sigma^\mu\Big\{n_\mu
\bar{\psi} U^{\gamma_5}\psi +i\frac{g_s^2}{16\pi^2}{K_5}_\mu ({\rm Tr}\
{\rm ln} U^\dagger-{\rm Tr}\ {\rm ln} U)\Big\}\, .
\label{surface}
\ee
Here $U$ is a unitary $U(3)$ matrix, $\psi$ is the quark field $\psi^T=(u\ d\ s)$, $G_\mu$ the octet
gluon field, $G_{\mu\nu}$ the gluon field tensor, $g_s$ the ``color"
gauge coupling~\footnote{We use ``tr" for color trace and ``Tr" for flavor trace. We use the normalization for the group generators Tr$(t^a t^b)=\frac 12 \delta^{ab}$ and likewise for the color.}  and $K_5^\mu$ is the (properly regularized)
Chern-Simons current \index{Chern-Simons!current}
\be
K_5^\mu=\epsilon^{\mu\nu\alpha\beta} (G_\nu^a
G_{\alpha\beta}^a-\frac{2}{3} g_sf^{abc} G_\nu^a G_\alpha^b
G_\beta^c),
\ee
the chiral field $U$ including $\eta^\prime$ field is
\be
U=e^{i\eta^\prime/f_0}e^{2i\pi/f_\pi}
\ee
and $U^{\gamma_5}$ is
\be
U^{\gamma_5}=e^{i\eta^\prime\gamma_5/f_0}e^{2i\pi\gamma_5/f_\pi}.
\ee
Note that due to the axial anomaly, $\eta^\prime$ is massive even if the quark masses are taken to be zero.  Note the boundary term in (\ref{cheshire}). It is crucial to match the inside given in QCD to the outside of the bag given in terms of hadronic variables. The outside is populated by the octet $\pi$ plus $\eta^\prime$ and also the massive degrees of freedom with the hidden symmetries implemented. For simplicity in discussions,  we will confine ourselves to the $U$ fields. The Chern-Simons current on the surface is needed to confine the color inside the bag. It is non-gauge-invariant and cancels the leaking color charge generated by the quantum anomaly on the boundary as shown in \cite{NRWZ}.

The Cheshire Cat Principle (CCP) formulated in Ref.~\cite{CHESHIRE} states that physics should be independent of the size of the bag $R$, the confinement size, provided that both strong dynamics and topology are properly taken into account.  Exactly formulated, this implies that the physics should be the same for the  bag picture in QCD and for the soliton picture with the bag shrunk to zero size, with the CC ``smile" implanted. In practice the CCP can only be approximate and at best applicable at low energy scale and  in particular at low density. Among others, there is the infamous difficulty of bosonizing at (3+1) dimensions for which exact bosonization does not exist. For this reason, testing the CCP has been limited.

The CCP has been shown rigorously for only two cases: The baryon charge and the color charge.

How the baryon charge is fractioned into the inside and the outside has been shown in (1+1) dimensions~\cite{CHESHIRE} and also in (3+1) dimensions~\cite{RGB,GJ}.\footnote{The proof was given for the magic angle -- equal fraction of baryon charge -- in Ref.~\cite{RGB} but was incorrect for other chiral angles. That the CC holds for any angle was proven by Ref.~\cite{GJ}. } This is not surprising for the rigorously conserved baryon charge at least within the Standard Model regime,  but how it comes out  is  quite intricate  involving symmetries and dynamics inside the bag and topology outside of the bag. The mechanism at work is that the boundary, a domain wall, for different vacua of quark/gluons and hadrons, breaks $U(1)$ symmetry, giving an anomaly which is compensated by an ``infinite hotel"~\cite{CHESHIRE,infinite-hotel}. The leaking baryon charge is taken up by the fractioned soliton outside.\footnote{ This raises the question as to how the anyonic property, i.e., the fractionation of the charge, can take place in  (3+1) dimensions. More on this below.}

As for the color charge, the situation is different.  Again due to the boundary, the QCD $U_A(1)$ anomaly generates a $U(1)$ color anomaly inside the bag. The anomaly induces the color charge to leak, but there is nothing outside to absorb the color charge. To preserve the color symmetry, this leaking color charge has to be stopped at the boundary.  The boundary condition involving the Chern-Simons current in (\ref{surface}) is put to  absorb the leaking color charge~\cite{NRWZ}. This is a case where the Lagrangian breaks a symmetry at the classical level which is  restored at the quantum level, a situation opposite to the usual anomalies in gauge theories.
\subsubsection{Cheshire Cat  in hadrons and nuclei: Chern-Simons term,\\
 $\mathbf{g_A^{(0)}}$ and fractional Quantum Hall droplet}
The Chern-Simons term on the surface of the bag turns out to have an extremely important role for two quantities. One is the flavor-singlet axial-charge of the proton $g_A^{(0)}$ and the other is the soliton structure of $\eta^\prime$.  Both are extremely subtle intricate matters involving hadron physics.

Because of the axial anomaly, the flavor singlet axial current $J_{5\mu}^{(0)}= \frac 12 g_A^{(0)} \bar{\psi}\gamma_\mu\gamma_5\psi$ is not conserved. Therefore the naive non-relativistic quark model prediction for the proton spin $J=\frac 12 g_A^{(0)}\simeq 1/2$ is violently at odds with the EMC experiment $g_A^{(0)} \approx 0.12 \ll 1$.   This led to what is now understood as a ``fake" proton-spin crisis. There is no crisis! Lattice QCD calculations indeed give $g_A^{(0)}\approx 0.21$~\cite{SESAM}. The Cheshire Cat with the boundary condition (\ref{surface}) gives a highly compelling explanation~\cite{FSAC}, disposing off the fake crisis: the measured $g_A^{(0)}$ is not directly related to the proton spin~\cite{leader}. The CC predicts $ 0\lsim g_A^{(0)}\lsim 0.2$. This is shown in Fig.~\ref{FSAC} from Ref.~\cite{FSAC}.\footnote{The calculation, where the bag contribution is treated in terms of the MIT bag and the baryon charge fractionation with the soliton structure and the $\eta^\prime$ are taken into account, could certainly be improved upon.}
\begin{figure}[h]
\vskip -5.cm
\includegraphics[width=8cm]{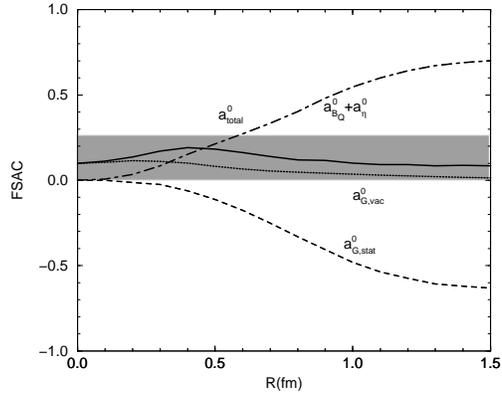}
\caption{Various
contributions to the flavor singlet axial charge of the proton as a
function of bag radius $R$ and comparison with the experiment: (a) quark
plus $\eta$ (or ``matter") contribution ($a^0_{B_Q} + a^0_\eta$) (dot-dashed line),
(b) the contribution of the static gluons due to quark source
($a^0_{G,stat}$) (dashed line), (c) the gluon Casimir contribution
($a^0_{G,vac}$) {(dotted line)}, and (d) their sum ($a^0_{total})$ (solid line). The shaded area
corresponds to the range admitted by experiments.}\label{FSAC}
\end{figure}

The crucial ingredient in this description is the Chern--Simons boundary term in the surface Lagrangian (\ref{surface}). While the first term of (\ref{surface}) mediates how the  baryon charge leaking from the bag is taken up by the soliton outside of the bag, the Chern--Simons term, which absorbs the leaking color charge,  controls how the flavor singlet axial charge is partitioned into the gluonic field  and the quark field inside the bag and meson fields, specially $\eta^\prime$ field, outside the bag. There is almost complete cancellation of the flavor singlet axial charge between the two components, giving $\sim 0$ net charge over all range of $R$ considered, thus providing a pristine proof of the CC principle.

What is noteworthy about this result is that it is precisely the Chern-Simons term that provides a long-standing validation of the skyrmion structure of baryons in QCD.  It was well-known since 1983 that when the flavor number is equal to 1, there was no skyrmion. This is because $\pi_3 (U(1))=0$. This led some to rule out the skyrmion approach as large $N_c$ QCD.  This problem was recently resolved first by the discovery that the baryon for $N_f=1$ is a fractional quantum Hall (FQH) droplet given in terms of Chern-Simons topological field theory~\cite{ZOHAR} and next a Cheshire Cat description of the FQH droplet~\cite{CC-MNRZ}.

Briefly Komargodski noted that the presence of stable superselection rules in the QCD  vacuum (instanton tunneling between vacua with different Chern-Simons number) implies  the existence of (2+1)-dimensional domain walls. These walls connect vacua with different Chern-Simons number and are observed to be stable at large $N_c$.
When these sheets are finite dimensional with a boundary, they can
carry massless edge excitations with baryon quantum numbers. They  are identified with high-spin  baryons. These sheets are  described by
a topological field theory through a level-rank duality argument~\cite{TFT}, much like in the fractional quantum Hall (FQH) effect~\cite{CMHALL}. The
 baryons are analogous to the gapless edge excitations in quantum Hall (QH) droplets.

In Ref.~\cite{CC-MNRZ}  it was shown that these baryonic QH droplets can be understood using the Cheshire cat principle (CCP)~\cite{CHESHIRE}. More specifically,  a chiral bag with a single quark species of charge $e$  (electric charge or fermion number) confined to a 1+2 dimensional annulus, leaks most quantum numbers. For all purposes the  bag radius is immaterial thanks to the CCP. In particular, when the bag radius is shrunk to zero, only the smile of the cat is left with spinning gapless  quarks running  luminally, explaining the edge modes and their chirality~\cite{ZOHAR}.
%

The upshot of what we  consider as a unified topological theory of baryons for {\it any} $N_f=1,2,3$, skyrmions and quantum Hall droplets,  is that finite nuclei, nuclear matter and dense nuclear matter could be all treated in terms of topology at large $N_c$.  We now argue that this notion can be extended  to compact-star matter encompassing the density regime up tp $n\lsim 7n_0$.

$\bullet$ { \bf Proposition V: \it The Cheshire Cat Principle is applicable to highly dense compact-star matter with topology change playing the role of hadron-quark continuity in QCD.}

\section{Topology Change in Dense Matter}
Given that QCD proper cannot, at present,  access dense matter,  we adopt the Cheshire Cat principle and resort to the generalized skyrmion approach (GSA) to treat compact-star matter. In this Section, we describe in detail the strategy of exploiting the topological inputs in nuclear and dense matter problems.

There are two practical difficulties in this endeavor. First is that quantizing the skyrmion matter for a given mesonic Lagrangian, even in the simplest form, at high density has not yet been worked out, and secondly fully accounting for relevant degrees of freedom in the Lagrangian from which solitons are obtained is yet to be formulated. Nonetheless there has been an impressive progress in the first with the Skyrme Lagrangian that contains only pions for light nuclear systems. As reviewed in Ref.~\cite{battye-manton-sutcliffe}, with some fine-tuning of the pion mass, the structure of light nuclei with mass number in the vicinity of 10 can be described rather satisfactorily.  Unlike the current numerical calculations in S$\chi$EFT approaches which mix in a variety of disparate ingredients, chiral dynamics, shell-model structure etc., one could say this is a genuine ``first-principle" calculation in nuclear physics.  However it is quite far from accessing to heavier nuclei, not to mention compact-star matter. As for the second, it is becoming clear from the recent developments in holographic QCD based on gravity-gauge duality for baryon structure~\cite{HOLO} that  mesons heavier than pions do play an important role for nuclear properties under extreme conditions. The prominent simple example is found that the lowest-lying hidden local symmetric  meson, namely, the $\rho$ meson, plays a key role in giving qualitatively correct binding energies and clustering structure of light nuclei~\cite{Sutcliffe-cluster}. How the $\omega$ meson and light scalar meson, crucial for phenomenological approach to nuclear structure, come in is yet to be worked out.

Our approach is, instead of attempting to quantize a generalized skyrmion Lagrangian, to map what is considered to be robust topological inputs to the Lagrangian, $bs$HLS, constructed above and treat the dynamics in a (Wilsonian) renormalization group (RG) flow with the EFT Lagrangian.  The parameters of the effective Lagrangian are to slide with density following  the vacuum change that takes place as density of the matter increases. This procedure which generalizes the approach first put forward in 1991, referred to in the literature as ``BR scaling"~\cite{BR91},  goes beyond the standard mean-field approach although it is not without certain ambiguities and shortcomings. They will be spelled out as clearly as possible.

To extract the intrinsic topological structure embodied in the Cheshire Cat Principle, we put  the skyrmions on crystal lattice.  This procedure with skyrmions was pioneered by Klebanov~\cite{Klebanov} and sharpened by Kugler and Shtrikman~\cite{Kugler}.  The application to nuclear and dense matter is extensively reviewed in Ref.~\cite{Park-Vento} and elsewhere~\cite{Ma-Rho}.

The principal merit of the skyrmion crystal approach is that in the large $N_c$ limit and at high density,  baryonic matter is a crystal in QCD. There are arguments in support of this picture in certain models at lower and higher dimensions than 4~\cite{QCDcrystal,popcorn}. We are not aware of a rigorous proof for this in 4D, but we shall assume that this lore is acceptable at higher density, typically, at $n\gsim 3n_0$. We will not take the crystal structure at lower densities to be reliable, but  we will consider all ranges above $n_0$ to be accessible by the method.
\subsection{Scale-symmetric Lagrangian at leading order}
Since we cannot for the moment do a reliable quantum calculation in terms of the generalized skyrmions, we will incorporate the topological information gained from the crystal lattice technique into an effective field theory, eventually the $bs$HLS Lagrangian introduced above. To give a general idea of how this can be done, we use the scale-symmetrized Skyrme Lagrangian for what follows. In this Lagrangian the Skyrme quartic terms represents {\it all} heavy degrees of freedom that are integrated out. The qualitative structure remains unchanged when heavy degrees of freedom are {\it explicitly} incorporated.  Approaching compact-star matter with full-brown $bs$HLS treated in a (Wilsonian) renormalization-group flow approach will be discussed in later chapters.

The Lagrangian we consider is the scale-symmetric Lagrangian (\ref{LOSS}) (implemented with the quartic Skyrme term)
\begin{eqnarray}
{\cal L}_{\rm loss} & = & \frac{f_\pi^2}{4} \left( \frac{\chi}{f_\chi}\right)^2 {\rm Tr}\left( \partial_\mu U \partial^\mu U^\dagger \right) +\frac{1} {32\epsilon^2}{\rm Tr}\Big([U^\dagger\del_\mu U,U^\dagger\del_\nu U]^2\Big)\nonumber\\
& &{}
 +  \frac{1}{2} \partial_\mu \chi \partial^\mu \chi+V(\chi).
\label{LOSS1}
\end{eqnarray}
This Lagrangian follows from the leading-order scale-symmetry (LOSS) approximation with the scale symmetry breaking -- both explicit and spontaneous -- put in the potential
\begin{equation}
V(\chi) = \frac{m_\sigma^2 f_\chi^2}{4} \left(\frac{\chi}{f_\chi}\right)^4 \left(\ln \frac{\chi}{f_\chi} -\frac{1}{4}\right).
\label{anomaly1}
\end{equation}
In dense matter, the dilaton mass drops with density, so the potential can be approximated by $V(\chi)\approx \frac 12 m_\sigma^2 + \cdots$. The LOSS approximation is expected to become more reliable the higher the density.
\subsection{Skyrmion-half-skyrmion transition}
This Lagrangian (\ref{LOSS1}) put on skyrmion crystal has been extensively studied in the literature~\cite{Park-Vento}. More recent developments are recounted in Ref.~\cite{HMLR}. The key observation that emerges from the series of analyses is that there is a robust topology change in dense medium from skyrmions to half-skyrmions at a certain density denoted $n_{1/2}$ which should lie above the normal nuclear matter density $n_0$. What is essential for the discussion made here is to note that the presence of the half-skyrmion ``phase"\footnote{The term ``phase" used here -- and in what follows -- strictly speaking is a misnomer. There is no order parameter in terms of a local field that characterizes the state involved, so does not belong to the usual Ginzburg--Landau--Wilson paradigm for phase transitions.} in dense matter is generic in the skyrmion description. In fact  the half-skyrmion structure is already present in the $\alpha$ nucleus (with four nucleons) as discussed by Battye, Manton and Sutcliffe in Ref.~\cite{multifacet}. Furthermore it turns out that its appearance is independent of what other degrees of freedom than that of pion are present. This can be understood by that the Lagrangian we will be using is strictly valid in the large $N_c$ limit, and in the large $N_c$ limit and at high density, baryonic matter is a crystal with the skyrmion fractionalized into two half-skyrmions.

The simple way to understand the phase change involved is in terms of the chiral $SO(4)$ coordinates, $(\sigma, \pi_1,\pi_2,\pi_3)$. There is an enhancement of the symmetry~\cite{manton-sutcliffe}
\be
(x^1,x^2,x^3)&\mapsto& (x^1+L,x^2,x^3),\nonumber\\
 (\sigma,\pi_1,\pi_2,\pi_3)&\mapsto& (-\sigma, -\pi_1,\pi_2,\pi_3),
\ee
as the lattice size $L$ is decreased (which corresponds to increasing the density $n$) in the system of skyrmions put in the face-centered-cubic (FCC) crystal. The symmetry enhancement lowers the energy, thereby inducing the phase change. Each cube in this configuration has the baryon number 1/2, i.e., half-skyrmion.

What is of the most importance is that  the quark condensate $\la\bar{q}q\ra$ -- which is local order parameter for chiral symmetry -- when averaged over space vanishes in the half-skyrmion phase\footnote{From here on the inputs from topology into effective Lagrangians will be labeled as ``Topological Input."}.

Formally this looks as if chiral symmetry is restored to Wigner phase. We will see, however, that this is not the case. Although the quark condensate is zero averaged over the unit cell, it is locally non-zero giving rise to a chiral density wave. The pion decay constant $f_\pi$ is non-zero with the hadrons gapped.\footnote{This is an analog to the pseudo-gap in high-temperature superconductivity~\cite{pseudogap}.}  Chiral symmetry is still broken.

$\bullet$ { \bf Proposition VI: \it The half-skyrmion phase in a solitonic description of dense baryonic matter is characterized by the quark condensate $\Sigma\equiv {\la\bar{q}q\ra}$ vanishing on average but locally nonzero with chiral density wave and  non-zero pion decay constant, resembling the pseudogap phase in condensed matter.}

We will develop the notion that this topology change from skyrmions to half-skyrmions captures the Cheshire Cat principle applied to a possible hadron-quark cross-over as density increases beyond $n_{1/2}$. In what is discussed below, given the qualitative nature of the discussion, the location of the density $n_{1/2}$ is not to be taken seriously. What is significant is that it exists and realistically  should lie above the normal nuclear matter density. Surprisingly, as we will see below,  the transition density can be pinned down by compact-star data.
\subsection{Cusp in the nuclear symmetry energy}
One of the key ingredients in the approach is the nuclear symmetry energy in the equation of state (EoS) in dense matter, particularly  compact-star matter.  Here the topology is found to play the most important role, dictating how the hidden symmetries can emerge in dense medium.

The symmetry energy $E_{sym}$ figures in the energy per baryon of the many-baryon system as the coefficient proportional to $\alpha^2$ where $\alpha=(N-Z)/(N+Z)$ with $N(Z)$ being the neutron (proton) number in the matter
\be
E(n, \alpha) & = & E(n, \alpha = 0) + E_{\rm sym}(n)\alpha^2 + O(\alpha^4) + \cdots .
\nonumber\\
\label{Esym}
\ee
As mentioned, it is unfeasible at present to fully calculate $E(n, \alpha)$ in the skyrmion-crystal approach even at high density. However the symmetry energy $E_{sym}$ can be calculated on crystal lattice because in the leading $N_c$ order, it is controlled by topology change~\cite{LPR-cusp}.

To obtain the symmetry energy, we consider an $A=N+Z$ system with $N\gg Z$ for $A\to \infty$. Rotate the whole matter through a single set of collective coordinates $a(t)$, $U(\vec{r},t)=a(t)U_0(\vec{r})a^\dagger (t)$, where $U_0 (\vec{r})$ is the static crystal configuration with the lowest energy for a given density. The canonical quantization leads to
\be
E^{\rm tot}=AM_{cl} +\frac{1}{2A\lambda_I} I^{\rm tot}(I^{\rm tot}+1) \, ,
\ee
where $M_{cl}$ and $\lambda_I$ are, respectively, the mass and the isospin moment of inertia of the single cell. The moment of inertia, calculated with (\ref{LOSS1}), is of the form
\begin{eqnarray}
&&\lambda_I = \int_{\mbox{Cell}} d^3 x \left\{
\frac{f_\pi^2}{6} \left(\frac{\chi}{f_\chi}\right)^2
\textstyle (3-\frac{1}{2}\mbox{Tr} (U_0 \tau_a U_0^\dagger \tau_a )) \right.
\nonumber\\
&&{}\qquad \qquad \qquad\quad\; \displaystyle +\frac{1}{24e^2} \left[
\textstyle (3-\frac{1}{2}\mbox{Tr} (U_0 \tau_a U_0^\dagger \tau_a ))
\mbox{Tr} (\partial_i U_0^\dagger \partial_i U_0 ) +\mbox{Tr} (\partial_i U_0 \tau_a \partial_i U_0^\dagger \tau_a )
\right.\nonumber\\
&&\qquad\qquad\qquad\qquad\qquad\quad\;{} \left.\left. \textstyle
+ \frac12 \mbox{Tr} (\partial_i U_0 U_0^\dagger \tau_a \partial_i U_0 U_0^\dagger \tau_a) + \frac12 \mbox{Tr} (\partial_i U_0^\dagger U_0 \tau_a \partial_i U_0^\dagger U_0 \tau_a)
\right]\right\}.\label{mom-of-inertia}
\end{eqnarray}
$I^{\mbox{tot}}$ is the total isospin which for neutron-star matter is approximately $I_3$. Thus for $\alpha\equiv (N-P)/(N+P)\lsim 1$,
\be
I^{\mbox{tot}}=\frac 12 A\alpha.
\ee
The energy per nucleon in an infinite matter ($A=\infty$) is
\be
E=E_0 +\frac{1}{8\lambda_I}\alpha^2.\label{E}
\ee
with $E_0=M_{cl}$.
This leads to the symmetry energy
\be
E_{sym}\approx \frac{1}{8\lambda_I}.
\ee

We have written down explicitly the expression for $\lambda_I$ in  Eq.(\ref{mom-of-inertia}) for two reasons. First it involves the conformal compensator (classical)  field and secondly there are two terms, the first involving only $U_0$ field and the second involving derivatives of $U_0$ appearing in the Skyrme quartic term. Now the Skyrme quartic term is to represent {\it all} higher degrees of freedom than the pions, such as the infinite towers that figure in holographic QCD,  integrated out from the chiral Lagrangian.

The numerical results obtained in Ref.~\cite{LPR-cusp} are in Fig.~\ref{LPR-fig}.
\begin{figure}[H]\centering 
\includegraphics[scale=0.3,angle=270]{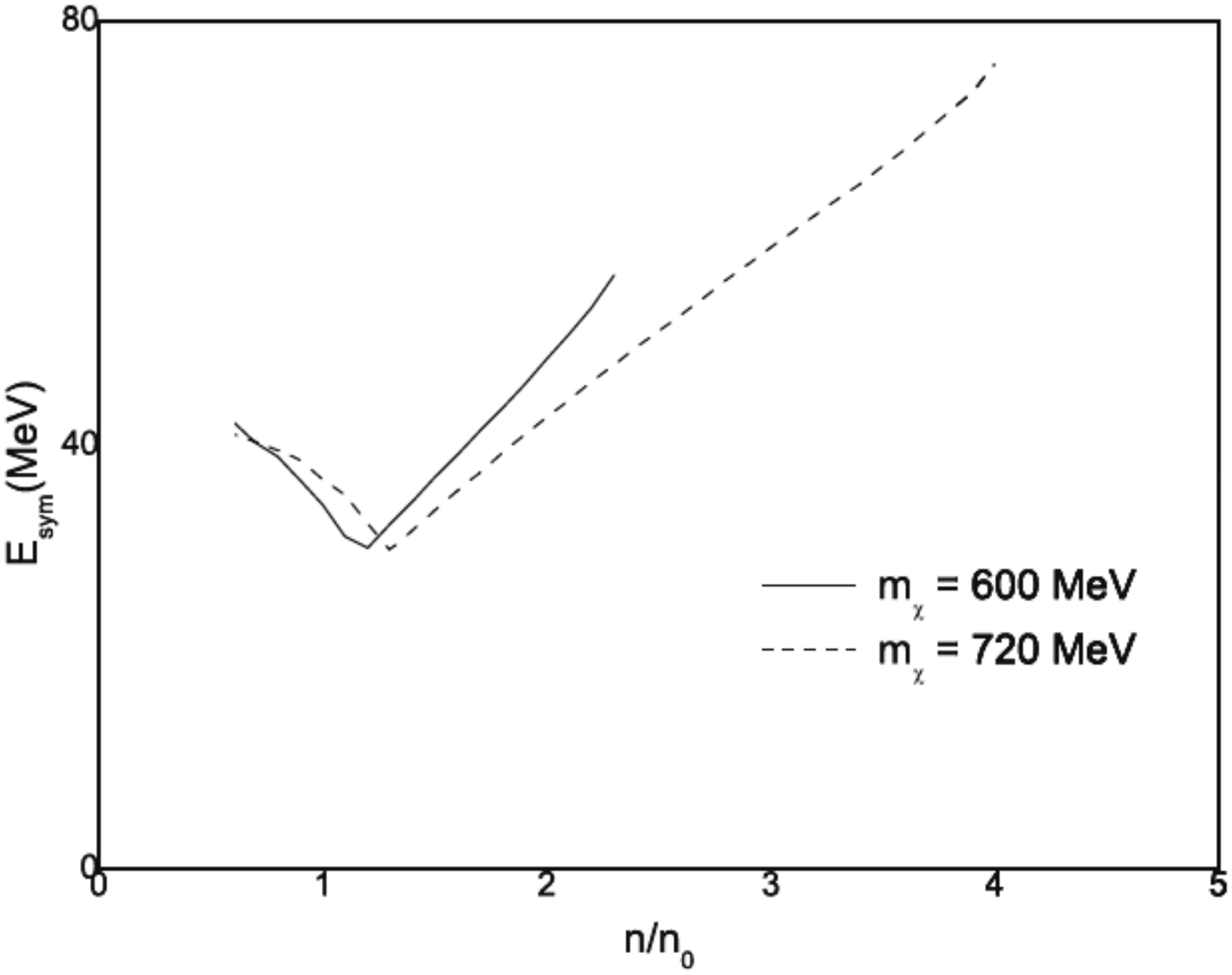}
\vskip -0.25cm
\caption{Schematic symmetry energy given by the collective rotation of the skyrmion matter with the parameters $f_\pi=93$ MeV, $1/e^2\approx 0.03$ and two vacuum values of dilaton mass. The cusp is located at $n_{1/2}$ which is not to be taken seriously. The lower density part in the regime $n<n_{1/2}$ is not reliable as the crystal description breaks down as the collective quantization method used is not applicable in that region.}
\label{LPR-fig}
\end{figure}
What is the most noteworthy is the cusp at $n_{1/2}$.  Also noteworthy are that (1) the symmetry energy is linear in density in the half-skyrmion phase and (2) that the cusp location is not sensitive to the dilaton mass\footnote{However the stability of the system depends sensitively on the (effective) dilaton mass. This accounts for the faster breakdown for $m_\chi=600$ MeV than for 720 MeV. As discussed in Section \ref{paritydoubling}, the  $\omega$ degree of freedom, breaking the $U(2)$ symmetry at high density, is indispensable for overcoming the repulsion and stabilizing  the matter. This also signals the importance of the effective value of the dilaton mass in the system for the dilaton-limit fixed point.}. The collective quantization of the skyrmion crystal is expected not to be reliable at low densities. However refined treatments with the Skyrme model (using the rational-map approximation) of mass splittings of nuclear isotopes show that the $E_{sym}$ in finite systems does tend to decrease as the mass number $A$ increases. This implies that the decrease in $E_{sym}$ seen just below $n_{1/2}$ in Fig.~\ref{LPR-fig} can be trusted, at least qualitatively. This will be confirmed later by other reasoning.

Now how can one understand the increase of $E_{sym}$ in density for $n > n_{1/2}$ giving the cusp structure at $n_{1/2}$?

To answer this question, we note in Eq.~(\ref{mom-of-inertia}) that were it not for the derivatives terms from the Skyrme quartic term, the $E_{sym}$ would either stay flat or increase only very slowly as density goes above $n_{1/2}$. This is because $\overline{\Sigma}=\overline{\la\bar{q}q\ra}\to 0$ as $n\to  n_{1/2}$. {It has been shown~\cite{LMR-cusp}  that the same cusp structure is obtained with the heavy vector degrees of freedom included in HLS Lagrangian}. Just the location of the cusp changes with additional degrees of freedom, with the location tending to go to higher densities the more heavy degrees of freedom are included. The linear density dependence for $n\gsim n_{1/2}$ remains independent of the degrees of freedom.  This feature is indicative of that the cusp is associated with the topology change with the emergence of quasiparticle structure with the half-skyrmions.


\subsection{The cusp and the vector manifestation}
We now describe how the cusp in the $E_{sym}$ influences the scaling with density of the hidden gauge coupling $g_\rho$ to nucleons, the effect first discussed in Ref.~\cite{BR91}. We do this by looking at the symmetry energy in terms of the nuclear tensor forces impacted by the scaling in the coupling $g_\rho$ as a function of density. For this we revisit the basic idea presented in 1991.
\subsubsection {The BR scaling for the nuclear tensor force}
One can address this problem with the full $bs$HLS Lagrangian, but that requires detailed discussions of how the scaling parameters of the Lagrangian are determined, which will be done below. However we do not need the whole battery of the full Lagrangian. For the purpose of illustration of the main idea, it requires only the pion and the $\rho$ meson as the relevant degrees of freedom. Dilaton will enter only indirectly providing the scaling relations for the masses and coupling constants in the Lagrangian. The $\omega$ meson can be considered as integrated out.  Imagine that all other fields than the $\pi$ are dropped from the $bs$HLS Lagrangian. This exercise can be done with the simplified Lagrangian (\ref{LOSS1}). As in Ref.~\cite{BR91}, nucleons are generated as skyrmions from the pion field and the $\rho$ meson is coupled in a gauge-equivalent way.

Given the effective Lagrangian with the bare parameters suitably defined, one can formulate an EFT by computing $n$-body nuclear potentials for $n \geq 2$ as described in Introduction. The {\it only} difference from the standard procedure S$\chi$EFT is that the parameters of the Lagrangian are sliding with density. Since the symmetry energy is dominated by the tensor forces, we focus on the latter. The scalar meson does not contribute at the leading order, hence there are only one-pion and one-$\rho$ contributions to two-body tensor forces. There can be three-body forces at the next power counting order or involving the $\omega$ meson to which we will return below. To see the qualitative feature of the tensor forces in medium,  we use the non-relativistic ($k^2/m_N^{\ast\, 2} \ll 1$) form of the tensor potential, valid in the region we are considering  as the in-medium nucleon mass stays heavy. The tensor potential is given by
\be
V^T_M \left( r \right) &=& S_M \frac{f_{NM}^{\ast\,2}}{4\pi} \tau_1\, \tau_2\, S_{12}{{\cal I}(m^\ast_{M} r)} \, ,\label{tensorM}\\
{{\cal I}(m^\ast_M r)}&\equiv& m_M^\ast \left[ \frac{1}{(m_M^\ast r)^3} + \frac{1}{(m_M^\ast r)^2} + \frac{1}{3m_M^\ast r} \right] e^{-m_M^\ast r} \,,
\label{radial}
\ee
where $M=\pi,\,\rho$, $S_{\rho(\pi)} = +1(-1)$ and
\begin{equation}
S_{12} = 3 \frac{\left(\vec{\sigma}_1 \cdot \vec{r} \,\right)\left(\vec{\sigma}_2 \cdot \vec{r} \,\right) }{r^2} - \vec{\sigma}_1 \cdot \vec{\sigma}_2
\end{equation} with the Pauli matrices $\tau^i$ and $\sigma^i$ for the isospin and spin of the nucleons with $i = 1,2,3$. The asterisk represents in-medium quantities, that is, the density dependence through the scaling parameters~\cite{BR91}.
The strength $f_{NM}^\ast$ scales as
\begin{equation}
R^\ast_M\equiv \frac{f_{NM}^\ast}{f_{NM}} \approx \frac{g_{M NN}^\ast}{g_{M NN}} \frac{m_N}{m_N^\ast} \frac{m_M^\ast}{m_M}\, ,
\end{equation}
where $g_{MNN}$ are the effective meson-nucleon couplings. What is significant in Eq.~(\ref{tensorM}) is that given the same radial dependence, the two forces (through the pion and $\rho$ meson exchanges) come with an opposite sign and this well-known fact plays a crucial role.

The property of the total tensor force -- that we shall simply refer to as ``tensor force" in the singular -- $V^T=V^T_\pi +V^T_\rho$ in medium will then depend on the ratio parameter $R^\ast$ and the masses $m^\ast$. The strength of the net force will crucially depend the cancelation between the two force contributions due to  the scaling of $R^\ast$ and $m^\ast$. If the ratio $R^\ast$ were approximately equal to 1, then the scaling of the masses will dictate how the net tensor force will fall with density. Now if one were to assume that the pion mass falls much less fast,  if at all,  than that of  $\rho$, a reasonable assumption given that the pion is fairly close to a Nambu-Goldstone mode while the $\rho$ is not, then one would expect that the tensor force strength would diminish in nuclei as the effective density of nuclei increases, say, from $^{12}$C to  $^{208}$Pb. In fact this turns out to be what happens in nature.

First we look at the tensor forces\footnote{When written in plural, we mean both the pion and $\rho$ components.} in the density regime of the skyrmion phase, $n < n_{1/2}$, --- that we will associate with the region R--I. There the master scaling relation written down in Ref.~\cite{BR91} and precisely defined in Refs.~\cite{PKLR,MR-PCM}
\be
\frac{m^\ast_N}{m_N} \approx \frac{m^\ast_\chi}{m_\chi} \approx \frac{m^\ast_V}{m_V} \approx \frac{f^\ast_\pi}{f_\pi} \approx \frac{\langle \chi \rangle^\ast}{\langle \chi \rangle}\equiv \Phi, \label{scaling}
\ee
where $V=(\rho, \omega)$, holds in R--I.  To the leading order in the counting involved with both scale and chiral symmetries~\cite{LMR}, the hidden gauge coupling $g_V$ and the dilaton--nucleon coupling $g_{\sigma N}$ do not scale in the skyrmion phase.

Because of the near Nambu--Goldstone structure of the pion with its ``small" mass, the scaling property of the quantities associated with the pion is difficult to pin down to high precision. For instance one would have to figure out the anomalous dimension of the quark mass operator  $\gamma_m$, which is not known. To the best we can ascertain~\cite{PKLR}, it seems reasonable that we can ignore density-scaling of the pionic quantities for all ranges of densities involved\footnote{It may very well be possible to do a highly precise chiral perturbation calculation in pionic atom systems for pionic properties in medium such as the pion mass $m_\pi^\ast$ and the pion decay constant $f_\pi^\ast$. However as discussed in Appendix A of Ref.~\cite{PKLR}, it would require certain fundamental issues -- apart from the $\gamma_m^q$ -- associated with the formalism employed for effective quantum field theory in nuclear systems. It is not clear whether such precision has been attained for the process concerned as discussed in Ref.~\cite{CNDIII} for electroweak processes in light nuclei.}.  We can therefore take
\be
R^\ast_\pi \approx \frac{m_\pi^\ast}{m_\pi}\approx 1\ {\rm  for}\   0\leq n\leq n_c.\label{Rpi}
\ee
where $n_c$ is the putative chiral restoration density. This means that the pion tensor does not change non-negligibly in dense medium. This has been checked numerically in Ref.~\cite{PKLR}.

Now the situation is quite different for the $\rho$ meson.

As mentioned above, if the Suzuki theorem and the vector manifestation (VM) were to hold --- that we are suggesting, then the $\rho$ mass would most likely drop to zero at some high density. As explained, the $\rho$ mass formula $m_\rho^2\sim f_\pi^2 g_\rho^2$ which holds to all loop orders\footnote{This theorem holds with the $O(p^2)$ HLS Lagrangian~\cite{HKY-KSRF}. With higher chiral order terms are added, then the theorem holds modulo $O(m_\rho^2/\Lambda_\chi^2)$ corrections. But if the $\rho$ mass drops in dense matter, then the theorem will remain valid.}, implies that the $\rho$ becomes massless without $f_\pi$ going to zero provided that $g_\rho$ drops to zero. This suggests that the gauge coupling switches from a constant value at $n < n_{1/2}$ to $g_\rho\to 0$ at $n>n_{1/2}$. Therefore while the $\rho$ mass drops in density, the coupling $g_\rho$ must also drop. This means that the ratio $R^\ast_\rho$ behaves
\begin{subequations}
\be
R^\ast_\rho&\approx& 1\quad \quad  \quad \;\;\;\; {\rm for}\ n\leq n_{1/2}, \label{RrhoI}\\
&\approx& \left(\frac{g_\rho^\ast}{g_\rho}\right)^2 \quad\ {\rm for}\ n> n_{1/2}.\label{RrhoII}
\ee
\end{subequations}

Since the nuclear symmetry  is dominated by the symmetry energy,  we first look at how it behaves in terms of the resulting tensor force. For qualitative aspect, we can use the closure approximation of  the two-nucleon diagram with iterated tensor forces~\cite{brown-machleidt}
\be
E_{sym}\approx c \, \frac{\la (V^T)^2\ra}{\Delta E}\, , \label{closure-sum}
\ee
where $c> 0$ is a known constant and $\Delta E\approx 200-300$ MeV is the energy of the intermediate state to which the tensor force connects dominantly from the ground state. This of course ignores other components of the nuclear force than the tensor and also fluctuations around the dominant intermediate state, so it should be taken as a sort of mean field approximation, giving a semi-quantitative estimate at near $n_0$.
\subsubsection{Symmetry energy from scaling hidden gauge coupling}
We begin by obtaining the tensor force that results from the scalings  (\ref{scaling})-(\ref{RrhoII}).
 \begin{enumerate}
 \item $\mathbf E_{sym}$ {\bf without half-skyrmions}: Suppose there is no topology change. Then  $R^\ast_\rho\approx 1$ will hold for all ranges of density.  The tensor forces will continue to cancel because of the dropping mass of $\rho$. This feature is shown in  Fig.~\ref{vtwo} with some exemplary parameters indicated therein.
 \begin{figure}[ht!]
\begin{center}
\includegraphics[width=8.5cm]{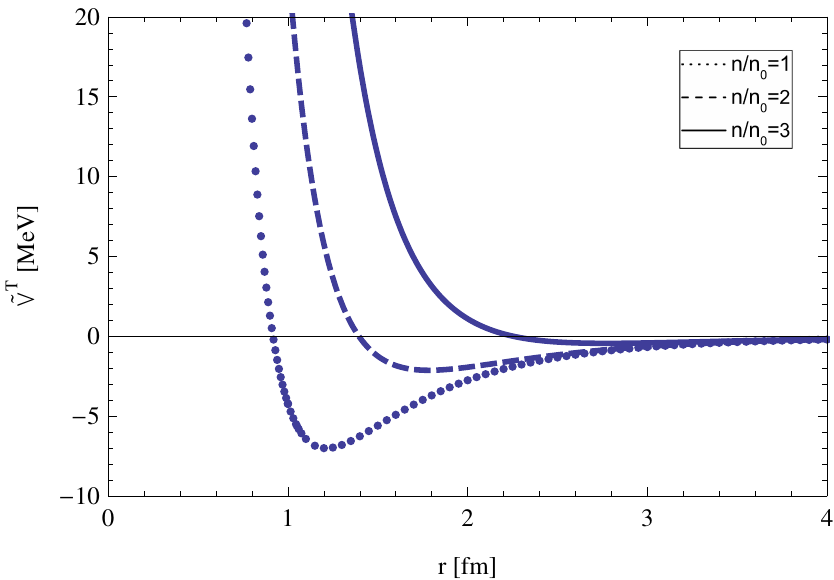}
\vskip -0.5cm
\caption{Net tensor force $\tilde{V}^T\equiv (\tau_1 \cdot \tau_2 S_{12})^{-1} (V_\pi^T +V_\rho^T)$ in units of MeV with   $\Phi \approx  1-0.15 n/n_0$ and $R^\ast_\rho\approx 1$ for all $n$. }\label{vtwo}
\end{center}
\end{figure}
We see that the strength of the tensor force attraction in the range  effective in nuclear interactions decreases as density increases, going to near zero at $n\sim 3n_0$ . Up to $n\sim n_0$,  this property can be confronted with nature and is seen to be consistent with observables in light nuclei.  A striking  case for  this feature of decreasing tensor force  is the famous long life-time for the C-14 dating, with the near vanishing of the Gamow--Teller matrix element at $n\lsim n_0$ explained by the interplay of the $\pi$ and $\rho$ tensor forces~\cite{C14}.

Thus with the tensor force so obtained without topology change,  $E_{sym}$ should continuously decrease. This scenario is clearly at odds with the cusp structure found in the skyrmion crystal prediction.

\item $\mathbf E_{sym}$ {\bf with skyrmion-half-skyrmion transition}: Now consider the topology change at $n_{1/2}$. Then we can take the scaling (\ref{RrhoII}) with $(g_\rho^\ast/g_\rho)^2\neq 1$ with all others the same as the case without topology change. Precisely how the ratio $R_\rho^\ast$ decreases from 1 for $n\geq n_{1/2}$ will be described below. For illustration let us take $R^\ast_\rho\approx \Phi^2$. The VM fixed point dictates that it goes to zero at some density $n_{VM}\sim n_c$. It will turn out for compact stars to drop differently from what is taken in this calculation.

\begin{figure}[ht!]
\begin{center}
\includegraphics[width=8.0cm]{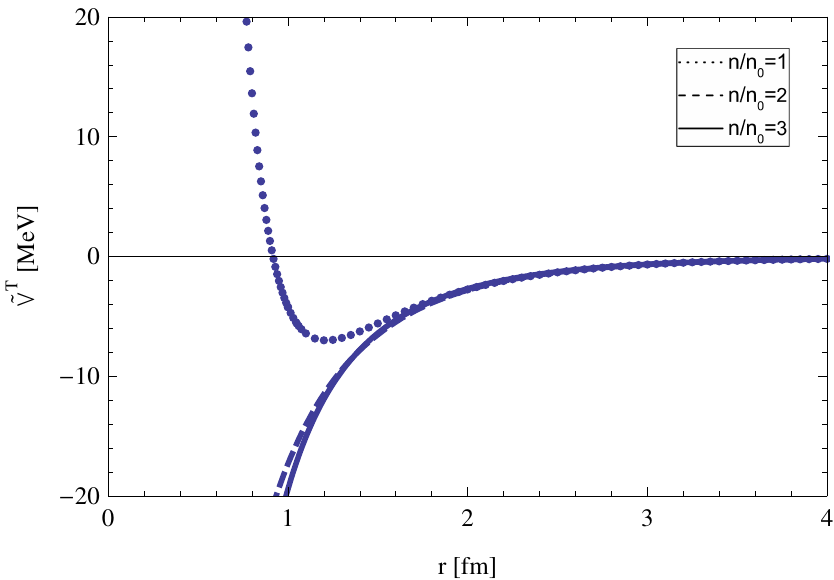}
\vskip -0.5cm
\caption{The same as Fig.~\ref{vtwo} with $\Phi \approx 1-0.15 n/n_0$  with  $R\approx 1$ for $n<n_{1/2}$ and $R^\ast_\rho\approx \Phi^2$ for $n>n_{1/2}$, assuming $ n_{1/2} \approx 2n_0$. }\label{vtnew}
\end{center}
\end{figure}

The result  shown in Fig.~\ref{vtnew} illustrates how the scaling of the gauge coupling affects the behavior of the tensor as the half-skyrmion phase sets in. The effect is dramatic. At the cross-over density, the $\rho$ tensor is almost completely suppressed, leaving the pion tensor to take over. The net tensor force, passed $n_{1/2}$, then is nearly the same as the pion tensor for $r\gsim 1$ fm. In reality one expects the changeover to be smooth, given that the topology change is not a phase transition. What is clear from Eq.~(\ref{closure-sum}), however, is that going toward to $n_{1/2}$ from below the symmetry energy is to drop and more or less abruptly turn over at $n_{1/2}$ and then increase beyond $n_{1/2}$. This gives precisely the cusp predicted in the crystal calculation.

The upshot of this result is that the cusp structure in $E_{sym}$, a consequence of topology change with the onset of the half-skyrmion phase, is signaling the changeover in the property of the gauge coupling from $n\leq n_{1/2}$ to $n>n_{1/2}$.

As a prelude to what is to come in confronting nature, we plot in Fig. \ref{Esym-vlowk} the symmetry energy obtained in the Wilsonian renormalization-group treatment~\cite{PKLMR} -- called $V_{lowk}$ RG approach to be explained blow -- with the $bs$HLS EFT Lagrangian.
\begin{figure}[ht!]
\begin{center}
\includegraphics[width=8cm]{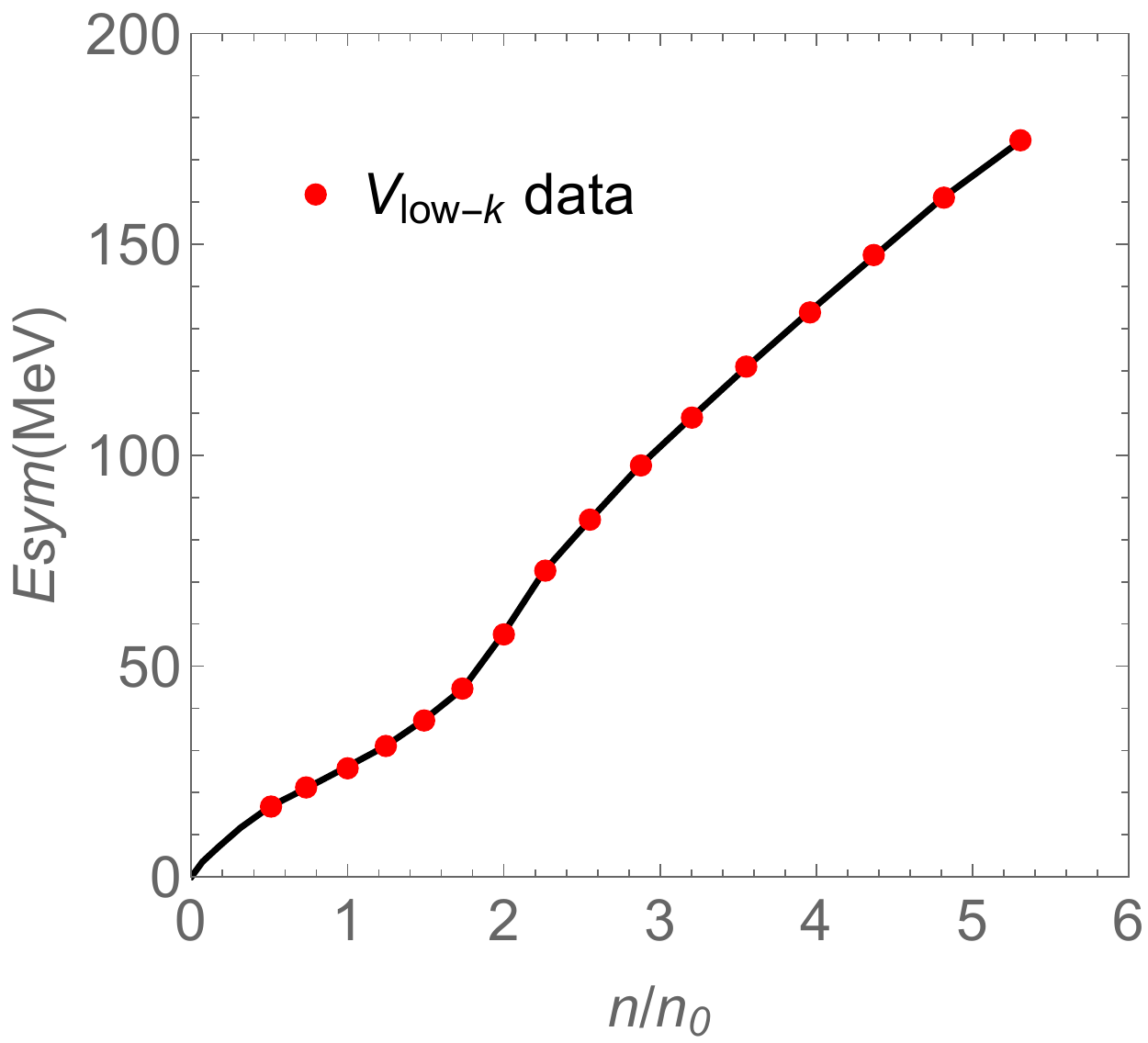}
\vskip -0.5cm
\caption{$E_{sym}$ (solid circle) obtained in the {\it full} $V_{lowk}$RG approach for $n_{1/2}= 2 n_0$ discussed in Section \ref{compactstar}.  This result is reproduced almost exactly by the pseudo-conformal model introduced below. The approach with higher order terms beyond the mean-field approximation described later smoothens the cusp singularity as well as  correctly treats the density regime $< n_{1/2}$. The solid line reproduces the linear density dependence of $E_{sym}$ for $n\gsim n_{1/2}$.}\label{Esym-vlowk}
\end{center}
\end{figure}
The cusp point given by the tensor-force structure is set, for illustration, at $n_{1/2}=2n_0$. The RG approach includes higher-order corrections in $1/\bar{N}\sim 1/k_F$ and hence smoothens the singularity structure present in both the crystal calculation and  the tensor force closure-sum calculation. It is noteworthy that after smoothing in the range $\sim (2-3)n_0$, the increase of $E_{sym}$ linear in density seen in the topology change present in the crystal calculation is reproduced exactly in this full RG calculation.

\end{enumerate}
\subsection{Quasiparticles in the half-skyrmion phase}\label{quasiparticles}
We  have thus far resorted to the mean-field approximation with the $bs$HLS EFT Lagrangian. We recall that given a well-defined EFT Lagrangian with the intrinsic inputs from QCD,  doing mean-field calculation can be considered to be equivalent~\cite{matsui} to doing Landau--Fermi liquid fixed-point theory~\cite{shankar}.  The question to ask is then whether the mean-field results obtained above can be reproduced in the skyrmion crystal calculation with the cusp structure at $n > n_{1/2}$. We now argue that the state in the region $n >n_{1/2}$ of the cusp supports half-skyrmions behaving as nearly non-interaction quasiparticles.

We show that the half-skyrmion phase in the skyrmion-crystal simulation is in a state that can be described almost entirely by mean fields, largely undistorted by strong interactions. This resembles Landau--Fermi liquid fixed point theory where the $\beta$ function for the quasiparticle interactions is suppressed. This striking feature was first found in the Skyrme model  with the Atiyah--Manton ansatz in Ref.~\cite{atiyah-manton-skyrmion}. Here we will show the phenomenon using  the HLS Lagrangian~\cite{PKLMR}.\footnote{Since what matters as in the structure of the tensor forces is the topology and the symmetry involved, largely independent of strong interactions mediated by non-topological fields, the same argument should apply to the dilaton in the half-skyrmion phase in $s$HLS.}

We write the chiral field $U$ as $U(\vec{x}) = \phi_0(x,\, y,\, z ) + i \phi^j_\pi(x,\, y,\, z )\tau^j$  with the Pauli matrix $\tau^j$ and $j=1,2,3$. Including $\rho$ and $\omega$, we write the fields placed in the lattice size $L$ as $\phi_{\eta,\, L}(\vec{x}\,)$ with $\eta =0,\, \pi,\, \rho,\, \omega$ (where the subscripts $0$ and $\pi$ represent the two components of the $U$ field and $\rho$ and $\omega$ the vector fields) and normalize them with respect to their maximum values denoted $\phi_{\eta,L,{\rm max}}$ for given $L$.  It can be shown, as in Ref.~\cite{atiyah-manton-skyrmion}, with HLS that in the half-skyrmion phase\footnote{The precise value of the half-skyrmion density which depends on the parameters is not important for our discussions.} with $L\lsim L_{1/2}$ where  $L_{1/2}\simeq2.9$ fm, the field configurations are invariant under scaling in density as the lattice is scaled from $L_1$ to $L_2$
\be
\frac{\phi_{\eta,\,L_1}(L_1\vec{t}\,)}{\phi_{\eta,\,L_1,\,{\rm max}}}= \frac{\phi_{\eta,\,L_2}(L_2\vec{t}\,)}{\phi_{\eta,\,L_2,\,{\rm max}}}. \label{scaling1}
\ee
Since other fields are quite similar with the pion field controlling the topology, we only show in  Fig.~\ref{scale_inv} the case of $\phi_{0,\pi}$ for $\phi_{0,\pi} (t, 0,0)$ vs. $t$ with $t\equiv x/L$. What is seen there is that density-scale invariance sets in for $L\lsim L_{1/2}$. One can see that the field is independent of density in the half-skyrmion phase with $L\lsim L_{1/2}$  whereas for the skyrmion phase with lower density with $L >  L_{1/2}$, it is appreciably dependent on density.
\begin{figure}[h]
\begin{center}
\includegraphics[width=6.0cm]{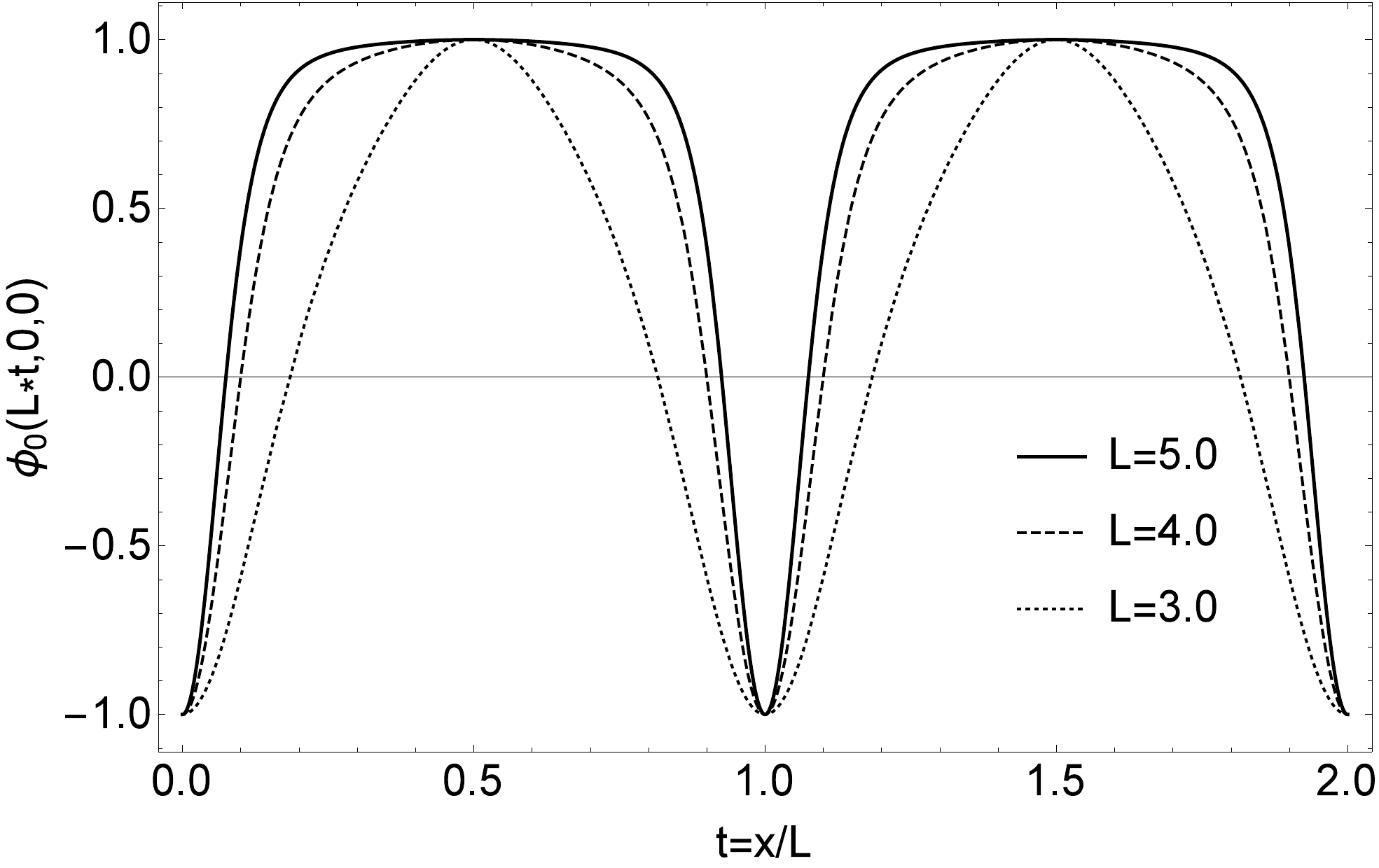}
\includegraphics[width=6.0cm]{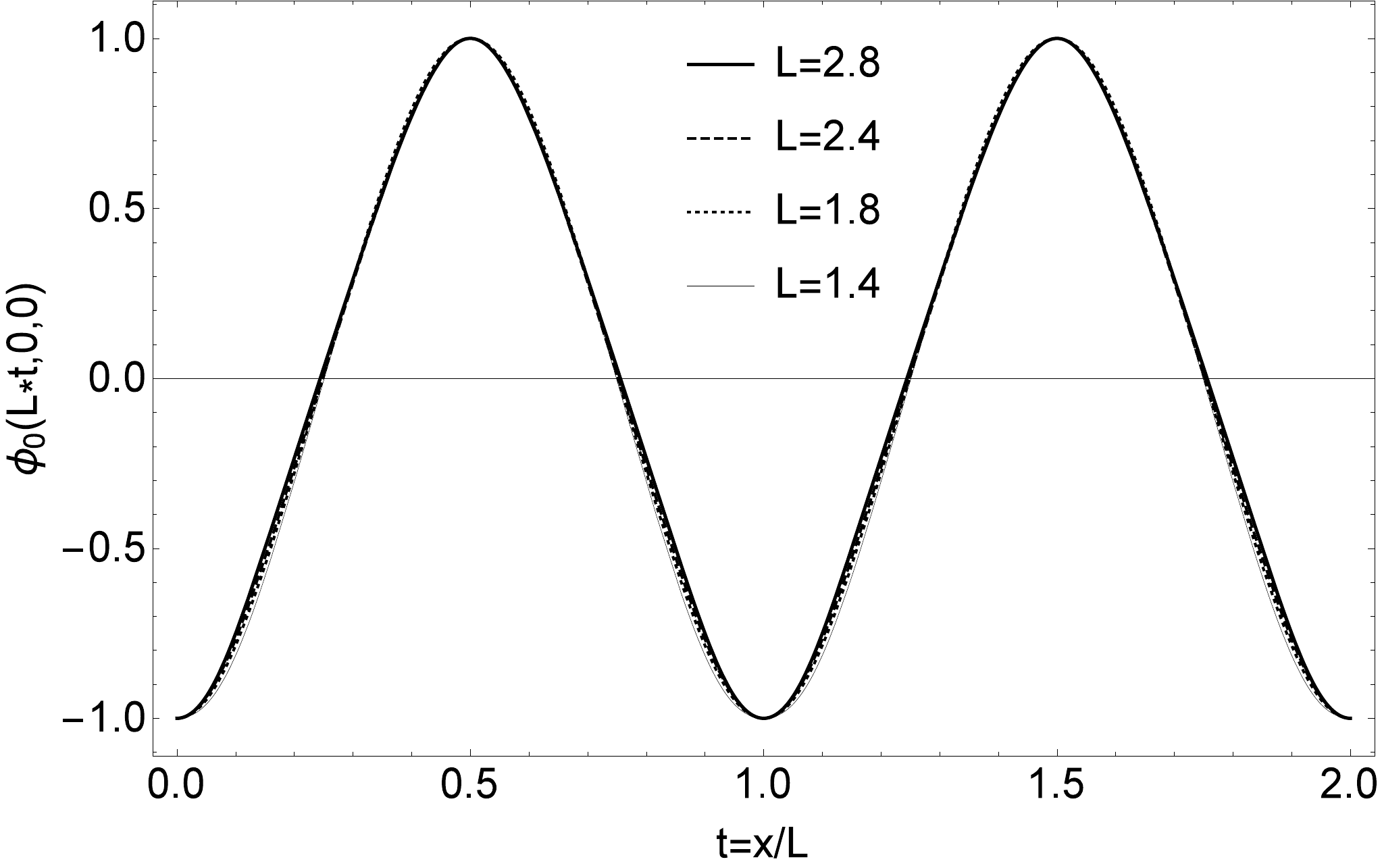}
\includegraphics[width=6.0cm]{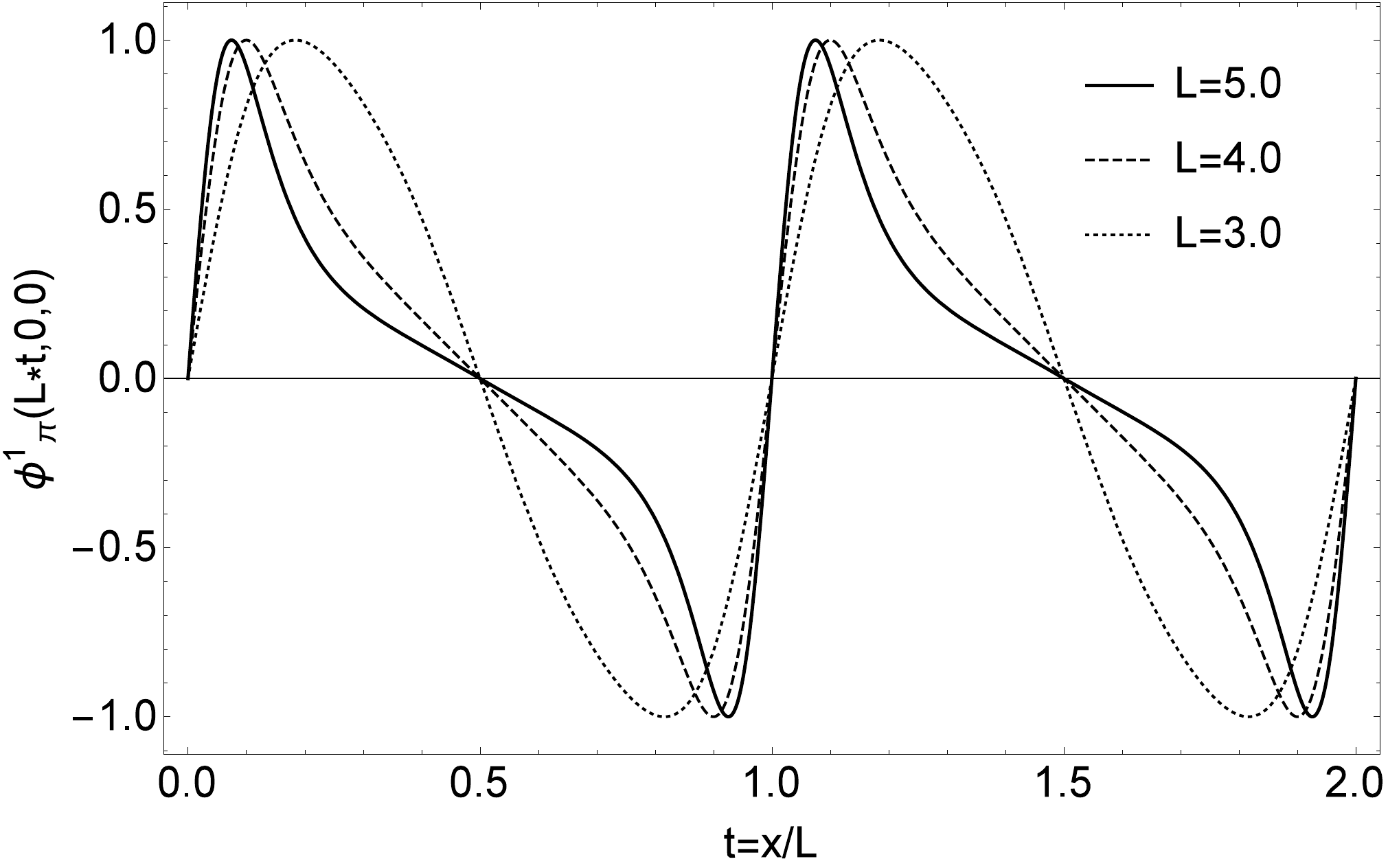}
\includegraphics[width=6.0cm]{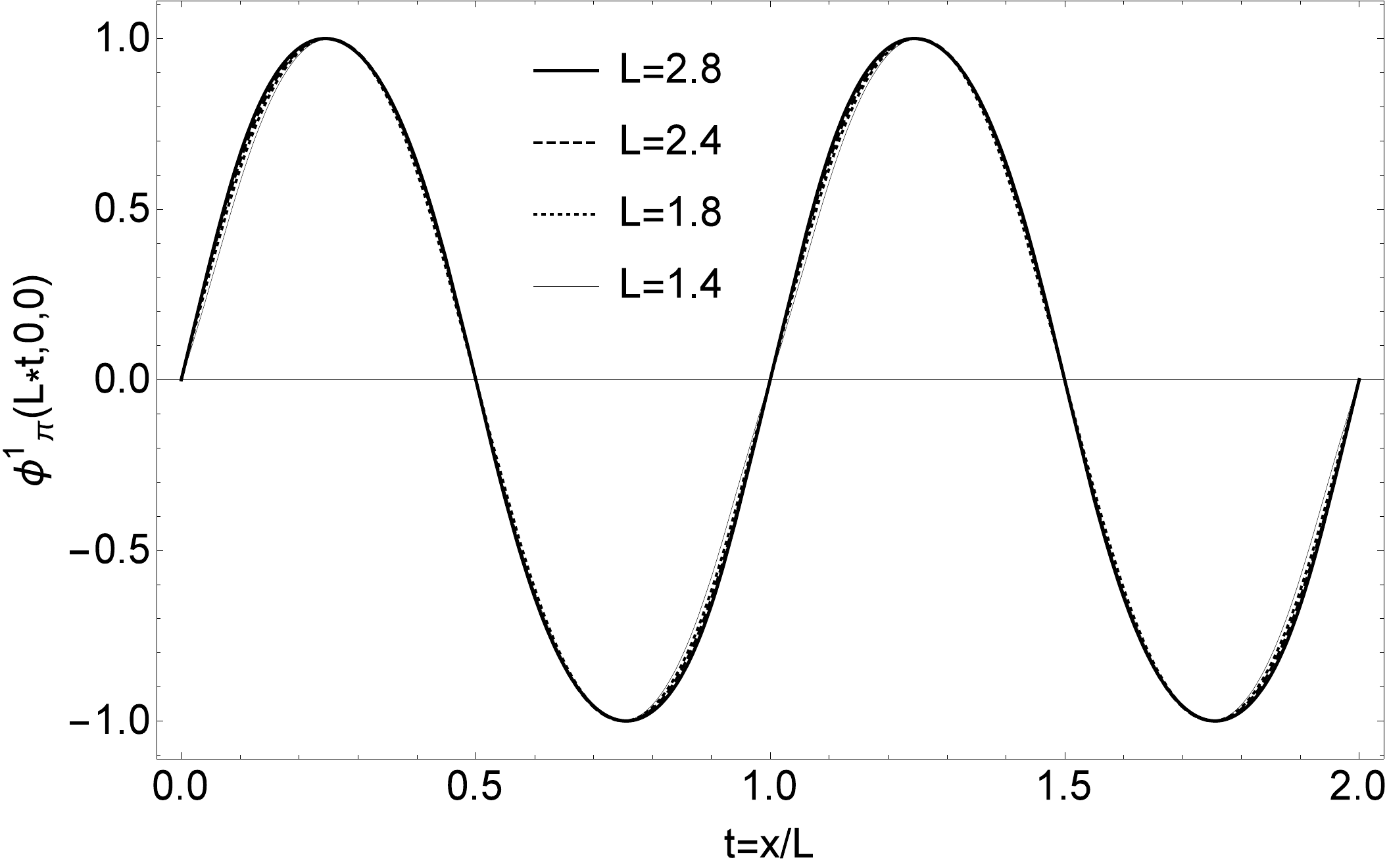}
\caption{  The field configurations $\phi_0$ and $\phi^1_\pi$  as a function of $t = x/L$ along the y = z = 0 line. The maximum values for $\eta=0,\pi$ are $\phi_{0,\,L,\, {\rm max}} = \phi_{\pi,\,L,\, {\rm max}} = 1$. The half-skyrmion phase sets in when $L=L_{1/2} \lsim 2.9\,{\rm fm}$.}
\label{scale_inv}
\end{center}
\end{figure}
What does this imply for the energy density?

The energy density for the skyrmion matter put on the lattice of lattice size $L$ can be written as
\begin{eqnarray}
\epsilon & = & E/A/V(=L^3) \nonumber\\
& = & \frac{1}{L^3} \int^L_0 d^3x \sum_{n,\,m} c_{n,m}\, f_{n,m}\left( \vec{\nabla}_x,\, \phi_{\eta,\,L}(\vec{x}\,)\right)\,,
\end{eqnarray}
where $c_{n,m}$ is the coefficient of $f_{n,m}$ which is the function of $\vec{\nabla}_x$ and $\phi_{\eta,\,L}(\vec{x}\,)$ having $n$th power of $\vec{\nabla}_x$ and $m$th power $\phi_{\eta,\,L}(\vec{x}\,)$ with $\nabla_{x,\,j} = \frac{\partial}{\partial\,x^j}$. One can reduce it to
\begin{eqnarray}
\epsilon
&=& \sum_{n,\,m} \left(\frac{1}{L} \right)^n \left(\phi_{\eta,\,L,\, {\rm max}} \right)^m \int^L_0 \frac{d^3x}{L^3}\, c_{n,m}\, f_{n,m}\left( L \vec{\nabla}_x,\, \frac{\phi_{\eta,\,L}(\vec{x}\,)}{\phi_{\eta,\,L,\, {\rm max}}}\right) \nonumber\\
&=& \sum_{n,\,m} \left(\frac{1}{L} \right)^n \left(\phi_{\eta,\,L,\, {\rm max}} \right)^m \int^1_0 d^3t\, c_{n,m}\, f_{n,m}\left( \vec{\nabla}_t,\, \frac{\phi_{\eta,\,L}(L\vec{t}\,)}{\phi_{\eta,\,L,\, {\rm max}}}\right) \nonumber\\
&=& \sum_{n,\,m} \left(\frac{1}{L} \right)^n \left(\phi_{\eta,\,L,\, {\rm max}} \right)^m  A_{n,\,m}\,, \label{energy_den}
\end{eqnarray}
 where $A_{n,\,m}$ is a constant independent of the lattice size $L$.

 Calculating the energy density (\ref{energy_den}) in skyrmion-crystal simulations involves field configurations satisfying their equations of motion. Hence (\ref{energy_den}) is a mean field expression. It captures all essential dynamics in terms of the mean fields of each degrees of freedom involved, with residual interactions suppressed. The density dependence lies, apart from the $(1/L)^n$ factor, in the maximum field configuration $\left(\phi_{\eta,\,L,\, {\rm max}} \right)^m$. This implies that in the half-skyrmion phase,  considered to set in at high density, the mean-field structure dominates. This agrees with the lore that at high density -- and in the large $N_c$ limit, the skyrmion crystal picture becomes valid  in QCD. In clear contrast, however, as one can see in Fig.~\ref{scale_inv}, the mean-field structure breaks down in the lower-density phase with $L > L_{1/2}$.  This also agrees with the understanding that the property of low-density baryonic matter -- including nuclear matter -- may be poorly captured in crystal.

In Ref.~\cite{HKL-MR}, the topology change was interpreted as change from a Landau Fermi-liquid state to a non-Fermi liquid state. It was considered in terms of normal baryonic degrees of freedom and as such one could think of the changeover as a breakdown of {\it baryonic} quasiparticle picture. However what we have here is different. The half-skyrmions are not {\it normal} baryons, with properties basically different from them. In terms of the structure discussed above as a conjecture in Section \ref{conjecture}, they could be dual excitations of quasi-quarks associated with a topological field structure of the fractional quantum Hall type with domain walls.  This picture may be related to the quarkyonic quasiparticles~\cite{quarkyonic,quarkyonic-jeong}. It should be recognized that as the DLFP is approached, the half-skyrmion quasiparticle picture must of course break down and go into  a non-Fermi-liquid state.

$\bullet$ { \bf Proposition VII: \it  The cusp singularity in  the symmetry energy $E_{sym}$ in the crystal description of dense matter at the leading $N_c$ order is caused  by the topology change at the transition density $n_{1/2}$, driven by ``heavy" hidden symmetry degrees of freedom. It leads to the appearance of half-skyrmion quasiparticles, reproduced by a drastic change of the nuclear tensor force driven by the approach to the dilaton limit fixed point $g_{\rho NN}\to 0$ and vector manifestation fixed point with $m_\rho\to 0$ at $n\gg n_{1/2}$.}

\subsection{Trading-in topology change for hadron-quark continuity}
Based on what has been developed above we will now argue that the topology change embodied in the Cheshire Cat phenomenon is the dual  to the hadron-quark continuity expected at some high density.
\subsubsection{Cheshire Cat for skyrmion-to-half-skyrmion transition}
The presently quoted density range for the crossover for compact stars is in the vicinity of $\sim (2-5)n_0$ as reviewed in Ref.~\cite{baym-kojo}. It may be that the microscopic QCD degrees of freedom figure to a much higher density where color-flavor locked  superconductivity may set in. We will not need to go that far for the phenomena we are interested in. Nevertheless we will suggest that there is no difficulty in principle in extending the Cheshire Cat Principle all the way to the color-flavor locking.

For the moment, we  focus on the topology change phenomenon with the cusp structure described in Section \ref{hidden-symmetries}.  We examine what the cusp implies for the ratio $m_N^\ast/m_N$ in the skyrmion crystal. Since the property of the symmetry energy in the half-skyrmion phase is dictated by topology we expect the effective nucleon mass to reach a constant $\sim m_0$. Indeed this is what is found in Ref.~\cite{MHLOR2014}. The result is shown in Fig.~\ref{solitonmass}. This result is obtained for the soliton mass in HLS with $(V_\mu=\rho,\omega)$ with $V_\mu\in U(2)$. As seen above, the $U(2)$ symmetry is broken as the DLFP is approached, hence the figure does not give the density regime where the scale symmetry is restored. The reason is that the dilaton $\chi$ is not included, therefore the interplay between the $\chi$ and the $\omega$ that enters in the nucleon mass is missing. Thus it is expected not to hold at high density beyond, say, at $n\gsim 7n_0$, much below the density  $n_{\rm vm} \sim 25n_0$, the density at which the vector manifestation fixed point is  located in the full renormalization-group formalism, ``$V_{lowk}$," to be defined below.

The feature seen in Fig. \ref{solitonmass} is generic, with the in-medium soliton mass representing the effective nucleon mass (and also the effective pion decay constant) going to a constant $\sim (0.7- 0.8)$ times the vacuum value at the topology change. This is predominantly, if not entirely, due to the space-averaged quark condensate going to zero at $n_{1/2}$.  We can see this simply as follows. With the chiral field written as $U=\phi_0 +i{\bm{\tau}\cdot{\bm\phi}}$,
\be
\left(\frac{f_\pi^\ast}{f_\pi}\right)^2 \approx1-\frac 23 (1-\la\phi_0\ra^2) \to \frac 13 \ {\rm as}\  \phi_0\to 0.
\ee
The same behavior holds for the nucleon mass.
\begin{figure}[h]
\begin{center}
\includegraphics[width=7.0cm]{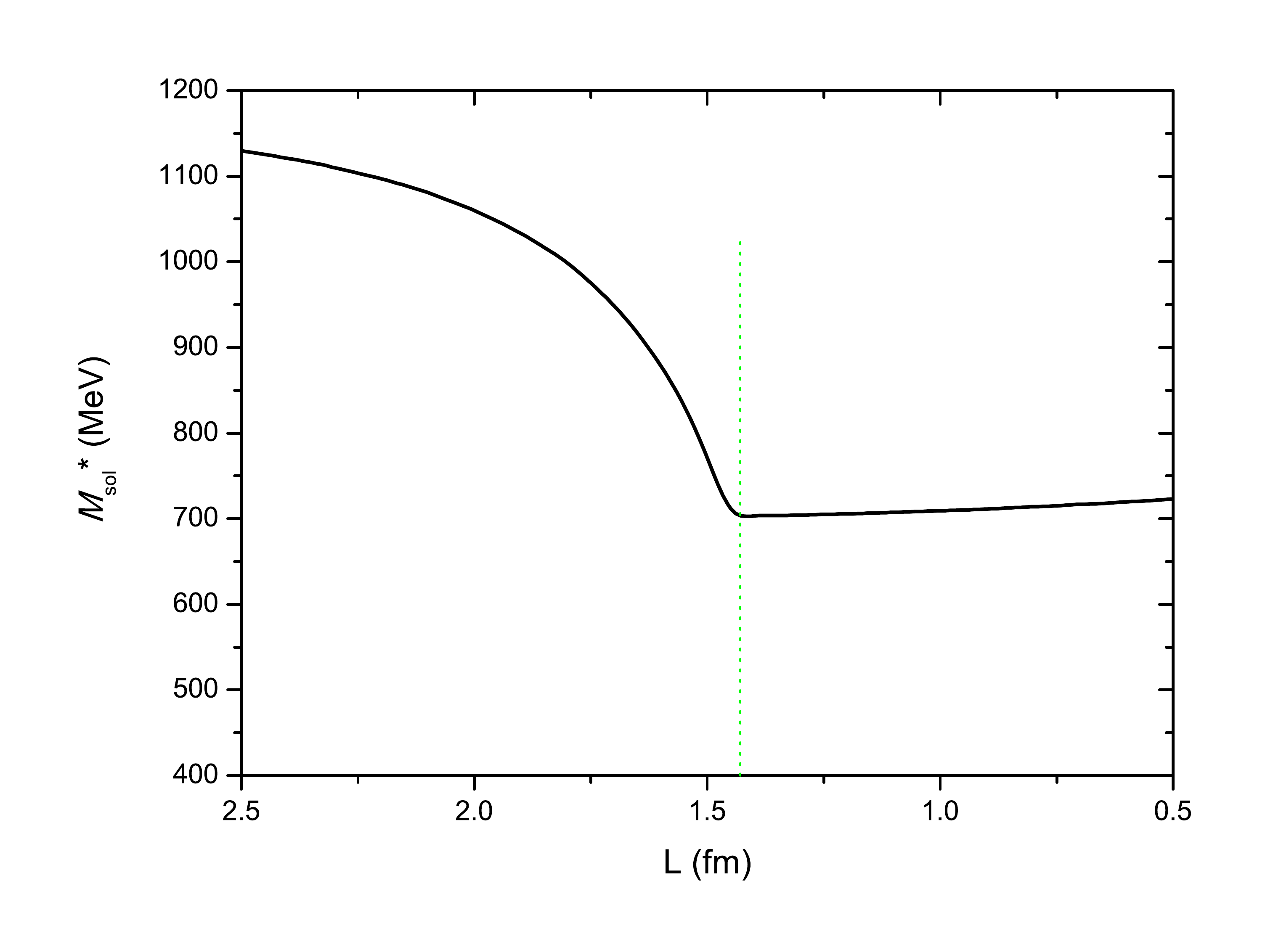}
\caption{The skyrmion soliton mass vs. lattice size $L$ in HLS. {The vertical line indicates the position of the topology change.}}\label{solitonmass}
 \end{center}
\end{figure}

In what follows in the application of the formalism summarized by the {\bf Propositions}, we adopt the thesis that the putative hadron-quark continuity expected to be operative for compact-star matter for $n\sim (2-7)n_0$ can be replaced by the topology change from skyrmions to half-skyrmions at $n_{1/2}\gsim (2-4)n_0$. How this can be justified in a general framework of the Cheshire Cat Principle has not yet been formulated. It will be presented as a conjecture below. Here we briefly suggest that this Cheshire Cat Principle could be extended to much higher densities than what is directly relevant for compact stars.
\subsubsection{Cheshire Cat for the color-flavor locking}
In a way to dramatize our assertion on the Cheshire Cat Principle, we skip all the intermediate densities and jump to the asymptotic density at which the color-flavor locking is to take place.

At asymptotic high densities,  the three light quarks, $u$, $d$ and $s$ which can be  taken to be massless, pair into diquarks with color and flavor locked
forming the color-flavor-locked (CFL) phase~\cite{CFL} and  the CFL
phase exhibits a spectrum with matching quantum numbers to that of
the zero density phase~\cite{SW-continuity}. We argue that this is a manifestation of the CC phenomenon at superhigh density. In the framework developed here, the CFL phase does not actually figure in the description of compact stars, but it is highly relevant to the notion developed here, namely, the notion of hadron-quark continuity from zero density to superhigh density. We illustrate this with a baryon in the CFL phase as described as skyrmion, that we call ``superqualiton"~\cite{HRZ}. The superqualitons carry quantum numbers $B=Y=(1\ {\rm mod}  2)/3$ and $S=1/2$.

We start with a brief review of  the CFL for the essence of the idea~\cite{CFL}.

At very high density, quarks with opposite Fermi momenta tend to
pair with color and flavor locked and get condensed.
The pertinent condensate can be taken in the
$(\overline{3},\overline{3})$ color-flavor representation
as
\be
\Big< q_{L\alpha}^{ia}q_{L\beta}^{jb} \Big>
={}-\Big<q_{R\alpha}^{ia}q_{R\beta}^{jb}\Big>
=\kappa \,\,\epsilon^{ij}\epsilon^{abI}\epsilon_{\alpha\beta I}\, .
\label{CFL}
\ee
Here $\kappa$ is some constant,
$i,j$ are $SL(2,C)$ indices, $a,b$ are color indices,
and $\alpha,\beta$ are flavor indices.
For finite $\kappa$, both global color $SU(3)_c$
and flavor $SU(3)_{f,L,R}$ symmetries are broken.
The flavor-color locking in Eq.~(\ref{CFL}) implies spontaneous breaking
through the color-flavor diagonal. Specifically:
$SU(3)_c\times\left( SU(3)_L\times SU(3)_R\right)\rightarrow
SU(3)_{c+L+R}$, with the emergence of 8 pseudoscalar
Nambu-Goldstone (NG) bosons, denoted below as $\Pi$,   together with 8 scalar NG
bosons $\Sigma$ to be ``eaten up"  by gluons through the Higgs mechanism.
There is an extra NB boson associated with
$U(1)_B\rightarrow Z_2$, but it has no relevance to our discussion. So we will ignore it. In an exact analogy to the octet pions at low density,  we introduce the chiral fields
\be
 U&=& \xi^\dagger_L\xi_R=e^{2i\Pi/F_\pi},\label{U}\\
\xi_{L,R}&=& e^{\mp i\Pi/F_\pi} e^{i\Sigma/F_\sigma}
\ee
where $\xi_{L}$ is a map from space-time to
$M_{L}=SU(3)_c\times SU(3)_{L}/SU(3)_{c+R}$
to describe the excitations of the right-handed
diquark condensate and likewise with $L\leftrightarrow R$.
Under  $SU(3)_c\times SU(3)_L\times SU(3)_R$
transformation by unitary
matrices $(g_c,g_L,g_R)$, $\xi_L$ transforms as $\xi_L\mapsto
g_c^*\xi_Lg_L^{\dagger}$ and $\xi_R$ transforms as
$\xi_R\mapsto g_c^*U_Rg_R^{\dagger}$.

The effective Lagrangian for the CFL phase
at asymptotic densities follows by integrating out the `hard' quark
modes at the edge of the Fermi surface.
The effective Lagrangian for $U_{L,R}$ is a standard non-linear sigma model in D=4
dimensions and should include the interaction of NB
bosons with colored but ``screened" gluons $G$.
The $SU(3)_c$ current of NB bosons in the CFL phase consists of two terms,
\begin{eqnarray}
{J_{cL}}^{A\mu} & = & {i\over 2}F^2{\rm Tr} \left(U_L^\dagger T^A\partial^{\mu}U_L \right) \nonumber\\
& &{} +
{1\over 24\pi^2}\epsilon^{\mu\nu\rho\sigma}
{\rm Tr} \left(T^AU_L^\dagger\partial_{\nu}U_LU_L^\dagger\partial_{\rho}U_L
U_L^\dagger \partial_{\sigma}U_L \right) ,
\end{eqnarray}
where the first term is the Noether current and the second one is from the Wess--Zumino term.
Expanding in powers of derivative, the effective Lagrangian for
the (colored) NB bosons is then
\begin{eqnarray}
{\cal L} & = & \frac {F_T^2}4 \,\,{\rm Tr} (\partial_0 U_L\partial_0 U_L^{\dagger})
-\frac {F_S^2}4 \,\,{\rm Tr} (\partial_i U_L\partial_i U_L^{\dagger})\nonumber\\
& &{}
+g_sG\cdot J_{cL}+n_L {\cal L}_{WZW} + (L\rightarrow R)\nonumber\\
& &{}+
{\cal O} \left( \frac {\partial^4}{(4\pi\Delta)^4}\right).
\label{2x}
\end{eqnarray}
The coefficient $n_{L,R}$ of the Wess--Zumino terms is fixed to 1 by color-flavor anomalies. Note that in the presence of a chemical potential, Lorentz symmetry is broken down to $O(3)$. In carrying out the derivative expansion, the (superconducting) scale  $4\pi\Delta$ is taken to be large.
The temporal and spatial decay constants $F_{T,S}$ can be fixed by the
`hard' modes at the Fermi surface.

As one can see from the $U$ field defined in Eq.~(\ref{U}), there is the same redundancy as with hidden local symmetry (HLS) at low density and hence  a local gauge symmetry. (Unlike in the case of low-energy gauge fields, however, here the local symmetry is  explicit, i.e., color symmetry, which is {\it not} hidden.) One can therefore generalize Lagrangian (\ref{2x}) to a local gauge invariant form in analogy to HLS.

Much like the effective Lagrangian for QCD at low density (giving rise
to skyrmions), the low-energy effective Lagrangian in the CFL phase admits a stable (static) soliton solution, with a
winding number given by the homotopy $\pi_3(M)=Z$. It has a baryon number $1/3$ and resembles the qualiton introduced by Kaplan~\cite{qualiton} for the constituent quark. Unlike the qualiton in the vacuum which was found unstable,  at superhigh density, the superqualiton is stabilized by the balance between the kinetic energy (attractive force) and the Coulomb energy (repulsive force).
The soliton quantum numbers are determined by the
Wess--Zumino  term upon quantization as in low-energy QCD. The usual collective quantization then gives a spin-half particle transforming under the fundamental representation of both the flavor group and the color group, which leads to a massive left-handed or right-handed quark in the CFL phase. The details of the picture are given in \cite{HRZ} that we do not go into here. The point stressed here is that the HLS structure of the low-density sigma model and the color gauge symmetric sigma model of high density are continuously connected by the skyrmion description, as a manifestation of a CC phenomenon.   There is a growing evidence this sort of continuity involving various aspects of topology in hadron physics.  Just to illustrate the point,  it has been shown  that the pion decay constant $f_\pi$ continuously evolves from low density to high density, say,  up to the CFL density~\cite{song-baym}.
\subsubsection{Cheshire Cat for hadron-quark continuity: A Conjecture}\label{conjecture}
It will be found later in the confrontation with Nature  that the cusp structure in the symmetry energy $E_{sym}$ -- predicted by the topology change -- provides a simple mechanism for the putative soft-to-hard change in the EoS for compact stars at $n\sim 2n_0$ needed to account for the observed massive $\sim 2 M_\odot$.
The question is whether or how the cusp structure in the topology change represents the ``quark deconfinement" process. We have no clear answer to this question. However we can offer a conjecture on how one can establish the connection.

First recall the Cheshire Cat mechanism~\cite{CC-MNRZ} for the fractional Quantum Hall droplet for the one-flavor topological baryon~\cite{ZOHAR}. The connection was made for the $N_f=1$ case and the argument can be extended to the case where $N_f=2,3$ that we are concerned with in this review. But it was not clear how the world for $N_f >1$ is connected to the world for $N_f=1$. There must be a connection since the $\Delta (3,3)^{++}$ that exists in the $N_f>1$ world  must also exist in the $N_f=1$ world.

Now suppose the $\eta^\prime$ becomes light as is expected at high density. Then the FQH pancakes could become relevant as density increases and figure in dense matter in a form of a stack of FQH pancakes. Interactions must then induce  the $N_c$ quarks with the fractional (1/$N_c$) baryon charge living on the boundary of the pancakes could tunnel between the pancakes. This could lead to sheets of fractional baryon-charged topological objects in (3+1) dimensions. In fact in recent developments of skyrmion crystal calculations of dense matter, one finds certain configurations unstable at low density but stabilized at high density of sheets with half-baryon charged objects called ``lasagnes"~\cite{PPV} and  also  with $1/q$-charged baryons in tube configurations with baryons living on the surface of the tube~\cite{canfora}. It seems not impossible that the layers of FQH droplets in (3+1) dimensions give rise to deconfined quasiparticles dual to quarks of fractional charges, e.g, half-skyrmions. Such deconfinement can take place in the presence of domain walls as in some condensed matter systems~\cite{sandvik}. This means that the half-skyrmions probed in the density regime $n> n_{1/2}$ as in Section \ref{quasiparticles} could be deconfined as in the N\'eel-VBS deconfined quantum critical transition~\cite{senthil,sandvik}.

It is however important to recognize that the half-skyrmion phase figuring in the lasagnes~\cite{PPV} or even in Section \ref{paritydoubling} cannot be the naive crystal configurations coming from the standard skyrmions because the half-skyrmions lodged in light nuclei, i.e., $\alpha$ particle, are  not deconfined, because,   separated,  their energies diverge~\cite{cho,nitta}. This implies that certain rearrangement of the vacua of the type associated with coupled FQH droplets must intervene for the deconfinement to take place.

There is a highly intriguing analogy between the fractional quantum Hall effect in condensed matter and in the Cheshire Cat mechanism for dense matter developed later. It is in the approach by Hu and Jain~\cite{jain} based on the Kohn-Sham theory of the FQHE.  The effective field theory, $Gn$EFT, constructed and applied to compact-star matter, belongs to the class of (covariant) density functional theory~\cite{piekarewicz}. Common in both is Chern-Simons topological field theory. How a connection can be made between the density functional theory developed in this work and FQH droplets brought in by the CCP is being investigated.

}
$\bullet$ { \bf Proposition VIII: \it The topology change  with the cusp singularity at $n_{1/2}$ is a dual, via Cheshire Cat, to the hadron-quark continuity in QCD responsible for  the  soft-to-hard change in the EoS.}

\section{Effective Field Theory for Baryonic Matter}
In this section, we specify the $Gn$EFT Lagrangian, $bs$HLS, that we will employ to discuss physics of compact-star matter.  For this we incorporate all the key ingredients obtained above with robust features provided by topological arguments, together with the hidden symmetries of QCD, into a single effective Lagrangian with the cutoff set above the vector meson mass. Thus the effective Lagrangian contains as relevant degrees of freedom, in addition to  the nucleons $N$, pions $\pi$, vector mesons {$V_\mu ~ ( = \rho, \omega)$} and dilaton $\chi$.
\subsection{$bs$HLS Lagrangian with ``intrinsic" QCD inputs}
The first thing to do is to incorporate certain inputs that can be obtained from QCD at the scale where the EFT is matched to QCD. The scale at which this can be done is typically set at the chiral scale $\Lambda_\chi\sim$ 1 GeV.  The effective theory Lagrangian then inherits nonperturbative quantities, such as quark, gluon and dilaton condensates, from QCD. They figure in the bare parameters of the EFT Lagrangian extracted from QCD, principally  through correlators, such as the vector and axial--vector. Given that the calculations are to be done with the cutoff set a scale below the chiral scale, the option that we adopt is to take into account the evolution of the condensates in the scale and the change of vacua as density is varied. Thus the EFT Lagrangian has the ``bare" parameters dependent on  those quantities inherited from QCD as the vacuum changes with density.  This type of density dependence is  called ``intrinsic density dependence (IDD)."  It will be precisely defined in Section \ref{IDD}. None of the standard effective $\chi$EFT approaches, i.e., S$\chi$EFT, seems to take into account this intrinsic density dependence.   Typically the parameters are picked at one scale, fit to experimental value in the vacuum, and are not evolved with the vacuum change. All density dependence in S$\chi$EFT comes from (``standard" many-body) nuclear interactions\footnote{In some cases, it is not totally clear how to distinguish the ``standard" dependence from the intrinsic one. A case that we will encounter is the short-range three-boy force effect. In our approach, this distinction can be made with little ambiguity.}. This approach could make sense at low densities where experimental information is available --- and with an astute fitting,  but it cannot be pushed beyond the normal density.

To construct the IDD-implemented EFT Lagrangian  $bs$HLS, we take the baryonic HLS Lagrangian and scale-symmetrize it. In doing this  it would be important to incorporate the dilaton in a systematic scale-chiral counting in the framework we follow, i.e., the Crewther--Tunstall (CT) formalism. That would allow processes where the dilaton can figure (such as scalar couplings to the nucleons, e.g., scale-exchange nuclear potential) at lower orders than in the usual S$\chi$PT\footnote{As in $K\to \pi\pi$ decay which is dominantly given by the tree order in the scale-chiral expansion instead of multi-loop order terms in standard $\chi$PT.}. The scale-chiral expansion, formulated up to date,  is however much too  cumbersome with too many unknown parameters and could not be exploited for systematic calculations~\cite{LMR}. It needs to -- and could very well -- be drastically tamed. Fortunately, however,  in dense medium where the dilaton mass drops so the scale symmetry breaking diminishes, the leading-order-scale symmetric (LOSS) approximation (\ref{LOSS}) is found to be simple and reliable enough.  The emergence of what is referred to as ``pseudo-conformal structure" could be evidencing this aspect. We shall therefore proceed with this approximation.

In the LOSS approximation, scale-symmetrizing HLS Lagrangians can be simply done by using the conformal compensator field $\chi=f_\chi e^{\sigma/f_\chi}$ as given in Eq.~(\ref{HLS}) for the meson sector
\begin{eqnarray}
{\cal L}_{\rm M} & = & \left(\frac{\chi}{f_\chi}\right)^2 \left(f_\pi^2 {\rm Tr}\left[\hat{\alpha}_{\perp \mu}\hat{\alpha}_{\perp}^\mu\right] + a f_\pi^2 {\rm Tr}\left[\hat{\alpha}_{\parallel \mu}\hat{\alpha}_{\parallel}^\mu\right]\right) - \frac{1}{2g^2}{\rm Tr}\left[V_{\mu\nu}V^{\mu\nu}\right] +\cdots\nonumber \\
& &{} + \frac{1}{2} \partial_\mu \chi \partial^\mu \chi +V(\chi)
\label{MHLS}
\end{eqnarray}
and in Eq.~(\ref{bHLS}) for the baryon-coupled sector
\begin{eqnarray}
\mathcal{L}_{\rm B}
&=& \bar{N}i\gamma^{\mu}D_{\mu}N - hf_{\pi}\frac{\chi}{f_{\chi}}\bar{N}N + g_{v\rho} \bar{N}\gamma^{\mu}\hat{\alpha}_{\parallel \mu}N\nonumber\\
&&
{}+ g_{v0} \bar{N}\gamma^{\mu}\mbox{Tr}\left[\hat{\alpha}_{\parallel \mu} \right]N
{}+ g_{A}\bar{N}  \gamma^{\mu}\hat{\alpha}_{\perp \mu}\gamma_{5} N .
\label{BHLS}
\end{eqnarray}
%
%

Here we give  a precise definition of how the IDDs in our EFT are extracted from the matching with QCD. The presence of the topology change requires considering two density regimes delineated by $n_{1/2}$, ``R--I" for
$n\leq n_{1/2}$ and ``R--II" for $n>n_{1/2}$. We will find a drastic change in the parameters in the two regions, in particular, with the $\rho$ gauge coupling as already seen for the tensor force.

As mentioned, we incorporate the QCD inputs, IDDs,  to the parameters of our $bs$HLS -- the sum of (\ref{MHLS}) and (\ref{BHLS}) -- by matching the EFT to QCD via correlators at the chiral scale $\Lambda_\chi\sim 1$ GeV. For this, we rely on  the path-breaking work of Harada and Yamawaki~\cite{HY:PR}. In our case, two additional degrees of freedom, the baryons and the dilaton,  need to be included as the relevant degrees of freedom in HLS Lagrangian.  Since our calculation is performed with the effective cutoff scale put  above the vector-meson scale, the dilaton with a mass $\lsim 500$ MeV should also figure explicitly.  The baryon mass is above the cutoff scale, but it must figure in one way or other in nuclear physics. It would be  natural to generate the baryons as solitons from mesonic theories. However, this, as mentioned, is not feasible at present at the density involved. They can alternatively be put in as explicit degrees of freedom as in S$\chi$EFT. The point is that in nuclear processes, at least at low energies, what is  involved are small (nuclear) fluctuations on top of the Fermi sea, so their role can be treated on the same footing as soft-pion processes in the sense of Weinberg's Folk Theorem, which of course limits the kinematics involved.  Furthermore having nucleons as explicit degrees of freedom makes incorporating the intrinsic condensate effects easier as we will demonstrate.

On the other hand, incorporating the scale symmetry with the dilaton degree of freedom in the EFT, given the controversial nature of hidden scale invariance in QCD with $N_f\leq 3$, is a more delicate matter, not treated in Ref.~\cite{HY:PR}.  We find however that  in the approach of Crewther and Tunstall~\cite{CT} expanding around the IR fixed point  and within the LOSS approximation, accounting for the matching with QCD is rendered easier in medium with the dilaton than in its absence.

It follows from the analysis of Ref.~\cite{HY:PR} that with the matching made at the chiral scale $\Lambda_\chi$\footnote{The matching should in principle be done at the lowest scale for QCD and the highest scale for EFT. Whether or not, $\Lambda_\chi$ is the optimal scale for this matching is not clear. It is assumed in Ref.~\cite{HY:PR}.},  the key parameters in HLS, namely, $f_\pi$, $g$ and the hidden local symmetry parameter $a$~\cite{HY:PR}, when evolved down to the effective cutoff scale $\Lambda_{\rm eff}$,  depend only negligibly on the quark condensate $\la\bar{q}q\ra$ and the gluon condensate $\la G_{\mu\nu}^2\ra$ --- and their mixed condensates. Therefore their vacuum changes can be safely ignored in evolving them.  This  is also the case in the baryon sector, so  the parameters $g_A$, $g_V$ etc. in the EFT Lagrangian can be taken to be unaffected by the vacuum change.

However with the dilaton present at the chiral scale in the CT theory, the situation is quite different. Once the vacuum is defined, the dilaton picks up the vacuum expectation value with the scale symmetry spontaneously broken via the potential, and as the vacuum is changed by density, the density dependence does enter transparently in various parameters and evolves in the EFT Lagrangian.  This is the principal IDD, holding globally in R--I and with the exception of the hidden gauge coupling (which will be addressed below), also in R--II. We should stress that this is the reasoning first introduced in Ref.~\cite{BR91}, which is now given a strong support by the  CT model~\cite{LM-BR}.

It is easy to see how this comes about from the Lagrangians (\ref{MHLS}) and (\ref{BHLS}). Apart from the potential term in (\ref{MHLS}), the Lagrangian is scale-invariant. In a given vacuum, however, the potential will break the symmetry, both explicitly and spontaneously, rendering the condensate $\la\chi\ra$ vacuum-dependent and hence density-dependent. This then leads to the scaling relation  (\ref{scaling}). Here low-energy theorems involving both pseudo-scalar and scalar Nambu-Goldstone bosons turn out to play a key role.

$\bullet$ { \bf Proposition IX:  \it In baryonic matter, scale symmetry is ``intrinsically" locked to chiral symmetry so that the pion decay constant scales in density as does the dilaton decay constant.}

The situation for the hidden gauge coupling $g_\rho$ is different because in the RG analysis,  $g_\rho$ (the HLS parameter $a$) is found to have the fixed point $g_\rho=0$ ($a$=1), namely, the  vector manifestation (VM) fixed point~\cite{HY:PR}.  The scaling (\ref{scaling}) for the $\rho$ meson still holds  in R--I, but because of the topology change at $n_{1/2}$  it does not in R--II. Because of the VM fixed point, the mass scaling for $\rho$ then deviates drastically from (\ref{scaling}) from $n_{1/2}$ on. This is the mechanism for the tensor force change from R--I to R--II responsible for the cusp discussed above.
\subsection{$\mathbf{V_{lowk}}$ Renormalization Group (RG) approach}\label{vlowkapproach}
Given the $Gn$EFT Lagrangian $bs$HLS matched to QCD at $\Lambda_\chi$ with the parameters suitably endowed with the IDDs determined as described, how does one go about calculating the equation of state for compact stars?

To answer this question, we adopt the $V_{lowk}$ RG approach implementing the strategy of Wilsonian renormalization group flow~\cite{Vlowk}.  There are more recent developments that in principle could improve on the $V_{lowk}$RG such as ``functional RG" or ``exact RG" etc. It is not obvious, however,  whether such an`` improvement" can actually be implemented in the framework anchored on the EFT with $bs$HLS. At present, they are mostly applied to ``toy" models. In any event, we find the $V_{lowk}$RG sufficiently versatile and easily amenable to further improvement over what has been done up to date.

We first need to bring the action down to  the scale at which the experimental data are available, say,  $E_{lab}\sim 350$ MeV or $\Lambda\sim 2.1$ fm$^{-1}$. This means that we will be integrating out -- in the sense of RG -- all meson fields with masses greater than $\sim 350$ MeV,  i.e.,  $\rho (770)$, $a_1 (1260)$, $\omega$ (782) and $\chi (\sim 600)$. The resulting action will consist of multi-baryon fields and pion fields. In a simplified form, it takes
\be
\L & = & \bar{\psi}(i\gamma_\mu \del^\mu -m_N^\ast)\psi + \frac{{C^\ast}_S^2}{f^2}(\bar{\psi}\psi \bar{\psi}\psi) - \frac{{C^\ast}_V^2}{f^2}(\psi^\dagger\psi\psi^\dagger\psi) +\cdots\, ,\label{4-fermi}
\ee
where the ellipsis contains coupling to the pion field and the IDDs are indicated by $\ast$. In principle the pion field cannot be integrated out for the processes involved in compact stars, but left out here just to compare with the well-known Walecka linear mean-field model~\cite{walecka-model}. The pion of course can contribute if one goes to the next corrections, namely, the Fock term.

In nuclear matter, treated in the mean field, the Lagrangian (\ref{4-fermi}) should give, with suitable choice of the parameters $C$, the same as the Walecka model. It would correspond to Landau Fermi liquid in the sense specified above. One can imagine that in the decimation down to the energy scale of nuclear processes, the Fermi sea is formed and the Lagrangian (\ref{4-fermi}) transforms to a Fermi-liquid fixed-point theory with an effective action -- built on Fermi-surface -- of the form~\cite{shankar}
\be
S=\int \bar{\psi}[i\omega-v^*k]\psi\frac{dk d\Omega d\omega}{(2\pi)^4}  +\frac{1}{2!2!}\int u\bar{\psi}\bar{\psi}\psi\psi \, ,
\label{fixedpointLag}
\ee
where $v^\ast=k_F/m_L^\ast$ with $m_L^\ast$ being the Landau quasiparticle mass. The Landau mass is to be at  the fixed point (which can be assured by putting a counter term). Here $u$ is a generic four-Fermi interaction representing all channels of quantum numbers. For symmetric nuclear matter, the leading terms are of the form of Eq.~(\ref{4-fermi}). RG analysis leads to the observation that the quartic interaction is marginal to leading order with the corrections suppressed as ${\bar{\Lambda}}/{k_F}\to 0$.
This is precisely the Landau (or, more properly in nuclear physics, Landau--Migdal) parameter $F(z)$.

Now how to go from $bs$HLS to Eq.~(\ref{4-fermi}) and then to the fixed-point action of the form  (\ref{fixedpointLag}) is the main task here. The approach we have followed -- and will use here -- is via $V_{\text{lowk}}$~\cite{Vlowk,SB-RG}. This approach has the advantage of being applicable to both finite nuclear systems and  infinite dense matter.

We next discuss  how to arrive at $V_{\text{lowk}}$ and then go to the fixed-point interactions $F$.  First we do this in the medium-free space. For this procedure,
we return to the integrating-out of the high momentum component of a half-on-shell $T$ matrix in the momentum space
\be
T(p^\prime,p;p^2)=V(p^\prime,p)+\frac{2}{\pi}{\mathcal P}\int_0^\infty\frac{V(p^\prime,q)T(q,p;p^2)}{p^2-q^2} q^2 dq.
\ee
Integrating out above the cutoff $\Lambda$, define the low-momentum $T$ matrix as
\be
T_{\text{lowk}}(p^\prime,p;p^2) & = & V_{\text{lowk}}(p^\prime,p) + \frac{2}{\pi}{\mathcal P}\int_0^\Lambda\frac{V_{\text{lowk}}(p^\prime,q)T_{\text{lowk}}(q,p;p^2)}{p^2-q^2} q^2 dq.
\label{Twithvlowk}
\ee
Requiring that $T_{lowk} (p^\prime,p;p^2)=T(p^\prime,p;p^2)$ for $p^\prime, p < \Lambda$, one  has the $T$ matrix given in terms of the low-momentum interaction $V_{\text{lowk}}$.
Since by RG invariance, $\frac{d}{d\Lambda} T_{\text{lowk}}=0$, we have the RG equation
\be
\frac{d}{d\Lambda} V_{\text{lowk}} (p^\prime,p;p^2)=\beta([V_{\text{lowk}}],\Lambda).
\ee

Now what is $V$, the ``bare" potential, computed from  $bs$HLS Lagrangian?

In the vacuum (i.e., $n=0$), two-nucleon (and three-nucleon if needed) potentials are calculated by irreducible diagrams (as in the S$\chi$PT with the suitable counting rule for the HLS fields~\cite{HY:PR} taken into account).
Putting this potential into Eq.~(\ref{Twithvlowk}), one can then determine $V_{\text{lowk}}$ by fitting phase shifts up to $E_{lab}\sim 350$ MeV. That would determine the vacuum parameters of the Lagrangian that figure in $V_{{lowk}}$ in the vacuum.

Now to apply to dense matter, say, the properties of nuclear matter and the EoS of compact-star matter, one incorporates the IDDs  in the Lagrangian, then incorporates higher-order irreducible diagrams in the driving term and sum ``reducible graphs" (corresponding to solving the Schr\"odinger equation for bound states or Lippmann-Schwinger equation for scattering).  This brings in the IDDs into the calculation of physical observables. By means of judicious -- and involved though transparent -- calculations in which only the forward-going terms contribute with non-forward-going terms suppressed  -- the so-called ``$\omega/Q\rightarrow 0$ limit" in Green's  function approach, 
one arrives at the Fermi-liquid fixed-point potential $V_{\text{FL}}=V_{\text{lowk}}+\delta V$
\be
\frac{d}{d\Lambda} V_{\text{FL}} (n,\Lambda)=0 \, ,\label{FLpotential}
\ee
which says that the $\beta$ function for the effective potential $V_{\text{FL}}$ is zero, modulo corrections of $O(({\Lambda}-k_F)/k_F)$.  Given such $V_{\text{FL}}$, one can write the Landau parameters (i.e., $f$, $f^\prime$, $g$, $g^\prime$ etc.) as linear combinations of $V_{\text{FL}}$ in appropriate quantum number channels~\cite{SB-RG}.

$\bullet$ {\bf Proposition X: \it The $V_{lowk}$ RG approach in the limit $(\Lambda-k_F)/k_F\to 0$ with $bs$HLS is equivalent to the mean-field theory with $bs$HLS Lagrangian, which in turn is equivalent to  Landau Fermi-liquid fixed point theory.}

What is found both remarkable and surprising with the $V_{\text{lowk}}$ approach in the matter-free vacuum (that is, $n=0$) is that various different bare potentials, both high-precision phenomenological and high-order effective field theoretic, yield for $\Lambda=2.1$ fm$^{-1}$, converge to a universal $V_{\text{lowk}}$. This would suggest that varying the cutoff around 2.1 fm$^{-1}$ -- corresponding to $E_{lab}$ to which accurate scattering data are available -- would not affect much the phase shifts and hence $\frac{d}{d\Lambda} V_{\text{lowk}}\approx 0$.
\subsection{The renormalization-group invariance of the tensor force}\label{tensor-force-fixed-point}
Given the precise definition of the $V_{lowk}$ RG approach, we can now describe the most remarkable property of the tensor force as given in $bs$HLS via-\`a-vis with the important role of the tensor force in compact-star matter. At present there is no rigorous proof,  but it turns out numerically that the tensor force is  a fixed-point quantity not only in free space but also in medium~\cite{MR-geb}, i.e., in exotic nuclei~\cite{otsuka}. In the case of the NN scattering in the $^3$S$_1$-$^3$D$_1$ channel in the matter-free space, the tensor force gets unaffected by the elementary strong interactions and in the case of exotic nuclei, the matrix element that singles out the tensor force component, namely the monopole matrix element, is unaffected by nuclear many-body correlations. This  means that the $\beta$ function for the NN interaction in the tensor channel both in the matter-free space and in medium is equal to zero. If it were not for the IDD, the matrix elements of two nucleon states of the tensor force would be strictly {\it density-independent}. This means that the density dependence in the matrix elements of the tensor force is uniquely given by the IDD.

There are two important consequences from this observation. First the cusp structure and the state of matter at $n\gsim n_{1/2}$, in terms of the tensor force, are controlled by the density dependence in IDDs inherited from nonperturbative QCD unscathed by mundane nuclear interactions, and secondly the renormalization-group invariant structure of the tensor force could be scrutinized in nuclear structure of exotic nuclei at densities lower than and at $n_0$, a subject of research relevant to  RIB (rare-ion-beam) machines~\cite{HIAF,RAON}.

$\bullet$ { \bf Proposition XI: \it The two-nucleon tensor force given in $bs$HLS is a RG-fixed point quantity  with the beta function $\beta(V^T)=0$  both in free space and in medium. The intrinsic density dependence  therefore is the unique -- and the only -- cause of the density dependence in the nuclear tensor force and hence in the nuclear symmetry energy as well as in the nuclear structure in exotic nuclei~\cite{MR-geb}.}

\section{Compressed Baryonic Matter and Compact Stars}
We shall now confront Nature with our approach summarized by the 11 {\bf Propositions} given above.
As stressed in Introduction, it is our aim to explore the domain that is not yet charted and make predictions. To do this we use one precisely -- and within the given framework, completely -- defined theoretical tool which comes out more or less consistent with the established results, both theoretical and experimental, but simple and powerful enough to explore new phenomena. Here our philosophy is totally different from the dominant trend in the field,  which is to arrive, mostly phenomenologically  with a mixed bag of models,  at an EoS that explains numerically ``everything" from nuclear matter to compact stars. In this way of approaching the problem, there is inevitably certain ambiguity in theoretical approaches and interpretations of the experimental data and hence lack of predictivity. We differ in that we do not adhere to what is considered to be {\it constraints} given in one known density region, i.e., nuclear matter, for arriving at a vastly different region, i.e., compact star matter. Our aim is to make predictions that could be unambiguously confirmed or falsified  by either reliable theories or trustful experiments.

In this section we will be concerned with the range of densities  $\sim (1-7) n_0 $. Apart from the VM fixed point $\gsim 25n_0$ which figures for certain observables, we will not be concerned with higher density ranges beyond $\sim 7 n_0$.  We will treat both nuclear matter and compact-star matter in one framework based on the $Gn$EFT using $bs$HLS.
\subsection{$\mathbf{V_{lowk}}$ renormalization group}
There are two quantities to specify for accessing quantitatively the matter denser than that of nuclear matter. One is the input parameter(s) that defines(define) the IDDs in the EFT Lagrangian and the other, thus far only mentioned without details, is the $bs$HLS Lagrangian with the parameters with the IDDs suitably incorporated.
\subsubsection{Intrinsic density dependence (IDD) in ``bare" parameters of $bs$HLS}\label{IDD}
We need to consider two density regimes R--I and R--II delineated by the topology change density. By the Cheshire Cat Principle, $n_{1/2}$ should correspond to the range $(2-4)n_0$ considered to capture the putative hadron-quark continuity.

In R--I, only one parameter $\Phi$ in Eq.~(\ref{scaling}) fixes all the IDDs. There is no first-principle -- QCD -- information on this quantity. It can however be fixed by nuclear experiments. For convenience, we take the form
\be
\Phi_I=\frac{1}{1+c_I\frac{n}{n_0}}
\ee
with $c_I$ a constant. The form is of course totally arbitrary, but for low density, say, up to $n_0$ it should be reliable enough. In fact the range of $c_I$ that gives a good fit to nuclear matter properties -- to be given below -- is found to be
\be
c_I\approx 0.13-0.20\label{range}
\ee with the upper value giving the measured pion decay constant~\cite{yamazaki-kienle}.  Given that the effective cutoff  used for the decimation is $\sim (2-3)$ fm$^{-1}$,  appreciably lower than the matching scale $\Lambda_\chi$, what enters in Eq.~(\ref{range}) is IDD$^\ast$ that includes small ``induced" density dependence.\footnote{This is what was called  DD$_{\rm induced}$ in Ref.~\cite{PKLMR}. It is an effect that is not inherited from QCD at the matching scale but renormalizes the IDDs due to decimations involving  nuclear interactions that are higher order in S$\chi$EFT power counting. An apt  example is  the case of the long C-14 life-time mentioned above.}  It is expected that there be small fine-tuning within the range (\ref{range}). This reflects the fine-tuning nature required for ground-state properties of nuclear matter, be that EFT or phenomenological.

In R--II, due to the topology change at $n_{1/2} > n_0$, some parameters do undergo drastic modifications. The most  important quantity is the hidden local gauge coupling $g_\rho$ and hence the $\rho$ mass related to $g_\rho$ by the low-energy theorem, i.e., KSRF relation. The precise form is of course unknown. We take the simplest form
\be
\frac{m_\rho^\ast}{m_\rho}\approx \frac{g_\rho^\ast}{g_\rho}\equiv \Phi_\rho\to \left(1- \frac{n}{n_{\rm VM}} \right)\ \ {\rm for } \ \ n > n_{1/2},\label{VMform}
\ee
where $n_{\rm VM}$ is the putative VM fixed-point density. Here we have  assumed a linear density dependence for simplicity.\footnote{How to join the $\Phi_\rho$ from 1 for $n\leq n_{1/2}$ to the linear form Eq.~(\ref{VMform}) is of course is known neither empirically nor theoretically. One thing that can be said with certainty is that the cusp structure that is given by the topology change requires  a sudden drop of  $\Phi_\rho$ at $n_{1/2}$. One should keep this uncertainty in mind in assessing the results. What matters is that $g_\rho$ drops toward the VM fixed point as we will see in numerical results.}  Where $n_{\rm VM}$ is located is not known in QCD. In compact stars, whether it is $\sim 6n_0$ or $\gsim 25 n_0$ does not make noticeable differences with one possible exception, namely, the star sound velocity as we will see below.

As for other parameters, apart from the properties of the $\omega$ meson, the scaling is very simple because in R--II,  as in the 1/2-skyrmion phase,  we learn from  {\bf Proposition IV}
that the parity doubling emerges giving rise to the chiral-invariant mass $m_0$ . It is locked to the dilaton condensate, which  becomes {\it density-independent} in R--II. Therefore we have the effective nucleon mass   becoming independent of density for $n>n_{1/2}$
\be
\frac{m_N^\ast}{m_N}\approx \frac{f_\chi^\ast}{f_\chi}\approx \frac{f_\pi^\ast}{f_\pi} \equiv \kappa\sim (0.6-0.9).\label{kappa}
\ee
The dilaton mass also goes proportional to the dilaton condensate. This follows from the partially conserved dilatation current (PCDC)~\cite{CT}
\be
\frac{m_\sigma^\ast}{m_\sigma}\approx\kappa.
\ee
The dilaton coupling to nucleon and other fields is unscaling to the leading order in scale-chiral symmetry, so it is a constant in R--II as in R--I.

As stated in {\bf Proposition III}, the role of $\omega$ meson is more involved from the point of view of scale-chiral symmetry. While the flavor $U(2)$ symmetry seems good in the vacuum and also  in R--I, it should be broken strongly in R--II. First of all it does not follow the $\rho$ meson toward the VM fixed point. This is already seen in mean-field theory~\cite{DLFP-PD} and confirmed in $V_{lowk}$ RG~\cite{PKLR}. It is however unquestionable that it has a crucial role  in dense matter in providing repulsion. Thus some sort of fine-tuning is  needed in the density-scaling of  its mass and  coupling constant. We take it as
\be
\frac{m_\omega^\ast}{m_\omega}\approx \kappa\frac{g_\omega^\ast}{g_\omega}
\ee
where $g_\omega$ is the $U(1)$ gauge coupling.  This is consistent with the procedure of generating the $\omega$ mass from ``Higgsed" $U(1)$ hidden gauge symmetry. It is found in compact-star structure that  $g_\omega^\ast$ must scale weakly in density in $n>n_{1/2}$, a signal for $U(2)$ symmetry breaking referred to above in going to the DLFP. It is taken in the numerical calculations as
\be
\Phi_\omega\equiv \frac{g_\omega^\ast}{g_\omega}\approx 1-d\frac{n-n_{1/2}}{n_0}
\ee
with $d\approx 0.05$.
\subsubsection{Double-decimation RG}
We are now completely equipped for doing full numerical calculations. The first decimation in the $V_{lowk}$ RG framework giving the Fermi-liquid fixed point potential $V_{\rm FL}$  is described in Section \ref{vlowkapproach}. Given the intrinsic density-dependent $V_{\rm FL}(n)$, one can then do the full Fermi-liquid calculation going beyond the fixed-point approximation described above. It consists of making higher order $1/\bar{N}$ corrections in the ring-diagram technique which has been checked to work fairly well~\cite{ring-diagram}.
\subsection{Nuclear matter in $bs$HLS}\label{nuclearmatter}
Once the vacuum parameters are fixed at the scale the vacuum $V_{lowk}$ is obtained then the only parameter in R--I is the scaling $c_I$ in Eq.~(\ref{range}). With the scaling taken into account to arrive at $V_{\rm FL}$, (\ref{FLpotential}), doing the second decimation with it is all there is to it for calculating equilibrium nuclear matter properties with appropriate $1/\bar{N}$ corrections taken into account. This calculation in R--I amounts to doing roughly N$^3$LO S$\chi$EFT including chiral 3-body potentials. The role of the three-body force, much heralded in the literature for the success of nuclear effective chiral field theory S$\chi$EFT that makes nuclear matter stabilized at the proper equilibrium density, is captured in our approach at a lower chiral order in two-nucleon potentials encoded with the IDD~\cite{dongetal}. Exactly the same mechanism is at work for the C-14 dating Gamow--Teller matrix element where the three-body potential effect in S$\chi$EFT is reproduced by the IDD.\footnote{What happens is that the short-range three-body due to $\omega$ exchange in $bs$HLS, when integrated out from $V_{lowk}$, goes  into the coefficient $c_I$ in the two-body tensor-force channel~\cite{Holt-Rho-Weise}. This is not, properly speaking, the IDDs inherited from QCD, but induced in the RG decimation as explained in Ref.~\cite{PKLMR}. This accounts for the range depending on different channels quoted in Eq.~(\ref{range}).}

There is really no big deal here as far as the normal nuclear matter properties are concerned. Given effectively only one free parameter, it would be too much to expect to achieve the accuracy obtained in the refined S$\chi$EFT results. Nonetheless the results are quite consistent with the available empirical  values. The values predicted by the theory~\cite{PKLMR} are: $E_0(n_0)/A-m_N\simeq{} -15.5 $ MeV, $n_0\simeq 0.154 $ fm$^{-3}$, $K(n_0)\simeq 215 $ MeV, $J\equiv E_{sym}(n_0)\simeq  26$ MeV, $L(n_0)\equiv 3n\frac{\del}{\del n} E_{sym} (n)|_{n=n_0}\simeq  49 $ MeV. They are to be compared  with the presently quoted empirical values $E_0/A-m_N={} -15.9 \pm 0.4$ MeV, $n_0= 0.164 \pm 0.007$ fm$^{-3}$, $K= 240\pm 20$ MeV or  $230\pm 40$ MeV, $30\lsim E_{sym} \lsim 35  $ MeV, $20 \lsim L\lsim 66$ MeV.  One possible caveat to note here is the symmetry energy at $n_0$ which is somewhat low compared with empirical values, somewhat underpredicted compared with N$^k$LO for $k\geq 3$ S$\chi$EFTs. This possible deficiency, if it is one,  may not be  serious.\footnote{There has been a suggestion~\cite{oezel-smallL} based on the analysis of gravitational wave events that the lower  value $L(n,0)\sim 20$ MeV is favored. This is less than 1/2 of our predicted value. If confirmed, this would be a serious discrepancy that may have a big impact on the tidal deformability predicted by our theory. We will return to this problem below.}  It could be easily remedied by fine-tuning the $c_I$ constants in appropriate channels involved. Furthermore we do not consider -- that we reiterate throughout this review -- that the properties of normal nuclear matter should  be taken as {\it the strict constraints} for the EoS for massive compact stars at a central densities $\sim (5-7)n_0$.

\subsection{Compact-star properties}\label{compactstar}
Due to the topology change at $n_{1/2}$, there is a drastic change in the scaling in the parameters of $bs$HLS leading to a qualitative impact on  the structure of the EoS.  This means that the location of the changeover density will play a crucial role. There is no theoretical way within the framework to pin down the location. Most encouraging however is that  the various astrophysical observations so far available, the maximum mass, the gravity-wave data and specially the star sound speed, do give the range where the topology change (a.k.a. hadron-quark continuity) is to set in.

We first discuss what one might call ``global properties" of massive stars, postponing the specific issues related to the recent developments to later.

For what follows in this subsection, we will pick $n_{1/2}=2n_0$, a sort of benchmark,  because a highly detailed analysis has been made for this transition density. This transition density  may be somewhat problematic because  the tidal deformability (TD)  $\tilde{\Lambda}$ predicted with that density comes only slightly below the upper bound $\tilde{\Lambda} < 800$ set by the LIGO/Virgo gravity wave data~\cite{TheLIGOScientific:2017qsa}. But otherwise all other properties seem to be fully consistent with data. The higher transition densities move  the TD somewhat lower, but not as low as near the lower bound 400 which other models seem to go down to.
\subsubsection{The VM fixed point,  $\mathbf{\la\theta_\mu^\mu\ra}$ and ``pseudo-conformal" structure}
Where the vector manifestation fixed point  $n_{\rm vm}$ at which the $\rho$-meson gauge coupling goes to zero is located is known neither theoretically nor empirically. While most of the global properties of compact stars do not seem to depend much on where $n_{\rm vm}$ lies as long as it is above $\sim 7n_0$ -- that we consider to be the possible central density of massive compact stars -- there is one quantity which is qualitatively affected by the location of $n_{\rm vm}$ and it is  the sound velocity of the stars. Now the sound velocity depends crucially on the trace of the energy--momentum tensor $\la\theta_\mu^\mu\ra$ of the system. We illustrate the situation by picking the possible vector manifestation densities $n_{\rm vm}=6n_0$ and $25n_0$. The former is considered to be what one expects for chiral restoration in S$\chi$EFT,  and the latter represents an ``asymptotic density" where perturbative QCD is expected to be applicable. The two choices with the parameters for  densities $n< n_{1/2}$ should give of course the same nuclear matter properties since nuclear matter is in R--I shared by both.\footnote{The numerical values given in Ref.~\cite{PKLR} have insignificant  differences from what is given in Section \ref{nuclearmatter} which are taken from \cite{PKLMR}. This is due to slightly different scaling parameters used there.}  For compact stars treated in $V_{lowk}$RG with the matter in the $\beta$ equilibrium, the predicted properties also come out to be more or less the same. For instance for the two values of $n_{\rm vm}= 6n_0~(25n_0)$, one finds the maximum star mass $M^{\rm max}/M_\odot =2.07~(2.05)$, the radius $R=11.7\ {\rm km}~(12.2\ {\rm km})$, the central density  $n_{\rm central}=5.6n_0~(5.1n_0)$.~\footnote{We quote the presently available value for $E_{sym}$ at  $n=2n_0$,  $39.2^{+12.1}_{-8.2}$ within 60\% confidence limit~\cite{BAL2019}. Our theory predicts
$E_{sym}\simeq  54 (52)\ {\rm MeV}$.  We will return to this matter in connection with the tidal deformability problem.}

There is however a dramatic difference in prediction between the two for the sound velocity of the star. This is seen in Fig.~\ref{Vs}.
\begin{figure}[h]
\includegraphics[width=7.9cm]{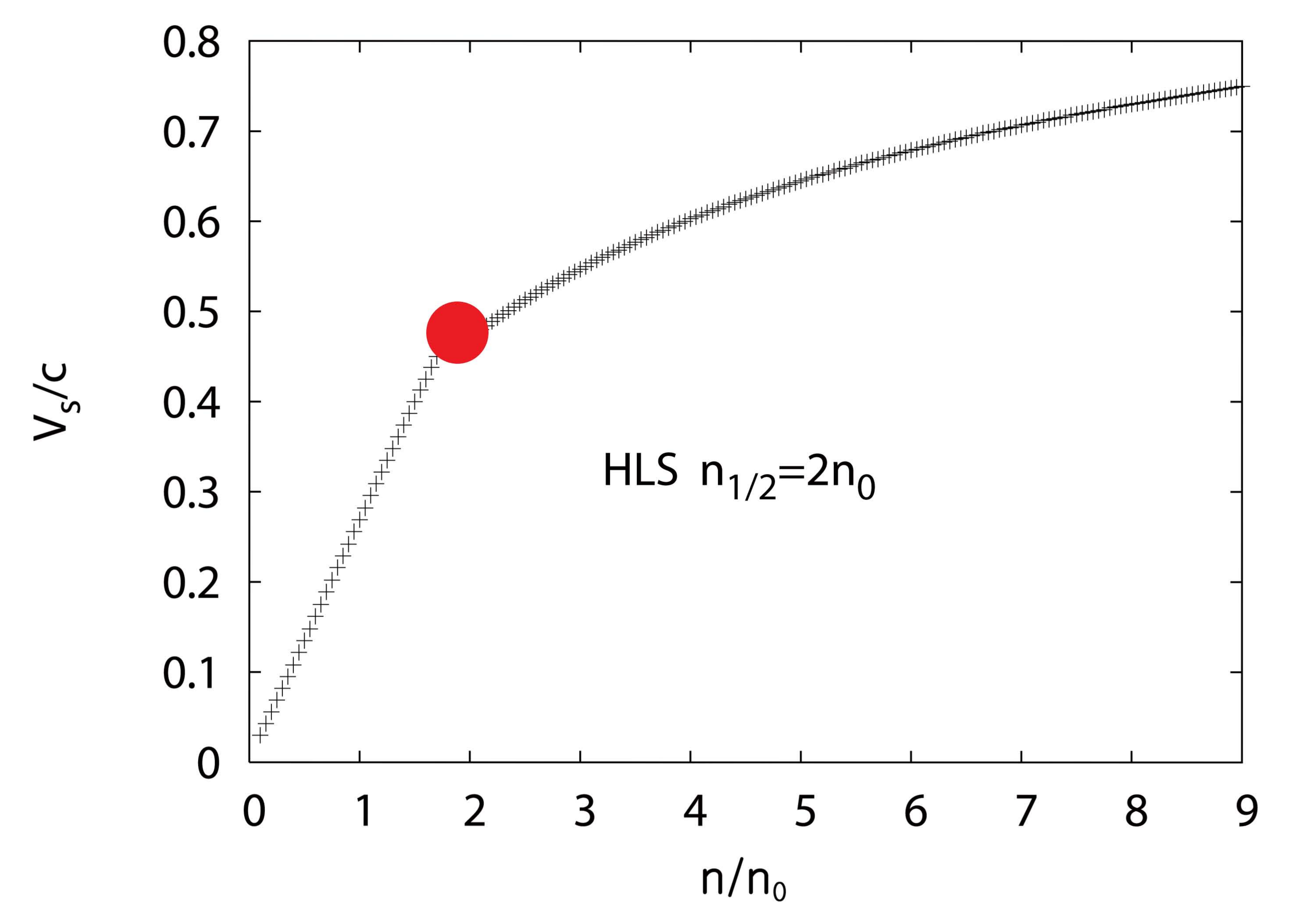}
\includegraphics[width=7.5cm]{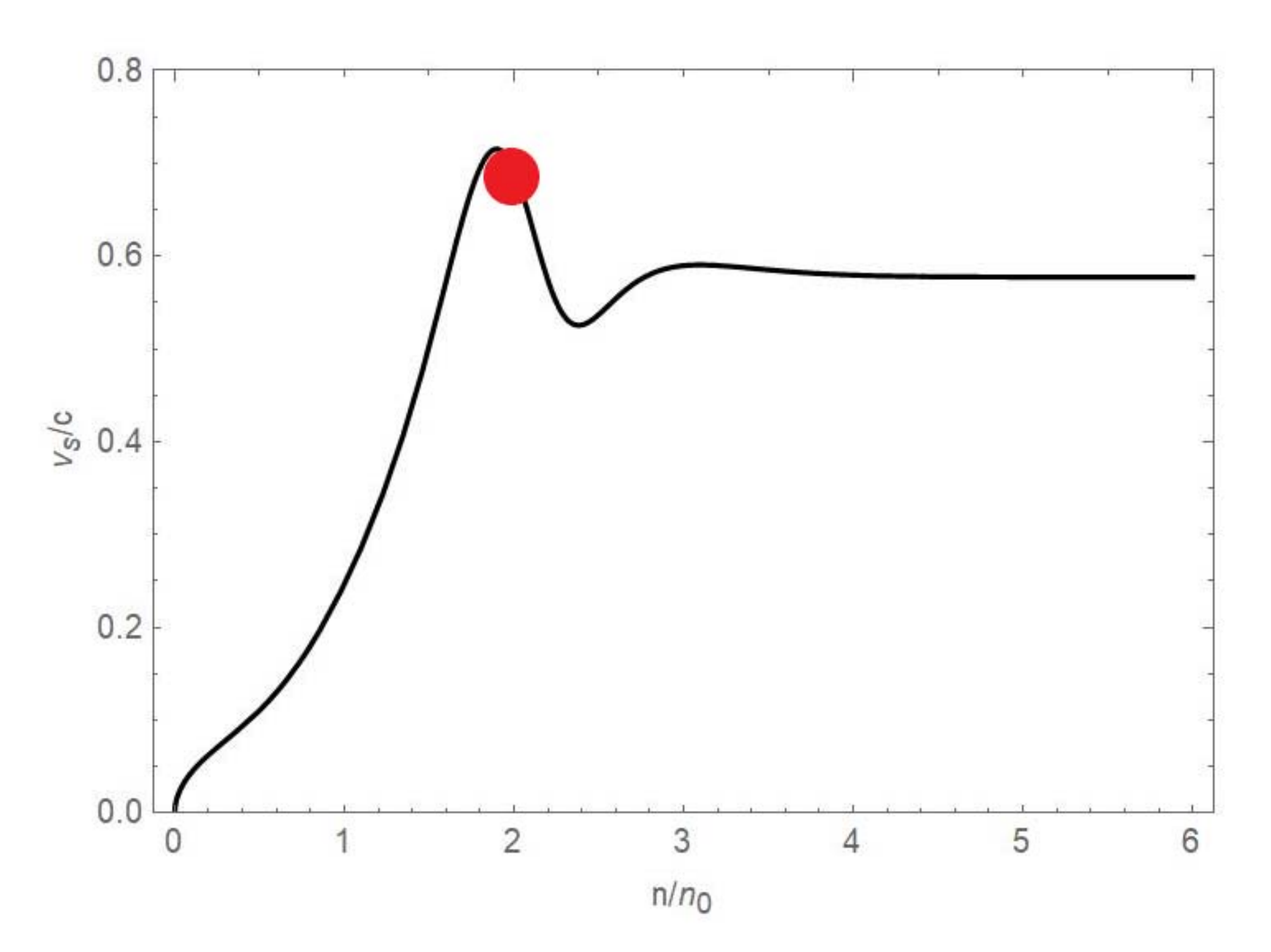}
\caption{Sound velocity for $n_{\rm vm}=6n_0$~\cite{PKLR} and $25n_0$~\cite{PKLMR}, both computed in $V_{lowk}$ RG with $n_{1/2}=2n_0$.  The solid circle indicates where point  the topology change takes place is.
}
\label{Vs}
\end{figure}
While the sound speed increases steadily passing the ``conformal velocity" $v^2_s=1/3$ at $\sim 3n_0$ when the $n_{\rm vm}$ is set at $6n_0$, it overshoots the conformal velocity at $\sim n_{1/2}=2n_0$, then comes down and converges to $v_s^2\approx 1/3$ when $n_{\rm vm}\gsim 25 n_0$. How this comes about is explained below. The former is similar to what is found in standard energy density functional approaches with no established causality constraint. A currently accepted scenario in sophisticated S$\chi$EFTs resembles, albeit remotely, the right-panel of Fig.~\ref{Vs}. But it shows a much broader and bigger bump not exceeding the causality bound $v_s=1$ before converging to the conformal speed  $v_s^2=1/3$ at an asymptotic density $\gsim 50n_0$~\cite{tews}. The convergence to the conformal speed at asymptotically high density is expected in perturbative QCD. But it is highly unorthodox that it converges to 1/3 so precociously at low density as our prediction does.\footnote{It is also counter-intuitive that the pseudo-conformal behavior sets in for higher $n_{\rm vm}$ than lower value. This may be associated with the $\rho$ decoupling from the nucleon in the approach to the DLFP (\ref{vmhere}). This issue remains to be clarified.}  We will argue below that this feature is due to the possible emergence of pseudo-conformal state at $n > n_{1/2}$.

It may very well be that the precocious conformal sound speed is due to the oversimplification of scale-chiral symmetry (LOSS). However it follows as a logical outcome of the {\bf Propositions} we have stated in this review.

To see this, let us first look at the problem in the mean-field treatment of the $bs$HLS constructed with the IDDs inherited from QCD. This mean-field argument will be reconfirmed in full $V_{lowk}$RG formalism.

We have shown ({\bf Proposition IV})  that, going toward the DLFP, the trace of  the energy-momentum tensor $\la\theta_\mu^\mu\ra$ in the mean-field approximation is a function of only the dilaton condensate. Now if the condensate goes to a constant $\sim m_0$ due to the emergence of parity-doubling, then the $\la\theta_\mu^\mu\ra$ will be independent of density. In this case, we will have
\be
\frac{\del}{\del n} \la\theta_\mu^\mu\ra=0.
\ee
This would imply that
\be
\frac{\del\epsilon(n)}{\del n}\left(1-3v_s^2\right)=0
\ee
where $v_s^2=\frac{\del P(n)}{\del n}/\frac{\del\epsilon}{\del n}$ and $\epsilon$ and $P$ are, respectively, the energy density and the pressure.
If we assume  $\frac{\del\epsilon(n)}{\del n}\neq 0$, i.e., no Lee--Wick-type states in the range of densities involved, we can  then  conclude
\be
v_s^2=\frac{1}{3}.
\ee

What we have shown here is in the RMF approximation with $bs$HLS. Now  in terms of the $V_{lowk}$RG approach, the RMF approximation amounts to the first RG decimation that corresponds to the Fermi-liquid fixed point approximation (\`a la {\bf Proposition X}). The corrections to the RMF result should be suppressed by $1/\bar{N}$ when $\bar{N}$ is large at high density. Now the two results given in Fig.~\ref{Vs} are treated in  full $V_{lowk}$RG which include important $1/\bar{N}$ corrections. Thus the drastic  difference between the two cannot be due to $1/\bar{N}$ corrections to the Fermi-liquid fixed point approximation. This means that while the dilaton condensate $\la\chi\ra^\ast $ goes to the density-independent constant $m_0$ due to the parity for $n_{\rm vm}\gsim 25 n_0$,  it does not when $n_{\rm vm}$ is lower, say,  at $\sim 6n_0$. This accounts for the sound velocity failing to converge to $v_s=1/3$ for $n_{\rm vm}\ll 25n_0$.
This suggests {\it the parity doubling at high density is linked to the $\rho$ decoupling from the nucleon together with the vector manifestation.}

The above chain of reasoning is confirmed in the full $V_{lowk}$ RG formalism specifically for the case of $n_{1/2} = 2 n_0$. In Fig. \ref{TEMT} is shown the trace of the energy--momentum tensor (left panel) that gives the conformal velocity for $n\gsim 3n_0$ (right panel).
 \begin{figure}[h]
\begin{center}
\includegraphics[width=7.0cm]{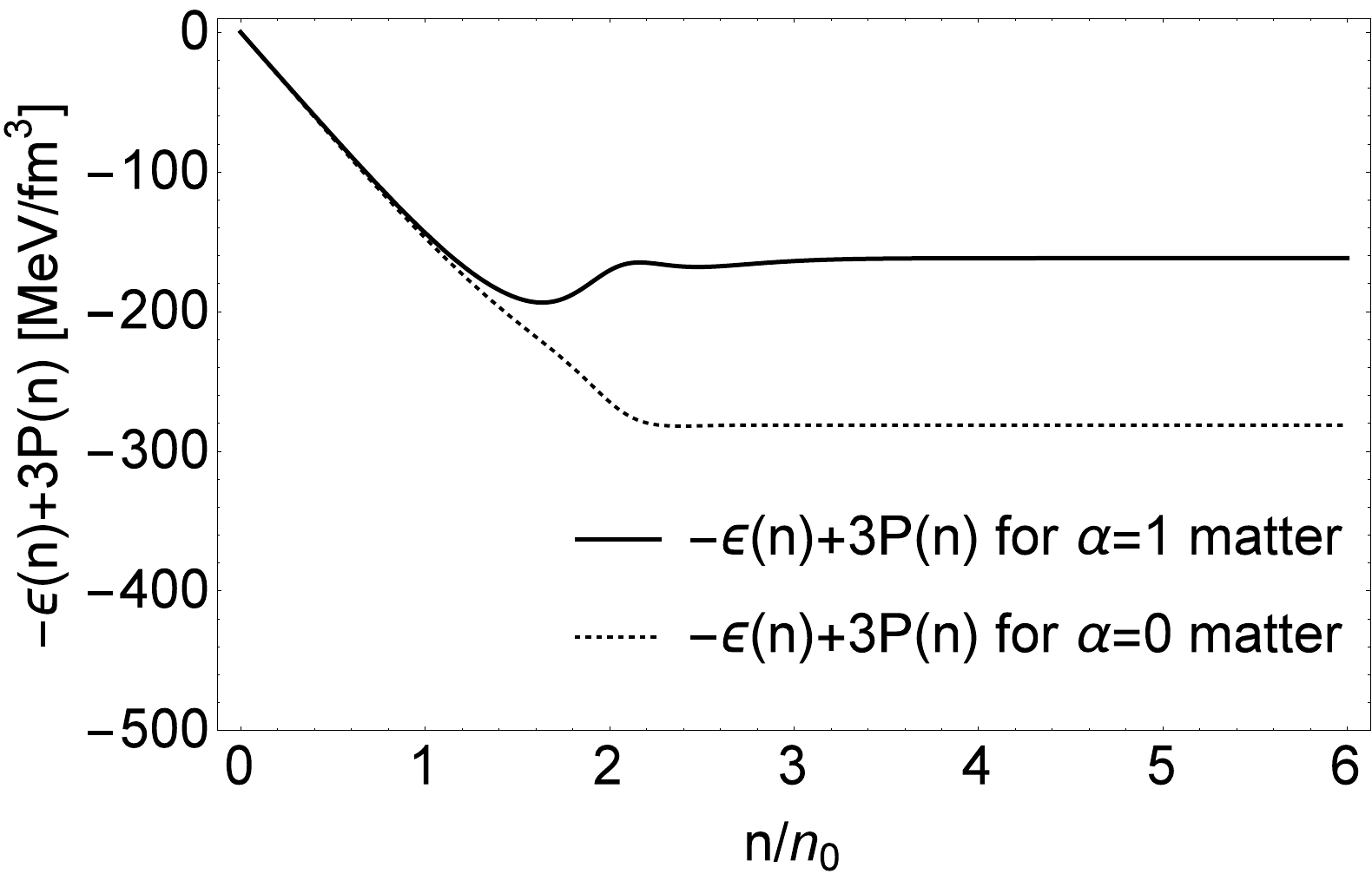}\;\;
\includegraphics[width=7.0cm]{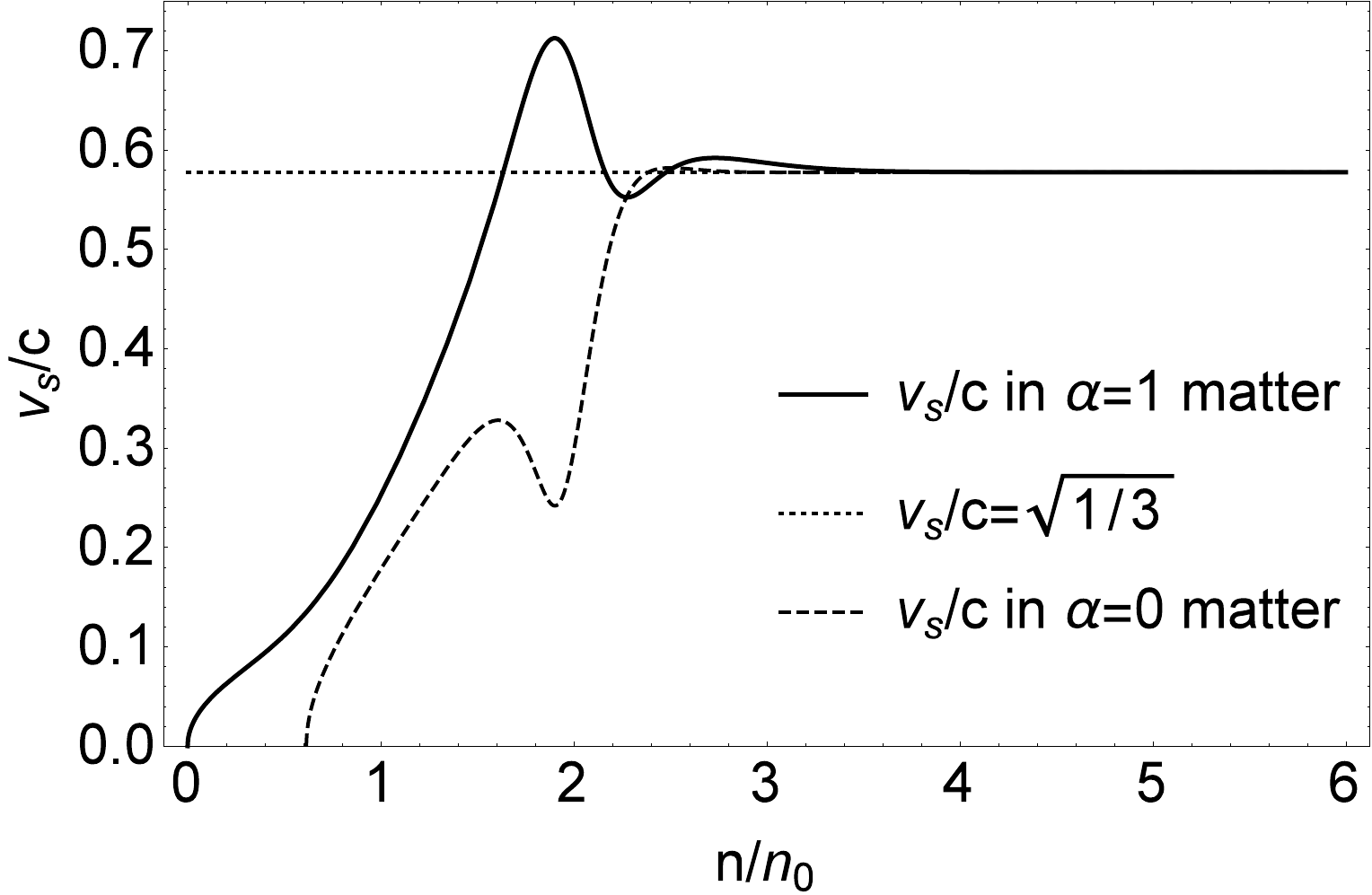}
\caption{$\la\theta_\mu^\mu\ra$ (left panel) and $v_s$ vs. density for $\alpha=0$ (nuclear matter) and $\alpha=1$ (neutron matter) in $V_{lowk}$ RG for $n_{1/2}=2 n_0$ and $v_{\rm vn}=25 n_0$.
 }\label{TEMT}
 \end{center}
\end{figure}
This feature of both the TEMT and the sound velocity are expected to hold for any $n_{1/2}$ at which the topology change sets in,  i.e., within the range $2\lsim n_{1/2}/n_0\lsim  4 $.

$\bullet$ { \bf Proposition XII: \it The parity doubling at $n\gsim n_{1/2}$ is associated with the movement toward the vector manifestation at high density $n_{\rm vm}\gg n_{1/2}$.}
\subsubsection{Pseudo-conformal model (PCM)}
We now focus on what we can say about the EoS of compact stars for the case of  $n_{\rm vm}\gsim 25 n_0$. The analysis will be made in the range $2 \lesssim n_{1/2}/n_0 \lesssim 4$ for the threshold density for topology change. The densities outside of this range in our framework are ruled out by the various star properties we shall now consider. In the preceding discussions, we did a full $V_{lowk}$ RG analysis taking $n_{1/2}=2n_0$. To simplify extending the analysis to higher $n_{1/2}$ densities, we will rephrase the full $V_{lowk}$ RG formalism in terms of what we call PCM (pseudo-conformal model).

This analysis is motivated, as stated above, by that the tidal deformability obtained for
{
$n_{1/2}=2.0n_0$,
$\Lambda\simeq 790$}~\cite{PKLMR,MR-PCM}, corresponds to the upper bound set by the gravity-wave data, and it seems likely that the bound will be tightened to a lower value. This gives us the lower bound for the topology change density
\be
n_{1/2}\gsim 2n_0.\label{lower}
\ee

It has  been verified that a higher $n_{1/2}$ could probably resolve this problem but perhaps not completely. Now the question  is: How far can one  increase $n_{1/2}$ without upsetting the good star properties we have obtained? In particular we are interested in how the range of the density allowed by  $n_{1/2}$ compares with the range of the  baryon--quark continuity as in the semi-phenomenological model of Ref.~\cite{baym-kojo}, an issue highly relevant to the possible applicability of the notion of Cheshire Cat to dense matter.

In order to examine how $n_{1/2}$ affects the star properties, particularly the gravity-wave data, the maximum mass etc., we proceed as follows. Whatever the topology change density $n_{1/2} >n_0$ is, the properties of ordinary  nuclear matter are fixed as stated already.  We assume that for $n_{1/2}\geq 2n_0$, slightly above that transition density, the sound velocity must be $v_s^2\approx 1/3$ as was found for the case of $n_{1/2}=2n_0$ and confirmed also at $2.6 n_0$. It turns out that this feature can be captured by a simple two-parameter formula for the energy per particle $E/A|_{n>n_{1/2}}$  in the form
\be
E/A= - m_N +X^\alpha x^b + Y^\alpha x^d\ \ {\rm with}\ x\equiv n/n_0 \, , \label{matchedE}
\ee
where $X$, $Y$, $b$ and $d$ are parameters to be fixed and $\alpha=(N-Z)/(N+Z)$. The sound velocity takes the form
\be
v_s^2=\frac{dP/dx}{d\epsilon/dx}=\frac{X^\alpha b(b+1)x^b+Y^\alpha d(d+1)x^d}{X(b+1)+Y(d+1)x^d}\, ,
\ee
where $P$ is the pressure and $\epsilon$ is the energy density. If we choose $d=-1$ and $b=1/3$, then the  $E/A$ given by
\be
E/A= - m_N +X^\alpha  x^{1/3} + Y^\alpha x^{-1}\ \ {\rm with}\ x\equiv n/n_0
\label{PC-RII}
\ee
has the sound velocity
\be
v_s^2=\frac{1}{3}\label{pc-sound}
\ee
independently of $X^\alpha$ and $Y^\alpha$.

What we refer to as the pseudo-conformal model (PCM for short)  for the EoS is   then $E/A$ given by the union of that given by $V_{lowk}$ in R--I ($ n<n_{1/2}$) and that given by Eq.~(\ref{PC-RII}) in R--II ( $n\geq n_{1/2}$)  with the parameters $X^\alpha$ and $Y^\alpha$ fixed by the continuity at $n=n_{1/2}$ of the chemical potential and pressure
\be
\mu_I=\mu_{II},\ P_I=P_{II}\ \ {\rm at} \ \ n=n_{1/2}.
\label{matchingE}
\ee
This formulation is found to work very well for both $\alpha=0$ and 1 in the entire range of densities appropriate for massive compact stars, say up to $n\sim (6 - 7)n_0$, for the case $n_{1/2}=2n_0$ where the full $V_{lowk}$RG calculation is available~\cite{PKLMR}. We apply this PCM formalism for the cases where $n_{1/2}> 2n_0$.

%
\subsubsection{Compact stars in PCM for $2 n_0 \leq  n_{1/2} \leq 4 n_0$}
First we show how the sound velocity comes out for $n_{1/2}/n_0=3\ {\rm and}\ 4$~\cite{MR-Vs}. (The case for $n_{1/2}=2n_0$ was given in Fig.~\ref{TEMT} (right panel).) The results are summarized in Fig.~\ref{Vs2-4} for neutron matter.

It cannot be over-stressed that the sharp matching with the extremely simple energy(-density) formula (\ref{matchingE}) at $n_{1/2}$ with the full $V_{lowk}$ result (that takes into account $1/\bar{N}$ corrections) below $n_{1/2}$ could well be a gross oversimplification of the truth. As conjectured in Section \ref{conjecture}, the cusp singularity is highly likely to involve domain-wall induced topological structures such as ``lasagne" sheets or FQH ``pancakes" with possible deconfinement of the constituents caused by domain walls. How the transition can take place  in the matching region is of course unknown in QCD just as in the hadron-quark continuity scenario. Therefore there is no reason to expect that the transition region is reliably captured in the extremely simplified PCM as given. What is relevant  is the onset of the PC sound velocity $v_s^2=1/3$  above the transition density indicating the emergence of PC symmetry.

This caveat, stated above, notwithstanding,  it is clear from Fig.~\ref{Vs2-4} that the sound velocity for the case of $n_{1/2}=4 n_0$  violates the causality bound $v_s^2 < 1$.  The spike structure could very well be an artifact of the sharp connection made at the boundary. What is however physical is the rapid increase of the sound speed at the transition point signaling the changeover of the degrees of freedom.  Significantly, this allows us to set the upper bound for $n_{1/2}$
\be
n_{1/2} \lsim 4 n_0.\label{upper}
\ee
Thus together with the lower bound (\ref{lower}), we can pinpoint the threshold density for topology change
\be
2 n_0 \lsim  n_{1/2} \lsim  4 n_0.\label{boundhalf}
\ee
We are not excluding the values at the bounds because there is no indication that they are clearly inconsistent with other global properties of the compact stars involved.
\begin{figure}[!h]
 \begin{center}
   \includegraphics[width=8cm]{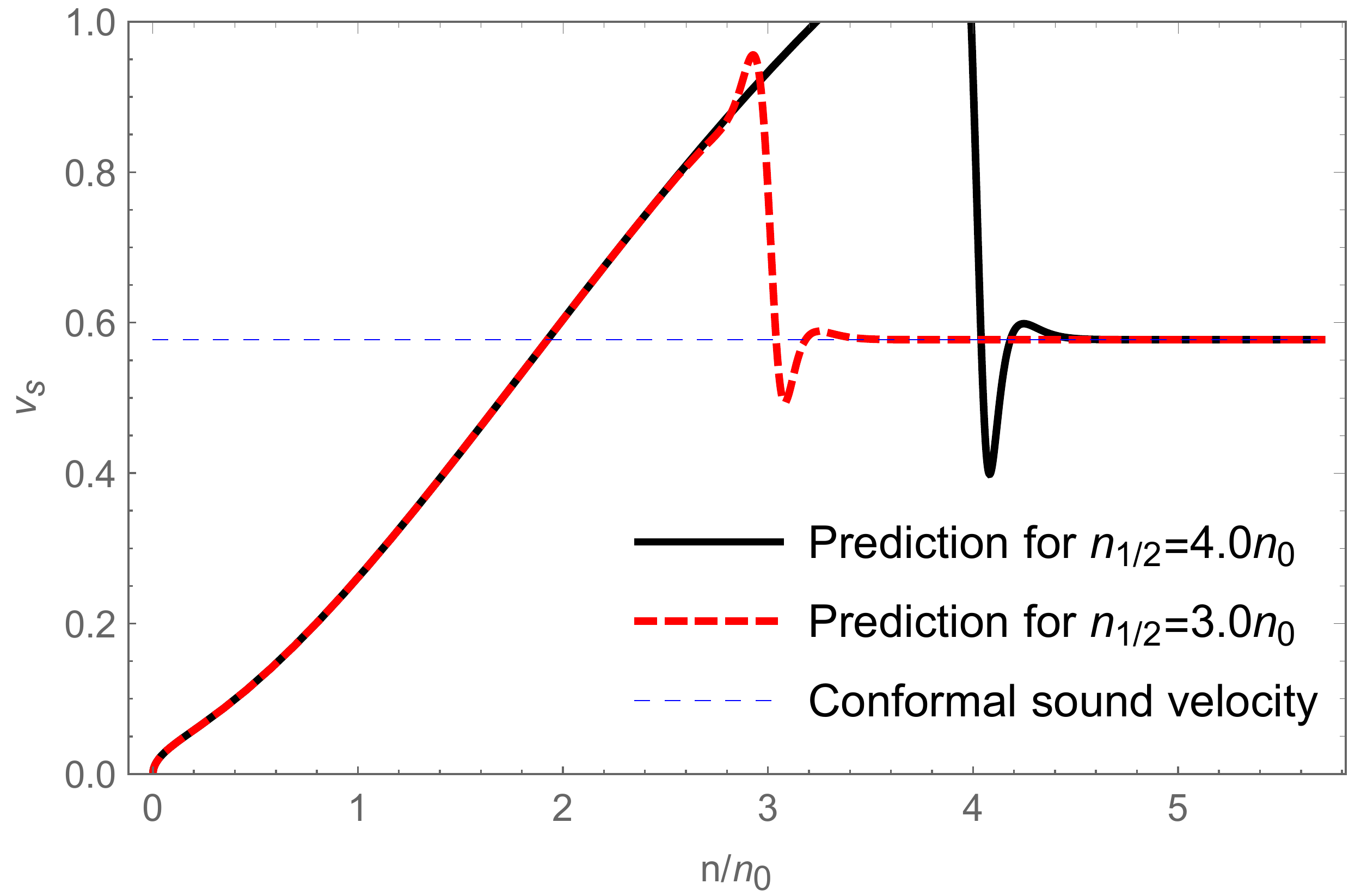}
  \end{center}
  \vskip -1cm
 \caption{Sound velocity as a function of density in neutron matter.  Proton contribution is the same as neutron's for $n > n_{1/2}$ and negligible compared with neutron's at $n< n_{1/2}$. }
\label{Vs2-4}
\end{figure}

$\bullet$ { \bf Proposition XIII: \it  Analyses on recently uncovered compact-star properties pin down the topology change threshold densities to the range ${2 \lsim  n_{1/2}/n_0 \lsim 4}$.}

Of course we cannot be more precise than the bound (\ref{boundhalf}). For simplicity, we just pick $n_{1/2} = 3 n_0$ as the threshold density for transition. Taking into consideration of the $V_{lowk}$RG result for Fig.~\ref{Vs} for $n_{1/2}=2 n_0$, one expects the sound speed  to converge to the conformal speed at  $\lsim 4n_0$. The important point is that this is an order of magnitude lower  than the asymptotic density $\gsim 50 n_0$ perturbative QCD predicts. This is a surprise, signaling the precocious emergence of pseudo-conformality in compact stars. However the robustness of the topological inputs figuring in the formulation convinces us that the precocious onset of the pseudo-conformal structure can be trusted at least qualitatively. In this connection, a recent detailed analysis of currently available data in the quarkyonic model does confirm that indicate the onset density of $v_c^2\approx 1/3$ at $\sim 4 n_0$~\cite{lattimer}.

Now given the extreme simplicity of the EoS for $n> n_{1/2}$, does the PCM still manage to capture all of  compact-star physics involving what is considered to be intricate nonperturbative processes?

We show that indeed with the exception of a possible tension with the tidal deformability ${\Lambda}$, which will be addressed below, there are no visible conflicts with the presently available empirical results. For this purpose we simply summarize the principal results for $n_{1/2}/n_0 = 3, 4$  given in Refs.~\cite{MR-PCM,MR-Vs} without detailed explanations. We note that apart from the conflict with the causality and slight tension in pressure, the case of $n_{1/2}=4n_0$ fares just as well as the case of $n_{1/2}=3n_0$. This is the reason why we cannot exclude within our framework $n_{1/2}=2n_0$ and $4n_0$.

\begin{figure}[!h]
 \begin{center}
   \includegraphics[width=7.3cm]{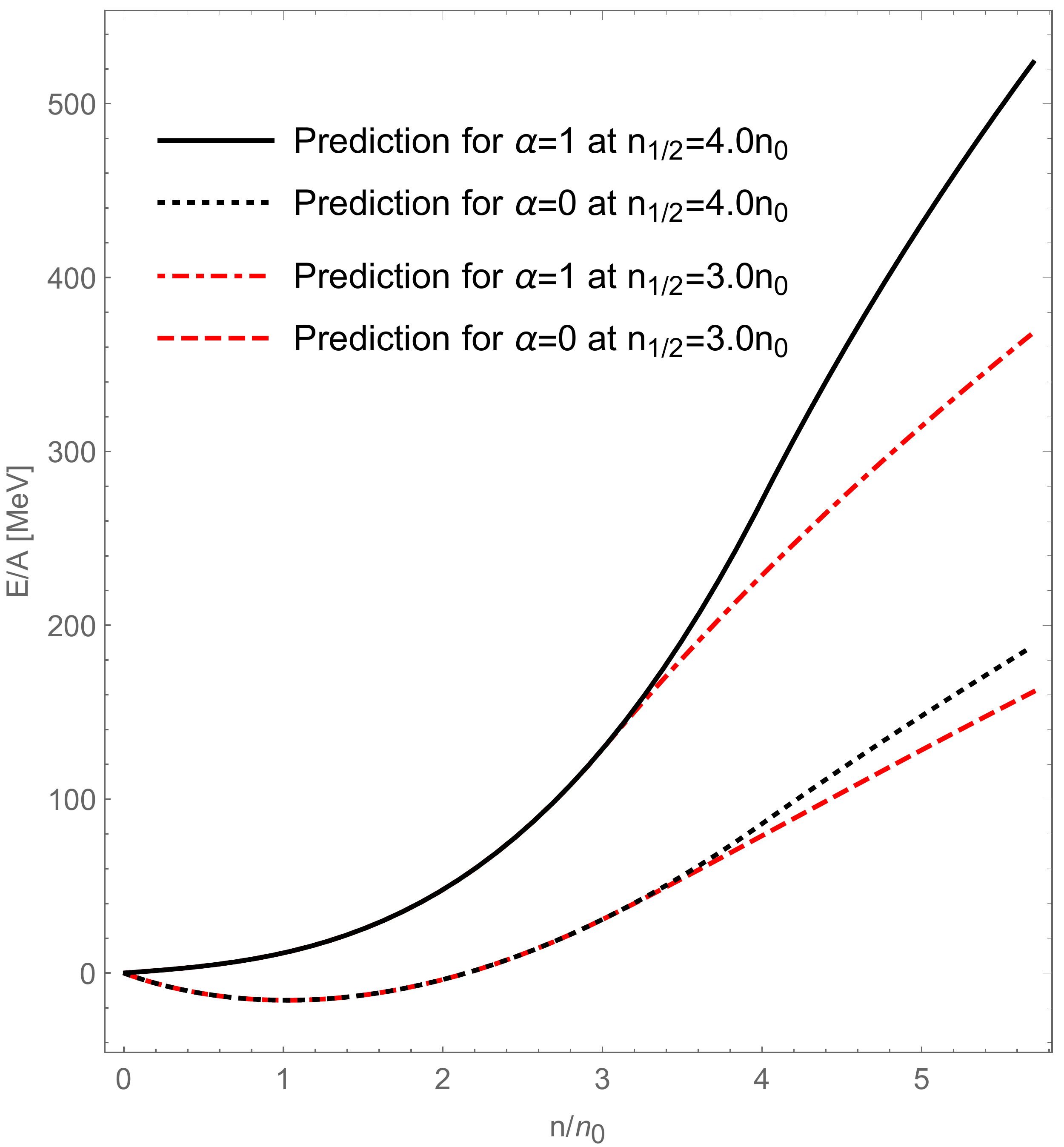}
   \includegraphics[width=7.3cm]{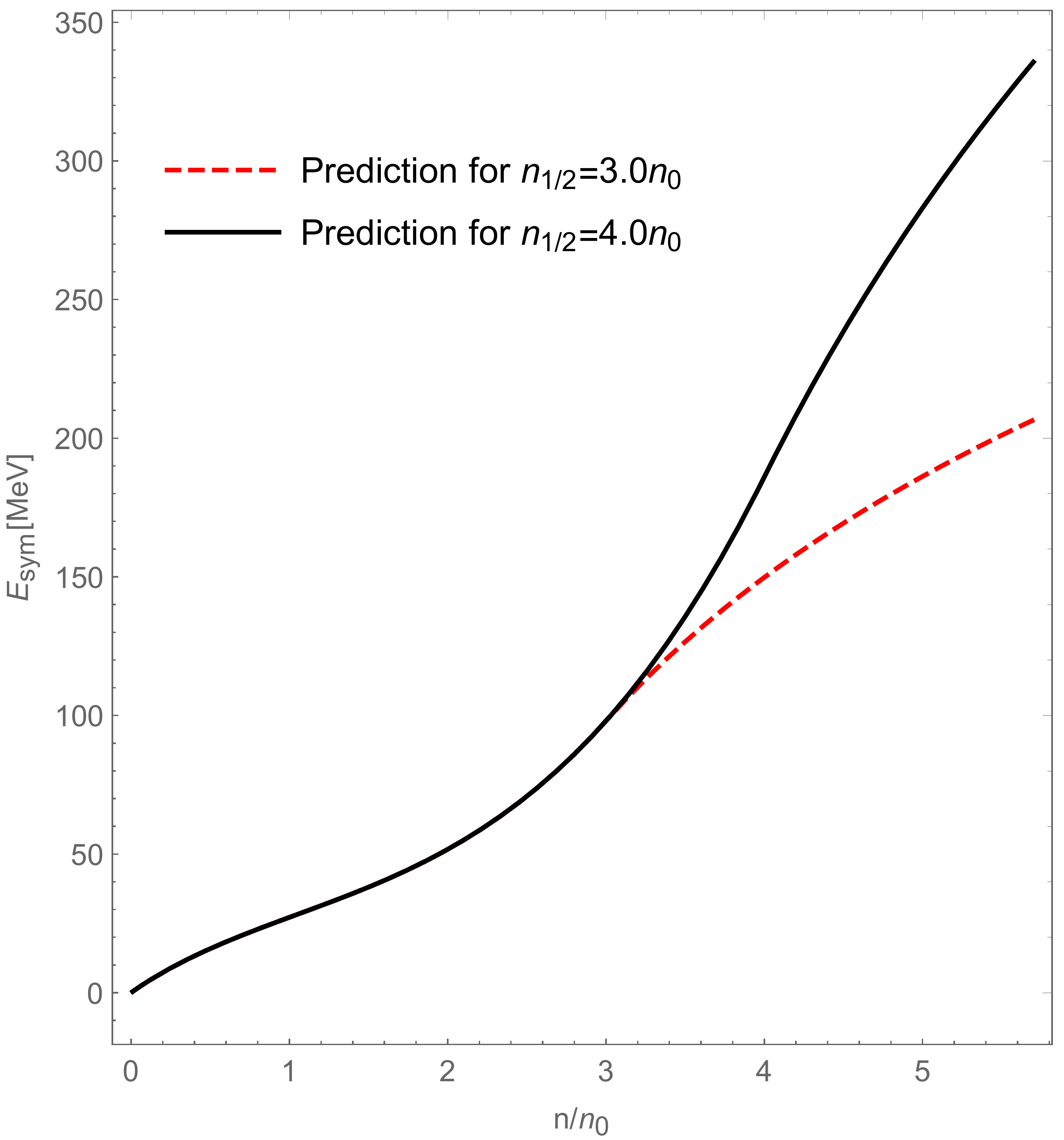}
  \end{center}
  \vskip -0.5cm
 \caption{Predicted $E/A$ and $E_{sym}$ vs. density. The upper (lower) curves of the left panel are for the pure neutron matter with $\alpha=1$ (symmetric nuclear matter with $\alpha=0$).}
\label{E-S}
\end{figure}

$\bullet$ {\bf Energy per particle and symmetry energy}:

We show in Fig.~\ref{E-S} the energy per particle $E/A$ for pure neutron matter and symmetric nuclear matter and the symmetry energy.  Recall that the normal nuclear matter properties are identical for all $n_{1/2}$. Note that the $n_{1/2}=4n_0$ case shows stronger repulsion for neutron matter and of course for the symmetry energy for $n\gsim 4n_0$. This repulsion may be in line with the causality violation for that density.  Otherwise, there are no conflicts with Nature.
 \begin{figure}[h]
 \begin{center}
   \includegraphics[width=8cm]{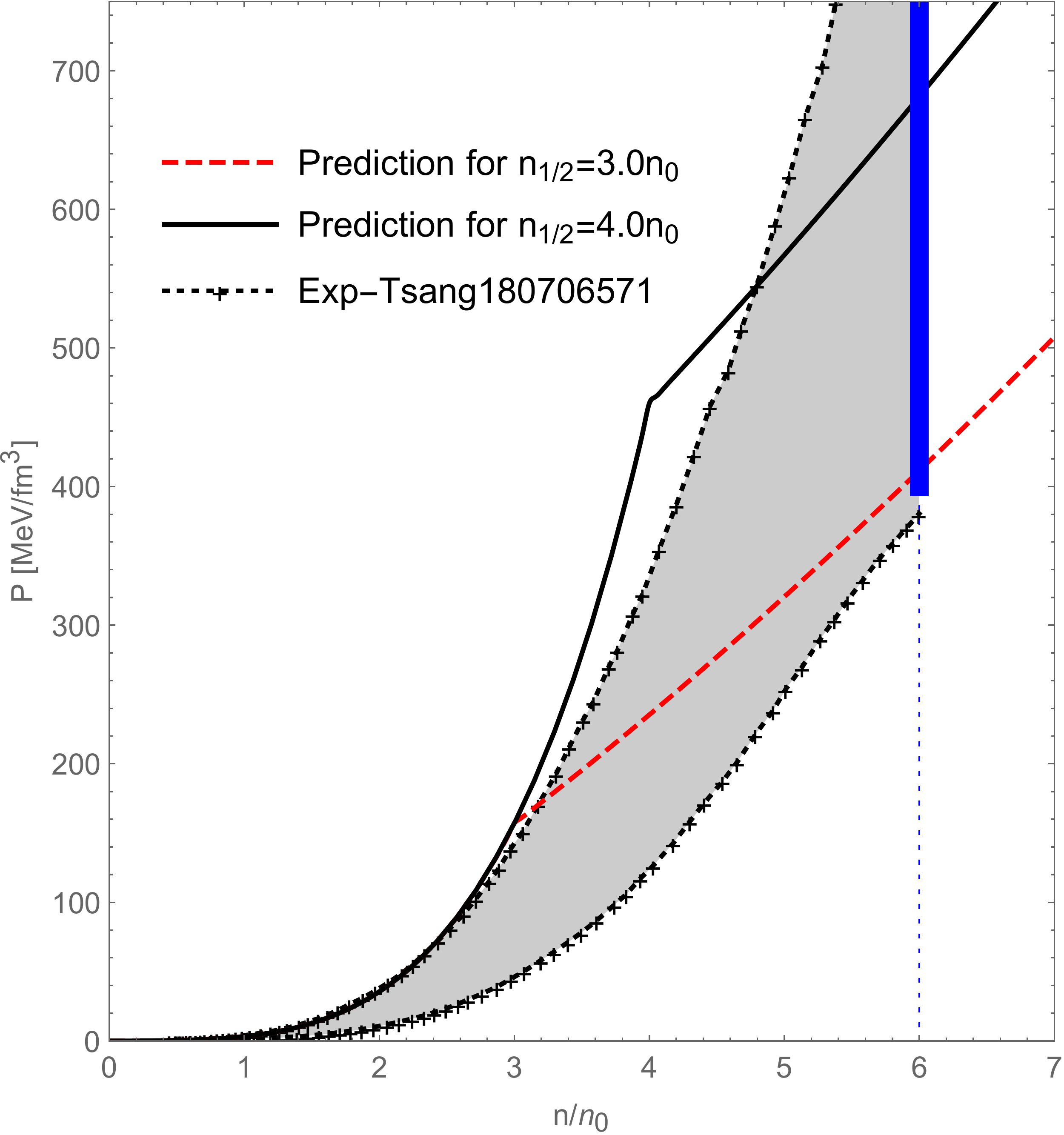}
  \end{center}
  \vskip -0.6cm
 \caption{Predicted pressure for neutron matter ($\alpha=1$) vs density compared  with the available experimental bound (shaded) given by Ref.~\cite{Tsang} {and the bound at $6n_0$ given by Eq.~\eqref{P2} (blue band)}.}
\label{Tsang}
\end{figure}

$\bullet$ {\bf Pressure for neutron matter ($\alpha=1$)}

Plotted below in Fig.~\ref{Tsang} is the predicted pressure $P$ for $n_{1/2}/n_0= 3,4$ compared with the presently available heavy-ion data~\cite{Tsang}.
The case of $n_{1/2}=4 n_0$, while consistent with the bound at $n\sim 6n_0$,  goes outside of the  presently available experimental bound at $n\sim 4n_0$. This may again be an artifact of the sharp matching, but that it violates the causality bound seems to put it in tension with Nature. Nonetheless it may  be too hasty to rule out the threshold density $n_{1/2}=4n_0$.  We need a better understanding of the cusp singularity present in the symmetry energy mentioned above before ruling this out.

 $\bullet$ {\bf Star mass $M$ vs. radius $R$ and central density $n_{\rm cent}$.}

\begin{figure}[h]
 \begin{center}
   \includegraphics[width=7.7cm]{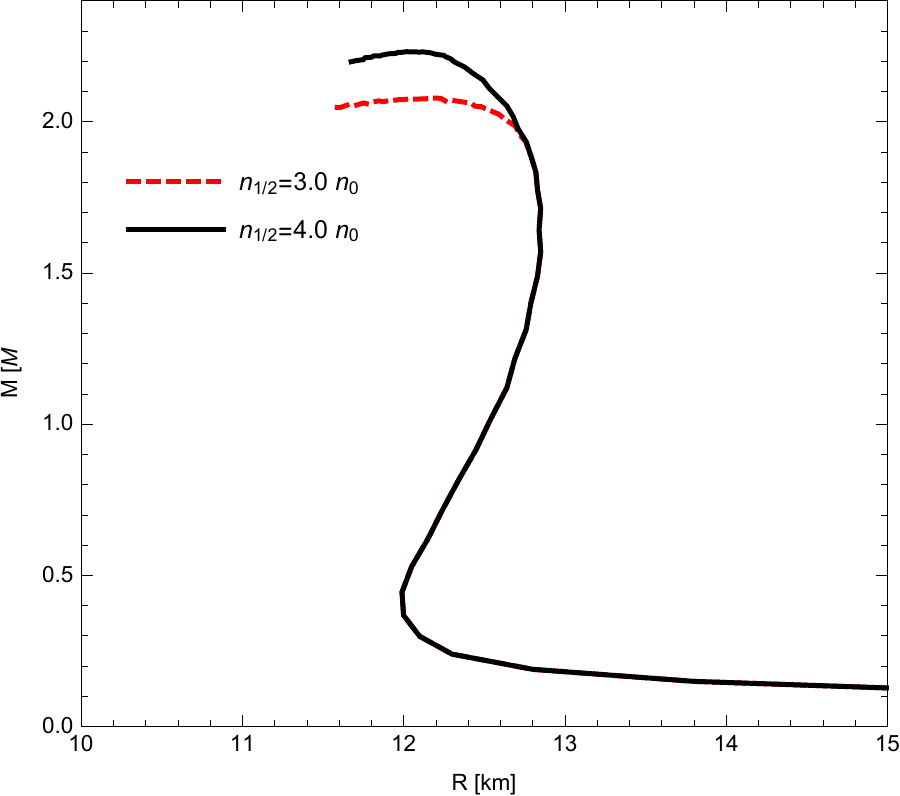}
   \includegraphics[width=7.5cm]{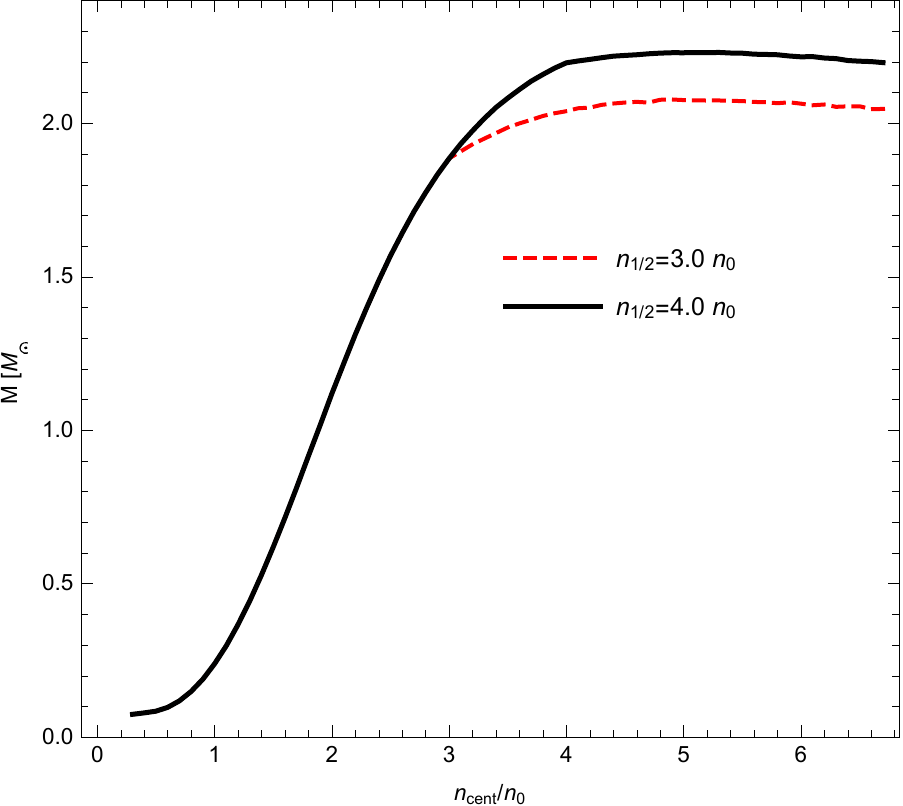}
 \end{center}
 \vskip -0.5cm
 \caption{Star mass $M$ vs. radius $R$ and central density $n_{\rm cent}$  with different choices of $n_{1/2}$. Note that below $M\approx 2M_\odot$, the curves for $n_{1/2}/n_0 =3.0\ {\rm and} \ 4.0$ represented in red with black dots are coincident.}
\label{star-mass}
\end{figure}

The solution of the TOV equation with the pressures of leptons in beta equilibrium duly taken into account as in Ref.~\cite{PKLMR} yields the results for the star mass $M$ vs. the radius $R$ and the central density $n_{\rm cent}$ as given in Fig.~\ref{star-mass}.
 The maximum mass comes out to be roughly  $2.04M_\odot \sim 2.23 M_\odot$ for $2.0 \leq n_{1/2}/n_0 \leq 4.0$, the higher the $n_{1/2}$, the greater the maximum mass. This bound is consistent with the observation of the massive neutron stars (\ref{M3})-(\ref{M1}). It is notable  that, when $n_{1/2} \geq 3.0 n_0$, changing the position of $n_{1/2}$  affects only the compact stars {with mass $\gtrsim 2.0 M_\odot$} although the mass-radius relation is affected by the topology change when $2.0 n_0 \leq n_{1/2} \leq 3.0 n_0$.

As for the central density of the stars,  it falls in the range  $\sim (4-5)n_0$, more or less independent of the topology change density.

$\bullet$ {\bf Gravitational wave observations}

Finally we turn to how our theory fares with what came out of the LIGO/Virgo gravitational observations. The relevant quantities that we will consider are the dimensionless tidal deformability $\Lambda_i$ for the star $M_i$ and $\tilde{\Lambda}$ defined  by
\begin{eqnarray}
\tilde{\Lambda} & = &  \frac{16}{13}\frac{(M_1 + 12 M_2)M_1^4 \Lambda_1 + (M_2 + 12 M_1)M_2^4 \Lambda_2}{(M_1 + M_2)^{5}}
\label{eq:tildeL}
\end{eqnarray}
for $M_1$ and $M_2$ constrained to the well-measured ``chirp mass"
\begin{eqnarray}
{\cal M} & = & \frac{(M_1 M_2)^{3/5}}{(M_1 + M_2)^{1/5}} = 1.188 M_\odot .
\label{eq:chirpmass}
\ee


There are many sophisticated issues on those quantities that have been the subject of current activities in the field, both theoretical and experimental. We shall not  go into those matters and focus only on those quantities that are crucial for the thesis that we have developed. It must be said rather generally that  what is available so far from GW170817 is not yet tight enough to give a verdict on the validity of our model. In fact at present it rules out only a few models available in the literature.

The predictions of our PCM for ${\Lambda}$ vs.$M$ and $n_{\rm cent}$ are plotted in Fig.~\ref{LvsM-nc}, and the correlations between $M$ vs. $(n_{\rm cent}, {\Lambda}, R)$ relevant to the LIGO/Virgo data associated are listed in Table \ref{Table}.
\begin{figure}[!h]
 \begin{center}
  \includegraphics[width=7.5cm]{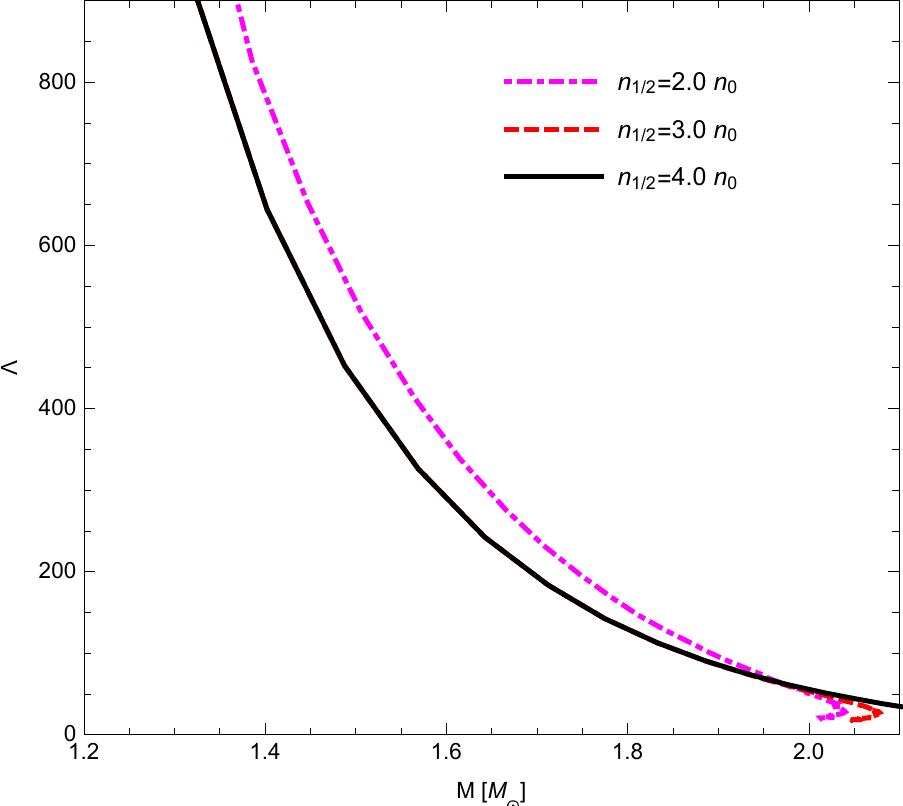}\quad \includegraphics[width=7.7cm]{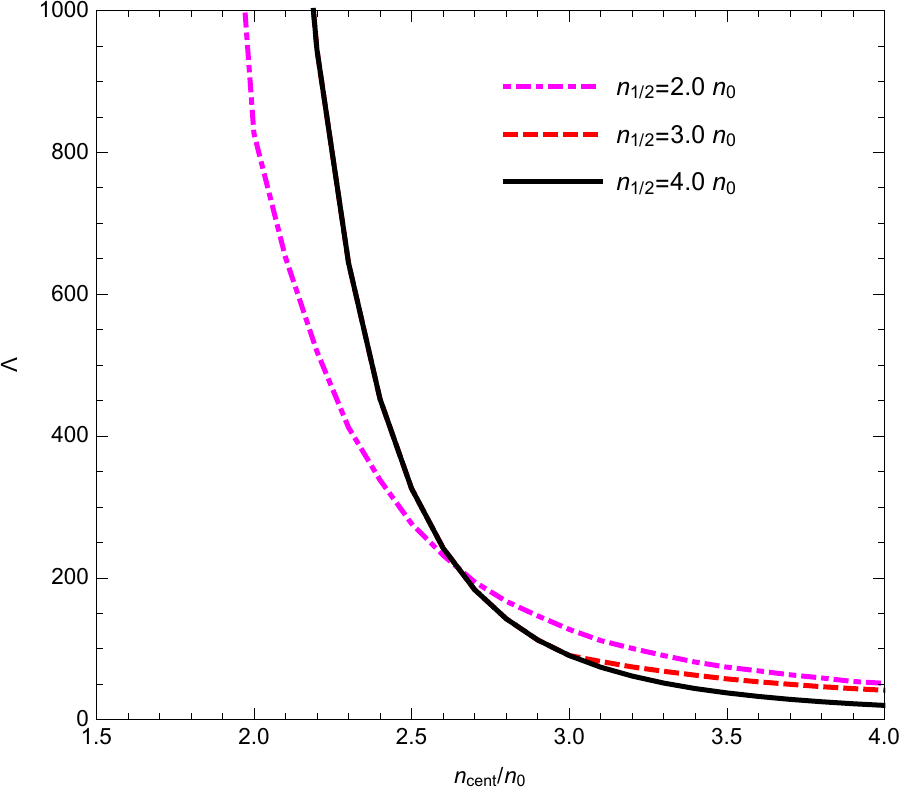}
  \end{center}
 \caption{The tidal deformability vs. star mass $M$ and the central density $n_{\rm cent}$. The results for $n_{1/2}=2n_0$ are given for comparison.}
\label{LvsM-nc}
\end{figure}
%
%
%
\begin{widetext}
\begin{table*}[!htp]
\centering
\caption{Properties of compact stars with different masses and $n_{1/2}/n_0=2,3,4$. The case of $n_{1/2}=2n_0$ is added for comparison.}
\label{Table}
\begin{tabular}{c|ccc|ccc|ccc}
\hline
\hline
\multirow{2}{*}{$M/M_\odot$}& \multicolumn{3}{c|}{$n_{cent}/n_0$}&\multicolumn{3}{|c}{ $\Lambda/100$}&\multicolumn{3}{|c}{ $R$/km}\cr\cline{2-10}
&$n_{1/2}=2.0$&$n_{1/2}=3.0$&$n_{1/2}=4.0$&$n_{1/2}=2.0$&$n_{1/2}=3.0$&$n_{1/2}=4.0$&$n_{1/2}=2.0$&$n_{1/2}=3.0$&$n_{1/2}=4.0$
\cr
\hline
1.12 &1.81&2.00&2.00&25.3&22.5&22.5&12.7&12.6&12.6\cr\hline
1.22 &1.88&2.10&2.10&16.7&14.2&14.2&12.8&12.7&12.7\cr\hline
1.31 &1.95&2.20&2.20&11.6&9.50&9.50&12.9&12.8&12.8\cr\hline
1.40 &2.02&2.30&2.30&7.85&6.52&6.52&13.0&12.8&12.8\cr\hline
1.49 &2.17&2.40&2.40&5.54&4.50&4.50&13.1&12.8&12.8\cr\hline
1.57 &2.31&2.50&2.50&4.00&3.25&3.25&13.1&12.8&12.8\cr
\hline
\hline
\end{tabular}
\end{table*}
\end{widetext}
We first make a general assessment of the predictions and then focus on the properties of the $1.4 M_\odot$ star of the GW170817 observation.

From Fig.~\ref{LvsM-nc} and Table~\ref{Table}, one can make the general observations relative to the LIGO/Virgo data  in terms of the pseudo-conformal model.  We will limit our considerations to the star mass range $1.2 \lsim M/M_\odot \lsim 1.6$.
\begin{enumerate}
\item
There are striking differences between the results of $n_{1/2}=2.0n_0$ and those of $n_{1/2} > 2.0 n_0$. The former, which might be disfavored by the bound $\Lambda_{1.4} < 800$, differs appreciably from the latter, all of which share nearly the same properties of $M$ vs. $n_{\rm cent}$ and $\Lambda$.
\item  The radius $R$ is remarkably independent of $n_{1/2}$ as well as of $M$ for $n_{1/2} > 2.0 n_0$.
\item The tidal deformability ${\Lambda}$ is extremely sensitive to $M$ and $n_c$ although $R$ is not. This means that  the star mass that fixes ${\Lambda}$ has to be determined very accurately  to impose  a strong constraint on the validity of the PCM.
\item For $n_{1/2} > 2n_0$,  the central density is located in R--I, i.e., below the cusp, hence ``soft,"  while for $n_{1/2}=2 n_0$ it is in R--II and hence ``hard." Thus the bound $\Lambda_{1.4} < 800$ requires that it be probed in R--I which makes $n_{1/2}=2 n_0$ disfavored.

 To confront the LIGO/Virgo data, we plot our predictions for $\tilde{\Lambda}$ in Fig. \ref{lambdabar} and for $\Lambda_1$ vs. $\Lambda_2$ in Fig.~\ref{L1vsL2}.
\begin{figure}[h]
\vskip 0.5cm
 \begin{center}
   \includegraphics[width=8cm]{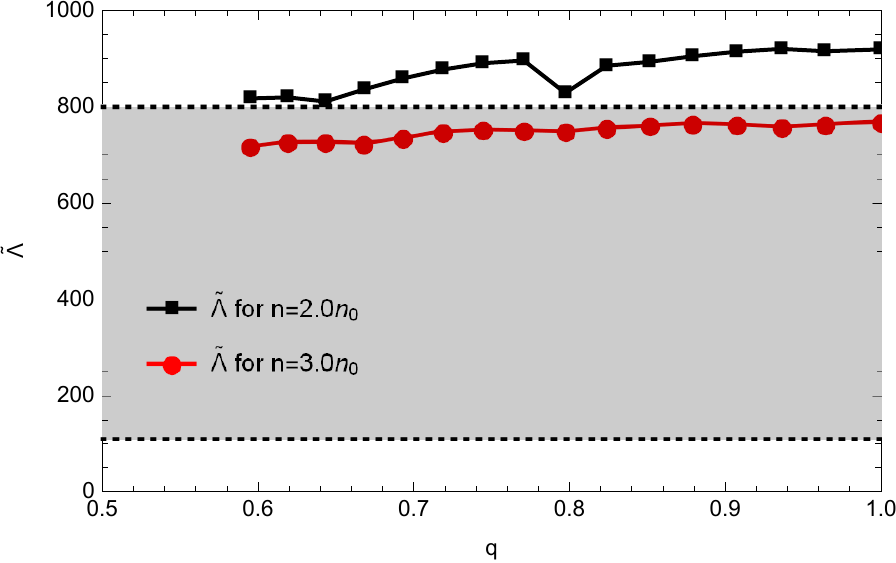}
  \end{center}
  \vskip -0.7cm
 \caption{The dimensionless tidal deformability $\tilde{\Lambda}$ for chirp
mass $1.188M_\odot$, as a function of the mass ratio q calculated from PCM with $n_{1/2} = 2.0 n_0$ (black line) and $n_{1/2} = 3.0 n_0$ (red line). The gray band is the constraint from the low spin
$\tilde{\Lambda} = 300^{+500}_{-190}$ obtained from GW170817.}
\label{lambdabar}
\end{figure}
\begin{figure}[h]
 \begin{center}
   \includegraphics[width=8cm]{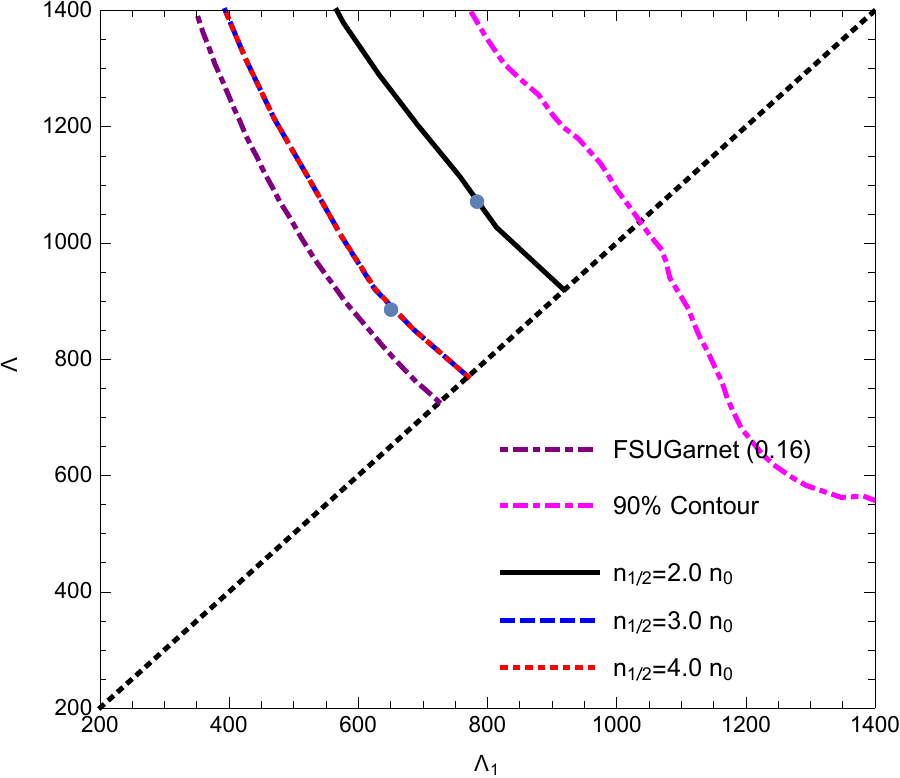}
  \end{center}
 \vskip -0.6cm
 \caption{Tidal deformabiliitiess $\Lambda_1$ and $\Lambda_2$ associated with the high-mass $M_1$ and low mass $M_2$ components of the binary neutron star system GW170817 with chirp
mass $1.188M_\odot$. The constraint from GW170817 at the 90\% probability contour is also indicated. We quote  ``FSUGarnet (0.16)"~\cite{Fattoyev:2017jql} as a presently available ``state-of-art" theoretical prediction.}
\label{L1vsL2}
\end{figure}
As it stands, our prediction is compatible with the LIGO/Virgo constraint for $n_{1/2}\gsim 2n_0$. Although there seems to be some tension with the pressure, the result for $n_{1/2}=4 n_0$ is of quality comparable to that of $n_{1/2}=3 n_0$.
\end{enumerate}

\subsubsection{Tension with ${\Lambda} <580$ for $M_{1.4}$?}

Let us now focus on a possible tension with a $\tilde{\Lambda}$ tightened to an appreciably lower value than $\sim 800$ MeV. For this discussion consider specifically $\Lambda_{1.4}$ discussed by many authors.

Taking $n_{1/2}\simeq 3 n_0$ as an appropriate threshold density for the onset of the half-skyrmion phase,  the PCM predicts for $M_{1.4}$
\be
{\Lambda}_{1.4}\approx 650, \ R_{1.4}\approx 12.8\ {\rm km}, \ n_{\rm cent}\approx 2.3 n_0.\label{PCM-prediction}
\ee
We are putting the numbers as approximate because the central density for $M_{1.4}$ is in the cusp point where the matching between the $V_{lowk}$ RG for $n \leq n_{1/2}$ and the two-parameter PCM formula (\ref{matchedE}) is made. It involves smoothing of the cusp singularity which is highly uncertain as stressed above.  The predictions are amply consistent with the bounds quoted in Section \ref{compact-star-bounds}, ${\tilde\Lambda} < 800$ and $R < 13.6$ km. Note that the central density is located in R--I, hence should be correlated with the properties of normal nuclear matter.

Since  the star mass for $M_{1.4}$ seems to have a central density $\sim 2 n_0$,  careful high-order S$\chi$EFT calculations could perhaps give a hint as to how far ${\Lambda}_{1.4}$ could be tightened downward. It is difficult to gauge how reliable such estimates could be,  but the indication is that it could be brought down~\cite{holt-lim} below 650 the PCM predicts or even below 400, the lower bound in (\ref{gwbound}).

In this connection,  the recent analysis of \cite{oezel-smallL} seems to favor $L(n_0) \sim 20$ MeV, much smaller than the PCM prediction $\sim 49$ MeV. On the other hand, the PCM predicts symmetry energy at $2n_0$,  $E_{sym} (2n_0) \sim 49$ MeV, which is essentially the upper bound of the range given by  \cite{BAL2019}, $39.2^{+12.1}_{-8.2}$ MeV. This means that the PCM symmetry energy increases with a slope greater by a factor of 2 to 3 times than what seems to be indicated by the gravity-wave data. This rapid increase, if confirmed,  could very well over-predict the tidal deformability compared with the data. If this turned to be the case, then the PCM could be seriously in tension with the bound $70 < \Lambda_{1.4} < 580$ given in \cite{Abbott:2018exr,holt-lim}.  It does not seem feasible within our PCM to lower much below $\sim 650$.
The reason for this possible tension is easy to understand. In the framework based on the topological structure of the cusp -- and since the ${\Lambda}_{1.4}$ probes just below the cusp density, it would be a difficult problem to figure out what is going on that boundary region. In the topology-based reasoning, the subtlety conjectured in Section \ref{conjecture} would be involved. Also in terms of the hadron-quark continuity, there would be an equally complicated mechanism at work. In both cases,  It could perhaps hide a fundamental new physics involving intricate topological effects in strong interactions.

We should also mention that the star crust, not given consideration in this review, could play a role. With certain crust profile and compactness, $\Lambda_{1.4}$ can vary from $\sim 620$ to $\sim 100$~\cite{crust}.
\section{Conclusion and further remarks}
With hidden gauge symmetry  and hidden scale symmetry made to emerge at high density implemented to an effective field theory ($bs$HLS), we have made extremely simple predictions of  what come out to be surprisingly consistent with all, except possibly the tidal deformability tightened to a much lower value than the present upper bound, of the observation of compact-star properties. The essential ingredient of the description is what is believed to be the robust topology change derived from skyrmions put on lattice, reliable at high density and with large number of colors in QCD. The changeover from skyrmions to half-skyrmions  at a density $n_{1/2}\approx 3 n_0$ is a Cheshire Cat mechanism for -- and a  duality to -- ``hadron-quark continuity" encoded in  QCD to go from low to high density  in compact star--star matter.  Intriguingly the mechanism shares a variety of topological phenomena actively being studied  in condensed matter systems.

It is found that the consistency with the presently available data, both terrestrial and space laboratories, requires the threshold density $n_{1/2}$ for the topology change to be pinned to the narrow range $2 \lsim  n_{1/2}/n_0 \lsim 4$. This is quite compatible with what is being taken for hadron-quark continuity in the literature.

A totally unexpected consequence of the  inevitability of the vector manifestation associated for the $\rho$ vector meson, intricately tied with parity-doubling symmetry, is to induce the emergence of pseudo-conformal symmetry with the (pseudo-)conformal sound speed converging at a precocious density $\gsim 4 n_0$, very far from the asymptotic density $\gsim 50 n_0$.  This prediction  is not shared by any other theoretical models available in the literature.  It is not impossible that the threshold density could be higher than what is found in this simplified PCM, but the prediction is definitely far from asymptotic. It would be of a great interest if this feature could be checked by a nonperturbative QCD calculation  or experimentally.
Another intriguing possibility is that this pseudo-conformal structure of dense medium is linked to the IR fixed point structure of scale-chiral symmetry.  That the sound velocity is conformal even though the dilaton condensate (and equally the quark condensate) is non-zero may be related with the existence of an IR fixed point where scale-chiral symmetry is spontaneously broken \`a la Crewther and Tunstall~\cite{CT}, which is drastically different from the current lore with large $N_f$ scale symmetry near conformal window. One highly intriguing possibility is its role in controlling the $\omega$ repulsion in stabilizing dense matter when the latter is treated as a skyrmion matter in scale-symmetrized hidden local symmetric Lagrangian. It is most likely connected to the role the $\omega$ plays in $bs$HLS in giving rise to the parity doubling discussed in Section \ref{paritydoubling}. A discussion on this matter is relegated to Appendix~\ref{appendix}.

Finally we point out an intriguing possibility for  ``continuity" in the role scale symmetry plays in nuclear physics from low density to high density.  At low density, the unitarity limit is in action with large scattering length in pionless EFT (with all meson fields {\it integrated out}) both in nuclear structure~\cite{vankolck-unitary} and in compact-star physics at the lower limit to the symmetry energy~\cite{lattimer-unitary}. And at high density, there emerges pseudo-conformality with the (pseudo)conformal sound velocity ( with all meson fields, pion as well as heavy fields, {\it integrated in}). In between the two opposite density regimes, there is the ordinary nuclear matter with possible ``hidden" scale symmetry or ``conformality lost."

A particularly important issue in mapping the hadron-quark continuity to the topology change that is left unaddressed in this work is the role of strangeness. In the approach anchored on hadron-quark continuity, strange quarks naturally enter. So the question is: How does the strangeness affect the compact star properties in the topology-based treatment? To address this issue, we would need to extend the $V_{lowk}$RG strategy to three flavors, which has not been done yet. However the preliminary analysis based on the Fermi-liquid fixed-point  approximation to the $V_{low}$RG~\cite{strangeness} indicates that strangeness could be banned to densities higher than what may be relevant for compact stars. A similar result was obtained sometime ago by Pandharipande et al~\cite{pandharipande} where strong short-range correlations were found responsible for pushing kaon condensation to high densities.  It seems that half-skyrmion coupling to strangeness needs to be carefully worked out to answer the question. This is an open problem for the future.

\subsection*{Note Added}
After the draft of this review was completed, there appeared a paper in which  the $\rho$ and $\omega$ mesons, which figured importantly in our work {comprehensively discussed above}, are proposed to be  the gauge bosons in a topological phase dual to  the gluon via the level-rank duality between the three dimensional gauge theories $SU(N_c)_{N_f}$ and $U(N_f)_{-N_c}$ for $N_f=2$~\cite{kanetal}. We touched on a similar duality in Section \ref{poweroftopology} for $N_f=1$ in connection with the Cheshire Cat principle. In  the scenario of \cite{kanetal},  both the $\rho$ and $\omega$ become massless at the chiral symmetry restoration, perhaps at high temperature as proposed by the authors. We would like  to point out that this scenario is in stark contrast to our scenario at high density where the $U(2)$ symmetry breaks down strongly, with the $\rho$ moving toward the vector manifestation $m_\rho\to 0$ while the $\omega$ remains massive.  We cannot say that the $\rho$ and $\omega$ will not go massless together, perhaps after a sort of a phase transition, at much higher density than that relevant to compact stars. But it appears that there is a definite difference between the two scenarios. It would be interesting to understand the difference.

\appendix

\section{Reparameterization of linear sigma model}

\label{app:LSM}

To see that the scale symmetry is hidden in the nonlinear sigma model, following Yamawaki, we consider the following $G = SU(2)_L\times SU(2)_R$  linear sigma model with two parameters $\mu$ and $\lambda$,
\begin{eqnarray}
{\cal L}_{L\sigma M} & = & \frac{1}{2} {\rm Tr} \left( \partial_\mu M\partial^\mu M^\dagger \right) - \frac{\mu^2}{2} {\rm Tr}\left(M M^\dagger\right)-\frac{\lambda}{4}  {\rm Tr}\left(M M^\dagger\right)^2 \,.
 \label{linear-sigma}
 \end{eqnarray}
The $2\times2$ matrix $M$ is defined as
\begin{equation}
M=\frac{1}{\sqrt{2}}\left({\hat \sigma}\cdot  1_{2\times 2} +2i {\hat \pi}\right)\, \quad \left({\hat \pi} \equiv {\hat \pi}_a \frac{\tau_a}{2}\right)
\label{MLinearSigma}
\end{equation}
which transforms under $G=SU(2)_L\times SU(2)_R$ as
\begin{equation}
M \rightarrow g_L \, M\, g_R^\dagger \,,\quad g_{R,L} \in SU(2)_{R,L}.
\end{equation}
Under the scale transformation of an operator ${\cal O}(x)$ with the scale dimension $d_{\cal O}$, $\delta {\cal O}(x)=  (d_{\cal O} +x^\nu\partial_\nu) {\cal O}(x)$,
the action $S=\int d^4 x {\cal L} (x)$ for a given Lagrangian  ${\cal L} (x)$ becomes
\begin{eqnarray}
\delta S 
& = & \int d^4 x (d_{\cal L}-4){\cal L}\,.
\end{eqnarray}
Thus the action will be scale-invariant only if $d_{\cal O}=4$ in the Lagrangian ${\cal L} =\sum_i {\cal O}_i$.  This means that the term $\propto \mu^2$ in Eq.~(\ref{linear-sigma}), i.e., the mass term, with scale dimension 2 is not scale-invariant while the other terms are. Now any complex matrix, such as $M$, can be decomposed into a hermitian matrix $H$ -- which is diagonalizable -- and a unitary matrix $U$ as $M=HU$. In the chiral symmetry spontaneously broken phase, i.e., Nambu-Goldstone (NG) phase, write\footnote{There can be a bit of confusion with notations.  The $\sigma$ field in Eq.~(\ref{Polar})  is not to be confused with the fourth component of the chiral four-vector $\hat{\sigma}$ in Eq.~(\ref{MLinearSigma}).}
\begin{equation}
M = H\cdot U\,, \quad H=\frac{1}{\sqrt{2}} \left(\begin{array}{cc}
 \sigma & 0\\
 0  &\sigma
 \end{array}\right)
 \,, \quad U= \exp\left(\frac{2i \pi}{f}\right) \,
 \label{Polar}
\end{equation}
 which under the chiral transformation, transform as
 \begin{equation}
 U \rightarrow g_L \, U\, g_R^\dagger\,,\quad H \rightarrow H\,.
 \end{equation}
In the chiral symmetry spontaneously broken phase, i.e., NG phase, we have $ \la \sigma \rangle\equiv f$ .
 It will be seen that the constant $f$ can be related to  the pion decay constant $f_\pi$ under certain condition. This identification will be important in what follows.
 Note that $H$ is a chiral singlet and $UU^\dagger=1$,  meaning $\la U\ra = \langle\exp\left(2i \pi/f\right)\rangle=1 \ne 0$, so the spontaneous symmetry breaking is properly taken into account. The Lagrangian then takes the form
 \begin{eqnarray}
  {\cal L}_{L\sigma M}
 &=& \frac{1}{2} \left(\partial_\mu \sigma\right)^2+ \frac{1}{4}{\sigma}^2\cdot {\rm Tr} \left(\partial_\mu U \partial^\mu U^\dagger\right) -V(M)\nonumber\\
 V(M) &=& \frac{\mu^2}{2} {\sigma}^2  + \frac{\lambda}{4}  \sigma^4\,. \label{V(M)}
 \label{LagM}
 \end{eqnarray}
 Minimizing the potential (\ref{V(M)}) at $\sigma=f$, we rewrite it as
 \be
 V(M)=\frac{\lambda}{4} \left[\left(\sigma^2 -f^2\right)^2-f^4\right] \,.
 \ee
 We therefore have at our disposal one coupling constant $\lambda$ that can be dialed to get at different limits.

\section{In-medium anomalous dimension of $G_{\mu\nu}^2$:  $0 <\beta^\prime\lsim 3$ in dense matter?}\label{appendix}
There is an interesting possibility that the $\beta^\prime$, the anomalous dimension of the gluon stress tensor $G_{\mu\nu}^2$, could play a role in dense matter described in terms of skyrmion crystal. In arriving at the LOSS approximation for the scale-chiral Lagrangian described in Section~\ref{loss-lagrangian}, we assumed that the coefficients $c_{1,2}\approx 1$ in the Lagrangian (\ref{LOSS}). Similar assumptions are  made for higher chiral order terms if they are included as in Ref.~\cite{LMR}. In this way, $\beta^\prime$ dependence is absent in the leading-order scale symmetry. Although highly successful, there is no rigorous justification for this assumption unless non-leading-order terms containing corrections $\delta_i$ terms in $c_i=1+\delta_i$ are checked and found to be small. This checking seems nearly impossible to perform at present given the number of unknown parameters involved~\cite{LMR}. In fact there is a case where the role  of $\beta^\prime$ could not be ignored when the $\omega$ meson is included in Eq.~(\ref{LOSS}) to describe dense matter in a skyrmion crystal description.

When dense matter is described in terms of the skyrmion crystal, the homogenous Wess--Zumino (hWZ) term plays a crucial role in bringing in the $\omega$ meson into the dynamics. It is essential for the stability of nuclear matter. This term is encoded in the Chern-Simons term in 5D holographic QCD~\cite{HOLO} which gives rise  -- when  KK-reduced to 4D --  to the hWZ term. There are three such terms in HLS Lagrangian~\cite{HY:PR}. For our discussion, one can reduce them into one term without losing the essential physics, say, $g\omega_\mu B^\mu$ where  $g$ is  the HLS coupling and $B_\mu$ is topological baryon current.  This term is of scale dimension 4, so scale-invariant in the action. Hence when scale-symmetrized, it does not get multiplied by the conformal compensator field $\chi$. Naively one would think that term would not be affected in medium by the dilaton condensate.

Now it is easy to see that this term contributes very importantly to the energy of dense system,
\begin{eqnarray}
 \left(\frac{E}{A}\right)_{WZ} & = & \frac 14\left(\frac{3g}{2}\right)^2\int_{\rm Box}
 d^3x\int d^3x^\prime B_0 (\vec{x})
 \frac{{\rm exp}(-m_\omega^\ast|\vec{x}-\vec{x}^\prime|)}
 {4\pi|\vec{x}-\vec{x}^\prime|} B_0 (\vec{x}^\prime) \, , \label{wzenergy}
 \end{eqnarray}
where ``Box" corresponds to a single FCC cell  and $m_\omega^\ast$ is the in-medium mass.
Note that while the integral over $\vec{x}$
is defined in a single (FCC) cell, that over $\vec{x}^\prime$ is
not. Thus, unless it is screened, the periodic source $B_0$
filling infinite space will produce an infinite potential $w$
which leads to an infinite $(E/B)_{WZ}$. The screening is done by
the $\omega$ mass, $m_\omega^*$. This means that  $m_\omega^*$ must go  up to control the integral~\cite{PRVI}. It is definitely at odds with the scaling we have deduced.
In fact it would bring havoc in nuclear interactions~\cite{PRVI}.  Among others it  would  block the system's flow, at high density, to the vector manifestation fixed point absolutely crucial for the framework.

This disaster can be resolved in the CT scheme. In this scheme, the hWZ term is modified to
\be
{\cal L}_{\rm hWZ}=g\omega_\mu B^\mu \left(c_h +(1-c_h)\left(\frac{\chi}{f_\chi}\right)^{\beta^\prime}\right)\, ,
\ee
where $c_h$ is an unknown constant like $c_{1,2}$. {In the LOSS approximation made in the normal parity Lagrangian (\ref{LOSS}),  we set $c_{1,2}\approx 1$ so the $ \beta^\prime$ dependence is present only in the dilaton potential. But in the anomalous parity Lagrangian there is nothing to suggest that one can set $c_h=1$. This is because in the skyrmion description, with $c_h=1$, there would be no influence of scalar channel on nuclear interactions that involve the $\omega$ meson.   This is in some sense  related to what would happen to Walecka's mean-field approach if the scalar attraction were suppressed. There would be no bound states.

In Ref.~\cite{PRVII}, an {\it ad hoc}\ solution was suggested by simply multiplying the hWZ tern by a scale-symmetry explicit-breaking coupling $(\chi/f_\chi)^3$, giving a suppression factor $(\la\chi\ra^\ast)^3$ in medium, which  made the integral (\ref{wzenergy}) harmless in having $\la\bar{q}q\ra^\ast$ go to zero (in the chiral limit) as density increased.

It was recognized in Ref.~\cite{MR-hwz} that the solution of Ref.~\cite{PRVII} was what corresponds to the CT model with $c_h\neq 1$ and $\beta^\prime >0$. A preliminary study indeed indicated  $c_h\sim 0.1-0.2$ and a $\beta^\prime\lsim 3$ would eliminate the divergence and make chiral and scale symmetry -- which are locked to each other in the CT theory~\cite{CT} -- restored at  high density.

However there is one subtlety that was not recognized in Ref.~\cite{MR-hwz}. It is the parity-doubling discussed in Section \ref{paritydoubling} that involves the interplay between the dilaton condensate and the in-medium nucleon mass in the Fermi-liquid fixed point approximation on the one hand and  on the other hand the quasiparticle structure in the half-skyrmion phase described in Section \ref{quasiparticles}. The two mechanisms must be inseparably related. This means that $c_h$ and  $\beta^\prime$ must be closely linked to the pseudo-conformal structure in the half-skyrmion phase. Understanding what takes place  here  would require a much more detailed study. It is intriguing that if correct, this problem would indicate that dense baryonic matter could provide a hint, as of now totally absent, for the anomalous dimension $\beta^\prime$ in QCD for $N_f\sim3$. With the PC structure being an emergent symmetry, this may be a medium-specific quantity, but it is a highly provoking issue in nuclear physics.

\section{The problem of ``quenched $g_A$" in nuclei}
Perhaps less prominent than in compact stars at high density discussed  in Appendix \ref{appendix}, an interesting possibility is  that the anomalous dimension $\beta^\prime$ could also figure at low density, say, at nuclear matter density, most likely in an extremely subtle way.  This is already hinted at in the ``continuity" from the unitarity limit at low density to the dilaton-limit fixed point at high density.  A case recently discussed is the problem of quenched $g_A$ in nuclear Gamow--Teller transitions~\cite{gA-MR}.

One of the longest-standing puzzles in nuclear physics is that the Gamow--Teller transitions in light nuclei seems to require a  ``universal quenching factor (UQF)"  $q^{UQF}_{\rm sm}\sim 0.75$ to the axial-vector coupling constant $g_A$  to describe the transition in ``simple shell model." This problem has recently been extensively reviewed,  properly updated in connection with neutrinoless double beta decay ($0\nu\beta\beta$) processes relevant for going beyond the Standard Model (BSM)~\cite{gAreviews}.

Briefly stated, the problem is as follows.

Consider the nuclear Gamow--Teller transition, say,  from a parent  state of $J^\pi=0^+, T=0$ to a daughter state of $J^\pi=1^+, T=1$. At zero momentum transfer, it is natural to expect that the process will be dominated by the single-particle Gamow--Teller operator $\sum_i \tau^{\pm}_i\sigma_i$ multiplied by the axial--vector coupling constant $g_A$. In the ``extreme shell model" description, the coupling constant must be ``renormalized" by both QCD effects (QCDe) and strong nuclear-correlation effects (SNCe). Since the current involved in nuclear physics is an effective current arising from QCD, the former would represent an {\it intrinsic} effect inherited from QCD in going to an effective field theory. The latter is given in higher-order many-body calculations in the class of S$\chi$EFT (without IDDs) and also in that of $Gn$EFT (with IDDs).  With $q_{\rm sm}=q^{UQF}_{\rm sm}\approx 0.75$, the effective $g_A$, denoted $g_A^{\rm eff}$,  comes out to be
\be
g_A^{\rm eff} &=&  q_{\rm sm} g_A\approx 0.75 \times 1.27\approx 1.
\ee

To see how this intriguing result comes, write
\be
q_{\rm sm}=1+\delta q_{\rm sm}^{QCDe} +\delta q_{\rm sm}^{SNCe}
\ee
corresponding to the quenching due to QCDe and that due to SNCe.
Drawing from the arguments developed in \cite{br91}, it was concluded in \cite{gA-MR} that
\be
  \delta q^{QCDe}_{\rm sm}\approx 0\label{QCDe}.
 \ee
 There are two elements involved in arriving at this result. First is the goodness of the LOSS approximation combined with that the nucleon coupling to the weak current is scale-invariant. There is a  possible caveat in the LOSS approximation which is connected to the role of $\beta^\prime$ at low density. This is an important  issue  to which we return below. The other element is the role of meson-exchange currents which figure in S$\chi$EFT (and also in $Gn$EFT). The key point of the discussions made in \cite{gA-MR,gA-changchun} is that the meson exchange currents figure at N$^m$LO for $m\geq 3$ in chiral expansion and should be strongly, if not completely, suppressed at zero momentum transfer.

Now how to calculate $\delta q_{\rm sm}^{SNCe}$? This is described in detail in \cite{gA-changchun}. When approached from Landau Fermi-liquid theory, what corresponds to the quenching factor in pure shell model, $\delta q_{\rm sm}^{SNCe}$, can be evaluated at the Fermi-liquid fixed point. It was found to be  $\delta q_{\rm sm}^{SNCe}\approx -0.25$ at nuclear matter density,  giving $q_{\rm sm}\approx 0.75$~ \cite{gA-MR,gA-changchun} . Although evaluated at the fixed point, i.e., at the nuclear matter equilibrium density, the result should be valid in finite nuclei since the result is more or less independent of the density in the vicinity of nuclear matter density .

What follows from the arguments given above is highly significant in that what is claimed to be the resolution of the quenched $g_A$ made with the  meson-exchange currents to N$^3$LO in chiral expansion -- heralded as a ``first-principle" approach -- in \cite{firstprinciple} is thrown in doubt.

There is however a caveat which is more of fundamental nature. A highly significant point, unrecognized in nuclear physics community, in connection with this observation is the role of emergent scale symmetry in nuclei. Following \cite{CT}, scale-chiral symmetry implies that $\beta^\prime$ should figure in the axial current as
 \be
 g_A\left[c_A +(1-c_A)\left(\frac{\chi}{f_\chi}\right)^{\beta^\prime}\right] \bar{\psi}\tau^\pm \gamma_\mu\gamma_5\psi
 \ee
 with $c_A$ an arbitrary constant that cannot be fixed by theory.  The LOSS approximation amounts to setting either $c_A\approx 1$ or $\beta^\prime\approx 0$.  There is no known theoretical argument as in the case of the hWZ term discussed in Appendix \ref{appendix} for why the LOSS approximation should be good in this case. Any deviation from the LOSS could affect $q_{\rm sm}^{QCDe}$ both in magnitude and in density dependence.

 On the other hand, if there were no or little deviation, then the ``continuity" in emergent scale symmetry mentioned in Conclusion would be vindicated since in the dilaton-limit fixed point $\gsim 25 n_0$ at which $g_A\to 1$ would be concordant with the vector-manifestation fixed point.

\acknowledgments

\label{ACK}

We are deeply grateful for discussions with Rod Crewther, Lewis Tunstall, and  Koichi Yamawaki on scale symmetry in QCD and with Maciek Nowak and Ismail Zahed on the Cheshire Cat phenomena. The earlier part of the review is based on a series of works done with Tom Kuo, Hyun Kyu Lee and Won-Gi Paeng,  whose invaluable collaboration is acknowledged. Y.~L. Ma was supported in part by National Science Foundation of China (NSFC) under Grant No. 11875147 and 11475071.



\end{document}